\documentclass[twocolumn,english,aps,showkeys,showpacs,eqsecnum,superscriptaddress]{revtex4}
\usepackage[T1]{fontenc}
\usepackage{color}
\usepackage{babel}
\usepackage{textcomp}
\usepackage{amsmath}
\usepackage{amssymb}
\usepackage{graphicx}
\usepackage[unicode=true,
 bookmarks=true,bookmarksnumbered=false,bookmarksopen=false,
 breaklinks=true,pdfborder={0 0 0},backref=false,colorlinks=true]
 {hyperref}
\hypersetup{
 pdfauthor={Francois Konschelle},
 citecolor=blue, linkcolor=magenta,urlcolor=blue}

\makeatletter
\@ifundefined{textcolor}{}
{%
 \definecolor{BLACK}{gray}{0}
 \definecolor{WHITE}{gray}{1}
 \definecolor{RED}{rgb}{1,0,0}
 \definecolor{GREEN}{rgb}{0,1,0}
 \definecolor{BLUE}{rgb}{0,0,1}
 \definecolor{CYAN}{cmyk}{1,0,0,0}
 \definecolor{MAGENTA}{cmyk}{0,1,0,0}
 \definecolor{YELLOW}{cmyk}{0,0,1,0}
}

\usepackage{eucal}

\DeclareMathOperator{\Tr}{Tr}
\DeclareMathOperator{\sgn}{sgn}

\DeclareMathOperator{\arcsinh}{arcsinh}

\def\i{{\mathbf{i}}}
\def\rv{{\bf r}}

\def\hv{{\bf h}}

\def\Sv{{\bf S}}

\def\Fcal{{\mathcal F}}
\def\Acal{{\mathcal A}}

\makeatother

\begin{document}

\title{Theory of the spin-galvanic effect and the anomalous phase-shift
$\varphi_{0}$ in superconductors and Josephson junctions with intrinsic
spin-orbit coupling}

\author{Fran\c{c}ois Konschelle}

\affiliation{Centro de F\'{i}sica de Materiales (CFM-MPC), Centro Mixto CSIC-UPV/EHU,
Manuel de Lardizabal 4, E-20018 San Sebasti\'{a}n, Spain}

\author{Ilya V. Tokatly}

\affiliation{Nano-Bio Spectroscopy group, Dpto. F\'{i}sica de Materiales, Universidad
del Pa\'{i}s Vasco, Av. Tolosa 72, E-20018 San Sebasti\'{a}n, Spain }

\affiliation{IKERBASQUE, Basque Foundation for Science, E-48011 Bilbao, Spain}

\author{F. Sebasti\'{a}n Bergeret}

\affiliation{Centro de F\'{i}sica de Materiales (CFM-MPC), Centro Mixto CSIC-UPV/EHU,
Manuel de Lardizabal 4, E-20018 San Sebasti\'{a}n, Spain}

\affiliation{Donostia International Physics Center (DIPC), Manuel de Lardizabal
5, E-20018 San Sebasti\'{a}n, Spain}
\begin{abstract}
Due to the spin-orbit coupling (SOC) an electric current flowing in
a normal metal or semiconductor can induce a bulk magnetic moment.
This effect is known as the Edelstein (EE) or magneto-electric effect.
Similarly, in a bulk superconductor a phase gradient may create a
finite spin density. The inverse effect, also known as the spin-galvanic
effect, corresponds to the creation of a supercurrent by an equilibrium
spin polarization. Here, by exploiting the analogy between a linear-in-momentum
SOC and a background SU(2) gauge field, we develop a quasiclassical
transport theory to deal with magneto-electric effects in superconducting
structures. For bulk superconductors this approach allows us to easily
reproduce and generalize a number of previously known results. For
Josephson junctions we establish a direct connection between the inverse
EE and the appearance of an anomalous phase-shift $\varphi_{0}$ in
the current-phase relation. In particular we show that $\varphi_{0}$
is proportional to the equilibrium spin-current in the weak link.
We also argue that our results are valid generically, beyond the particular
case of linear-in-momentum SOC. The magneto-electric effects discussed
in this study may find applications in the emerging field of coherent
spintronics with superconductors. 
\end{abstract}



\date{July 31, 2015}

\maketitle

\section{Introduction}

Over the past decades, superconductor-ferromagnetic (S-F) structures
have been studied extensively \cite{buzdin.2005_RMP,Bergeret2005}.
The spatial oscillatory behavior of the superconducting condensate
induced in the ferromagnet leads to interesting effects as oscillations
of the density-of-state in F/S \cite{SanGiorgio2008,Boden2011,Kontos2001}
and F/S/F \cite{Cottet2007} structures, oscillations of the Josephson
current in S/F/S Josephson junction \cite{Bulaevskii1977,Buzdin1982,Ryazanov2001,Kontos2002},
and oscillations of the critical temperature \cite{buzdin.2005_RMP}.
Moreover, in the case of multi-domain ferromagnets or artificial multilayer
structures with inhomogeneous magnetization, the singlet Cooper pairs
from a superconductor can be transform into long-range triplet pairs
that may explain the long-range Josephson coupling observed in S/F/S
structures \cite{Bergeret2001,Bergeret2001a,Robinson2010,Khaire2010,Anwar2010,Anwar2012a,Usman2011,Klose2012,Gingrich2012,Witt2012,Robinson2012a,Kalcheim2012,Kalcheim2014,Alidoust2015b,Banerjee2014}.
Triplet correlations also leads to a dependence of the critical current
on the magnetic configuration of diverse S/F structures \cite{Leksin2012,Banerjee2014a,Wang2014d,Chiodi2013,Gu2014,Jara2014,Rusanov2004,Rusanov2006}.
Such phenomena suggest interesting perspectives of exploiting triplet
correlations for the emerging field of coherent super-spintronics
\cite{Eschrig2011,Linder2015}. Also promising applications might
be found by using superconducting materials in combination with ferromagnetic
insulators that may act as spin-filters \cite{Bergeret2012,Pal2014,Massarotti2015}.
In particular several thermal effects related to these material combination
have been studied in recent works \cite{Bergeret2013a,Bergeret2014b,Ozaeta2014a,Giazotto2014a,Giazotto2014,Nasti2015}. 

All the above mentioned phenomena in S/F structures originate from
the interaction between the superconducting correlations and the exchange
field of the ferromagnet. However it has recently been shown that
 spin-orbit coupling (SOC) in S/F structures will also lead to, for
example, a long-range triplet component \cite{Bergeret2013,Bergeret2014}
and peculiarities in the density of states \cite{Jacobsen2015,Arjoranta2015,Alidoust2015a}.
On the other hand, transport properties of non-superconducting structures
with strong SOC are being intensively studied because of their potential
application in a novel direction of spintronics, which exploits the
coupling between spin and charge currents \cite{Maekawa2012,Kuschel2014,Manchon2014,Ciccarelli2014}. 

In particular, the SOC in semiconductors and normal metals is at the
root of a number of interesting phenomena that originate from the
coupling between the charge and spin degrees of freedom. Prototype
of these phenomena is the spin Hall effect (SHE) \cite{Dyakonov1971,Dyakonov1971a,Chazalviel1975,Hirsch1999,Dyakonov2008,Mishchenko2004,Kato2004,Kato2004a,Wunderlich2005,Raimondi2006,Raimondi2011}
which consists in the creation of a spin polarized current by an electric
field. Reciprocally, by means of the inverse SHE a spin current can
create an electric field \cite{Valenzuela2006,Morota2010,Isasa2014}.
These effects allow to generate and detect spin polarized currents
in non-magnetic materials \cite{Saitoh2006,Ando2008,Kimura2007,Uchida2010,Kajiwara2010}.

There is also another relevant effect in normal systems related to
the SOC. It consists in creating a stationary spin density $S^{a}$,
along the $a$-direction in response to an electric field $E_{k}$
applied in $k$-direction \cite{Aronov1989,Edelstein1990}. Within
linear response, this effect is described by 
\begin{equation}
S^{a}\left(\omega\right)=\sigma_{k}^{a}\left(\omega\right)E_{k}\left(\omega\right)\;,\;\label{eq_EE}
\end{equation}
where the sum over repeated indexes is implied here, and throughout
this paper. In particular, in 2D systems with Rashba SOC, the applied
electric field and the generated spin density are perpendicular to
each other. This magneto-electric effect, also called the \textit{Edelstein
effect} (EE), has been observed in experiments \cite{Kato2004,Silov2004}.
The Edelstein conductivity $\sigma_{k}^{a}\left(\omega\right)$ in
Eq.(\ref{eq_EE}) is related to the Kubo correlator $\chi_{k}^{a}\left(\omega\right)=\left\langle \left\langle \hat{S}^{a};\hat{j}_{k}\right\rangle \right\rangle _{\omega}$
of the spin and current operators via $\sigma_{k}^{a}\left(\omega\right)=\chi_{k}^{a}\left(\omega\right)/\mathbf{i}\omega$
\cite{Shen2014a}. Because of the gauge invariance in normal systems
the function $\chi_{k}^{a}\left(\omega\right)$ should vanish in the
limit $\omega\to0$ reflecting the fact that there is no response
to a static vector potential. Therefore the $\sigma_{k}^{a}\left(0\right)=\sigma_{k}^{a}$
remains finite and describes the dc EE. It has been pointed out in
Ref.\cite{Shen2014a} that this property, together with the Onsager
reciprocity principle, implies that the inverse dc EE, also referred
to as the spin-galvanic effect, consists in generating a charge current
$j_{k}$ by a steady spin generation induced by a time-dependent magnetic
field via the paramagnetic effect: 
\begin{equation}
j_{k}=\sigma_{k}^{a}\left[g\mu_{B}\dot{B}^{a}\right]\;,\label{eq_iEE}
\end{equation}
with the Land\'{e} $g$-factor, $\mu_{B}$ the Bohr magneton, and
$\dot{B}^{a}$ the time derivative of the magnetic field along the
$a$-axis. The inverse EE effect has also been observed in experiments
\cite{Ganichev2002,Sanchez2013}.

Similar magneto-electric and spin-galvanic effects should also exist
in superconductors \cite{Edelstein1995,Edelstein2005}. However, there
the physical situation is different because in the presence of the
superconducting condensate the gauge invariance does not forbid the
existence of a finite static current-spin response function $\chi_{k}^{a}$.
In contrast to the normal case, in a superconductor an equilibrium
electric (super-)current can flow in the absence of an external electric
field. The supercurrent $\boldsymbol{j}=n_{s}\mathbf{v}_{s}$ (here
$n_{s}$ is the density of superconducting electrons and $\mathbf{v}_{s}$
the superfluid velocity) is proportional to the gradient of the macroscopic
gauge-invariant phase $\nabla\tilde{\varphi}=\nabla\varphi-e{\bf A}\sim\mathbf{v}_{s}$,
which is the physical field coupled to the current operator in the
Hamiltonian of a superconductor. The existence of such a gauge-invariant
field implies that the static response function $\chi_{k}^{a}=\left\langle \left\langle \hat{S}^{a};\hat{j}_{k}\right\rangle \right\rangle _{\omega=0}$
can be nonzero without violating the gauge invariance. In principle,
a supercurrent can thus generate an equilibrium spin polarization
according to the general linear response relation: 
\begin{equation}
S^{a}=\chi_{k}^{a}\partial_{k}\tilde{\varphi}\;,\label{eq_SEE1}
\end{equation}
where $\partial_{k}=\partial/\partial x_{k}$. This effect has been
indeed theoretically demonstrated by Edelstein for a 2D superconductor
with Rashba SOC, who calculated the proportionality tensor $\chi_{k}^{a}$
at temperatures $T$ close to the critical superconducting temperature
$T_{c}$, in both pure ballistic \cite{Edelstein1995} and diffusive
\cite{Edelstein2005} limits.

Because in the superconducting state the response function $\chi_{k}^{a}$
at $\omega\to0$ is finite, the reciprocity of the EE effect becomes
complete. In contrast to the normal case, in superconductors a static
Zeeman field $\mathbf{B}$ can induce a supercurrent $j_{k}$. Therefore,
instead of Eq.~(\ref{eq_iEE}) the following relation holds 
\begin{equation}
j_{k}=e\chi_{k}^{a}h^{a}\;,\label{eq_SEE2}
\end{equation}
where $h^{a}=g\mu_{B}B^{a}$. An explicit expression of this type
has been obtained in a particular case of a 2D ballistic superconductor
with intrinsic Rashba SOC \cite{Yip2001,Bauer2012a}.

It is then clear that the free energy of a superconductor with a SOC
must have a term of the Lifshitz-type 
\begin{equation}
F_{L}=h^{a}\chi_{k}^{a}\partial_{k}\tilde{\varphi},\label{Lifshitz-gen}
\end{equation}
and equations (\ref{eq_SEE1})-(\ref{eq_SEE2}) follow directly from
the general thermodynamic definitions of the spin and current densities,
$\boldsymbol{S}=\delta F/\delta\mathbf{h}$ and $\boldsymbol{j}=-\delta F/\delta\mathbf{A}$.

In principle, equations (\ref{eq_SEE1}) and (\ref{eq_SEE2}) apply
for bulk superconductors, but one can expect similar effects to occur
also in a S-X-S Josephson junction, between two massive superconductors
(S) and a normal or ferromagnetic bridge X with an intrinsic SOC.
In a Josephson junction the supercurrent depends on the phase difference
$\varphi$ between the superconducting electrodes. In the particular
cases of a weak proximity effect between the S and the X, or in the
high-temperature regime ($T\lesssim T_{c}$), the current phase relation
is given by $j=j_{c}\sin\varphi$, where $j_{c}$ is the critical
Josephson current. 

When the SOC competes with a Zeeman effect, the natural conjectures
following Eqs.~(\ref{eq_SEE1})-(\ref{eq_SEE2}) are: (i) In accordance
with Eq.~(\ref{eq_SEE1}), the flow of a supercurrent may generate
a spin polarization in the X bridge (the Edelstein effect); (ii) In
turn, from Eq. (\ref{eq_SEE2}), a Zeeman (spin-splitting) field may
induce a supercurrent through the junction, even if the phase difference
between the electrodes vanishes (the inverse Edelstein effect).

In other words, the inverse EE is presumably the cause of an anomalous
phase $\varphi_{0}$ which modifies the current phase relation according
to $j=j_{c}\sin\left(\varphi-\varphi_{0}\right)$, with a non-trivial
(i.e., non equal to $0$ or $\pi$) equilibrium phase $\varphi_{0}$.
This defines the so called $\varphi_{0}$-junctions, a subject that
has been extensively studied in the past years in different systems,
including conventional superconductors with SOC \cite{krive_kadigrobov_shekhter_jonson.2005,buzdin:107005.2008,Liu2010,Malshukov2010,Reynoso2012,Yokoyama2012,Yokoyama2014,Yokoyama2014b,Campagnano2014,Bergeret2014a,Konschelle2014a,Kulagina2014,Mironov2014},
with triplet correlations \cite{Braude2007,Tanaka2007,Eschrig2007,Grein2009,Liu2010a,Margaris2010a}
or in contact with topological materials \cite{Tanaka2009,Dolcini2015},
and also in hybrid systems with non-conventional superconductors \cite{Geshkenbeim1986,Yip1995,Tanaka1997,Sigrist1998,Kashiwaya2000,Asano2003,Asano2005,Brydon2008,Rahnavard2014},
quantum dots \cite{reynoso_etal:107001.2008,Zazunov2009,Brunetti2013},
and hybrid $\left(0-\pi\right)$-structures \cite{buzdin_koshelev.2003,goldobin_koelle_kleiner_buzdin.2007,Goldobin2011,Sickinger2012}.
$\varphi_{0}$-junctions may produce a self-sustained flux when embedded
in a SQUID geometry \cite{Mints1998}, act as phase batteries in coherent
circuits \cite{buzdin.2005.J,buzdin.2005}, present a current asymmetry
and act as a supercurrent rectifiers \cite{reynoso_etal:107001.2008}.

In the present work we develop a complete theory of the magneto-electric
and spin-galvanic effects in hybrid superconducting structures and
confirm the above conjectures. We focus on systems with linear in
momentum SOC that can be conveniently described in terms of an effective
background SU(2) gauge field. This allows us to use the SU(2) covariant
quasiclassical equations for the Green's functions derived in Ref.\cite{Bergeret2013,Bergeret2014,Konschelle2014}.
We establish a connection between the tensor $\chi_{k}^{a}$ in Eqs.
(\ref{eq_SEE1}-\ref{eq_SEE2}) and the equilibrium spin-current ${\cal J}_{j}^{a}$
\cite{Rashba2003,Tokatly2008b}. We show that in a generic S-X-S Josephson
junction the condition for a nontrivial anomalous phase $\varphi_{0}$
to appear is that ${\cal J}_{j}^{a}h^{a}\neq0$, where $h^{a}$ can
be either an external Zeeman field or the internal exchange field
of a ferromagnet. Our SU(2) covariant formulation results in a simple
and tractable system of equations to describe hybrid structures with
arbitrary linear in momentum SOC, temperatures, degree of disorder,
and quality of the hybrid interfaces. We also show that qualitatively
our results are generically valid beyond the particular case of the
linear in momentum SOC.

The structure of the paper is the following: In the next section we
present a qualitative discussion of the superconducting proximity
effect in structures with SOC and its connection with the spin diffusion
in normal systems. This qualitative analysis allows us to guess the
form of the quasiclassical equations for superconducting structures
in the presence of generic spin fields, and in particular to explicitly
show the analogy between the charge-spin coupling in normal systems
and the singlet-triplet coupling in superconducting ones. In section
\ref{sec:Model} we present our model, discuss the associated symmetries,
and derive microscopically the quasiclassical equations for generic
linear in momentum SOC. In section \ref{sec:Edelstein-effect} we
use the derived equations to explore the magneto-electric effects
in bulk superconductors. We generalize the previously known results
for the EE and its inverse obtained for 2D Rashba SOC \cite{Edelstein1995,Edelstein2005,Yip2001,Agterberg2006}
to generic linear-in-momentum SOC, and relate them to the spin current
and the SU(2) gauge-fields. In section \ref{sec:Diffusive-limit}
we explore the Josephson effect through a S-X-S diffusive junction
and in section \ref{sec:Ball-limit} through a ballistic one. In both
cases we show that the anomalous phase $\varphi_{0}$ is proportional
to $\mathcal{J}_{i}^{a}h^{a}$ and determine its dependence on other
parameters of the structure, like temperature and length. We finally
present our conclusions and discuss possible experimental setups to
verify our predictions in Section \ref{sec:Discussion-and-Conclusions}.

\section{Diffusion of superconducting condensate in the presence of spin-orbit
coupling: Heuristic arguments\label{sec:Diffusion-of-superconducting}}

Before presenting the full quantum kinetic theory it is instructive
to discuss at the qualitative level the main features of the proximity
induced superconductivity in the presence of an intrinsic SOC. For
this sake we present a simple heuristic derivation of the equations
describing the coupled motion of the singlet and triplet components
induced in a ferromagnet from a bulk $s$-wave superconductor.

Let us consider a S-X-S junction, where X is a diffusive ferromagnet.
We assume that the system is at equilibrium, and that the proximity
effect between S and X is weak. In such a case the junction is fully
described by the quasiclassical anomalous Green function $\hat{f}(\rv)$,
which describes the superconducting condensate in X. In general $\hat{f}(\rv)$
is a $2\times2$ matrix in the spin space $\hat{f}=f_{s}\hat{1}+f_{t}^{a}\sigma^{a}$.
Here the scalar $f_{s}$ and the vector with components $f_{t}^{a}$
describe the singlet and the triplet components of the condensate,
respectively. In this section we show, that the functions $f_{s}(\rv)$
and ${\bf f}_{t}(\rv)$ are reminiscent of the charge and spin density
in the normal systems.

In the absence of SOC, but in the presence of the exchange field $\hv$
the diffusion of the condensate is described by the well known linearized
Usadel equations (see e.g. Ref.~\cite{Bergeret2005}), 
\begin{eqnarray}
 &  & D\nabla^{2}f_{s}-2\left|\omega_{n}\right|f_{s}+2\i\sgn(\omega_{n})h^{a}f_{t}^{a}=0\;,\label{Usadel-s1}\\
 &  & D\nabla^{2}f_{t}^{a}-2\left|\omega_{n}\right|f_{t}^{a}+2\i\sgn(\omega_{n})h^{a}f_{s}=0\;,\label{Usadel-t1}
\end{eqnarray}
where $D$ is the diffusion constant, and $\omega_{n}$ is the Matsubara
frequency. The terms proportional to $2\left|\omega_{n}\right|$ are
responsible for the decay of the superconducting correlations in the
normal metal. The last terms in the left hand sides of Eqs.~(\ref{Usadel-s1})-(\ref{Usadel-t1})
describe the usual singlet-triplet coupling coming from the exchange
field. It is worth emphasizing the presence of imaginary unit $\i$
in the exchange field terms, which reflects the breaking of the time
reversal symmetry. Because of this, the singlet-triplet conversion
due to the exchange field is always accompanied with a phase shift
of $\pi/2$. This point will be of primary importance in the following
for understanding the origin of the anomalous phase $\varphi_{0}$.

To understand how the Usadel equations (\ref{Usadel-s1})-(\ref{Usadel-t1})
are modified in the presence of SOC we recall the description of the
diffusion of spin $\Sv(\rv)$ and charge $n(\rv)$ densities in normal
systems. The general spin diffusion equation in a normal conductor
with SOC takes the form, 
\begin{equation}
\partial_{t}S^{a}-D\nabla^{2}S^{a}=\mathcal{T}^{a},\label{S-diffusion1}
\end{equation}
where $\mathcal{T}^{a}$ is a so called spin torque. In the absence
of SOC, $\mathcal{T}^{a}=0$ and hence spin is a conserved quantity
which satisfies the usual spin diffusion equation. In non-centrosymmetric
materials SOC acts as an effective momentum-dependent Zeeman field
that causes precession of spins of \textit{moving} electrons. This
precession breaks conservation of the average spin, and shows up formally
as a finite torque $\mathcal{T}^{a}\ne0$ in Eq.~(\ref{S-diffusion1}).
In the diffusive regime the motion of the electrons consists of a
random motion superimposed on an average drift caused by the density
gradients. The spin precession related to these types of motion generate
the corresponding contributions to the spin torque. To the lowest
order in gradients the general expression for the torque can be written
as follows \cite{Malshukov2005,Stanescu2007,Duckheim2009}, 
\begin{equation}
\mathcal{T}^{a}=D\left[-\Gamma^{ab}S^{b}+2P_{k}^{ab}\partial_{k}S^{b}+C_{k}^{a}\partial_{k}n\right].\label{torque}
\end{equation}
Here the first term describes the Dyakonov-Perel (DP) spin relaxation
that originates from the spin precession of randomly moving electrons
\cite{Dyakonov1971}. The positive definite matrix $\Gamma^{ab}$
is the DP relaxation tensor with the eigenvalues equal to the inverse
squares of the DP spin relaxation lengths. The other two contributions
to the torque are related to the average motion of spins. In particular,
the second term in the right hand side of Eq.~(\ref{torque}) originates
from the diffusive motion of spins caused by inhomogeneities of the
spin density distribution. The corresponding spin precession is described
by antisymmetric (spin rotation) matrices $P_{k}^{ab}=-P_{k}^{ba}$
with $\|\hat{P}\|\sim1/\ell_{{\rm so}}$, where $\ell_{{\rm so}}$
is the spin precession length.

The last term in Eq.~(\ref{torque}), which is proportional to the
charge density gradient, can be called the spin-Hall torque. The charge
density gradient generates the charge current which is then transformed
to the spin current via the spin Hall effect. Precession of the spins
driven by the charge density gradient, via the spin Hall effect, is
the origin of the spin-Hall torque in Eq.~(\ref{torque}). The spin-Hall
torque is parameterized by the tensor $C_{k}^{a}$ which is proportional
to $\theta_{{\rm sH}}/\ell_{{\rm so}}$, where $\theta_{{\rm sH}}$
is the spin Hall angle -- the conversion coefficient between the charge
and the spin currents.

Equation (\ref{S-diffusion1}) with the spin torque of Eq.~(\ref{torque})
is commonly used in spintronics context to describe spin dynamics
in semiconductors with intrinsic SOC \cite{Malshukov2005,Stanescu2007,Duckheim2009}
(for a discussion between intrinsic and extrinsic SOC, see \textit{e.g.}
\cite{Raimondi2011}). In the stationary case the diffusion equations
for the spin and charge densities reduce to 
\begin{eqnarray}
 &  & \nabla^{2}n+C_{k}^{a}\partial_{k}S^{a}=0,\label{C-diffusion2}\\
 &  & \nabla^{2}S^{a}-\Gamma^{ab}S^{b}+2P_{k}^{ab}\partial_{k}S^{b}+C_{k}^{a}\partial_{k}n=0.\label{S-diffusion2}
\end{eqnarray}
It is important to emphasize here that spin-charge coupling mediated
by the spin-Hall torque ($C_{k}^{a}$) is responsible for the EE.
This can be seen directly from Eq.~(\ref{S-diffusion2}): A uniform
charge density gradient produces a uniform spin density given by $S^{a}=(\hat{\Gamma}^{-1})^{ab}C_{k}^{b}\partial_{k}n$.

We can now construct the Usadel equations in the presence of SOC in
analogy to the normal case. Since SOC does not violate the time reversal
symmetry it acts in exactly the same way on the time-reversal conjugated
states composing the Cooper pair. Therefore the diffusion of the singlet
and the triplet condensates should be modified by SOC in complete
analogy with the diffusion of the charge and spin densities in normal
systems. The formal connection between the diffusion of the triplet
condensate function $f_{t}^{a}$ in superconductors and the spin density
$S^{a}$ in normal metals has been discussed recently in Ref.~\cite{Bergeret2014},
and it has been also noticed in Ref.~\cite{Malshukov2010}. Hence,
in order to include the effects of SOC in the Usadel equations all
we need to do is to replace the diffusion operators (the Laplacians)
in Eqs.~(\ref{Usadel-s1}) and (\ref{Usadel-t1}) with the diffusion
operators entering Eqs.~(\ref{C-diffusion2}) and (\ref{S-diffusion2}),
respectively. The result is the following system of equations describing
a coupled diffusion of the singlet and triplet condensates in the
presence of SOC, 
\begin{eqnarray}
\nabla^{2}f_{s} & - & \kappa_{\omega}^{2}f_{s}+\sgn\left(\omega_{n}\right)\Big[\i\frac{2h^{a}}{D}f_{t}^{a}+C_{k}^{a}\partial_{k}f_{t}^{a}\Big]=0\label{Usadel-s2}\\
\nabla^{2}f_{t}^{a} & - & \big(\kappa_{\omega}^{2}\delta^{ab}+\Gamma^{ab}\big)f_{t}^{b}+2P_{k}^{ab}\partial_{k}f_{t}^{b}\nonumber \\
 & + & \sgn\left(\omega_{n}\right)\Big[\i\frac{2h^{a}}{D}f_{s}+C_{k}^{a}\partial_{k}f_{s}\Big]=0.\label{Usadel-t2}
\end{eqnarray}
In contrast to the normal case, in addition to the DP relaxation,
both the $f_{s}$ and ${\bf f}_{t}$ experience an additional decay
proportional to the inverse decay length $\kappa_{\omega}=\sqrt{2|\omega_{n}|/D}$,
due to the finite lifetime of the superconducting condensate in the
normal metal.

The most important novel feature of Eqs.~(\ref{Usadel-s2})-(\ref{Usadel-t2})
is the presence of two mechanisms for the singlet-triplet coupling
which are described by the two terms in the square brackets. The first
mechanism is the above discussed Zeeman coupling related to the modification
of the internal structure of the Cooper pair by the spin-splitting
field $\mathbf{h}$ {[}see Eqs.~(\ref{Usadel-s1})-(\ref{Usadel-t1}){]}.
The second channel of singlet-triplet coupling comes from the spin-Hall
torque, which converts the gradient of $f_{s}$ into ${\bf f}_{t}$
and vise versa, in a complete analogy with the EE in normal systems.
The corresponding singlet-triplet ``conversion amplitudes'' have
a relative phase shift of $\pi/2$, which is related to the different
transformation properties of the Zeeman and spin-orbit fields with
respect to the time reversal. We will see in the next sections that
the interference of these two singlet-triplet conversion channels
is indeed responsible for the magneto-electric/spin-galvanic effects
in superconductors, and, in particular, for the appearance of the
intrinsic anomalous phase $\varphi_{0}$ in Josephson junctions.

Although the present heuristic derivation of Eqs.~(\ref{Usadel-s2})-(\ref{Usadel-t2})
may seem imprecise, it uncovers a simple, but deep connection between
the physics of inhomogeneous superconductors with SOC and the well
known spintronics effects, such as the spin Hall effects and direct
and inverse magneto-electric effects (EE). In Sec.~\ref{sec:Model}
we present a rigorous derivation of the quasiclassical kinetic equations
for superconductors with a linear in momentum SOC, which in the diffusive
limit confirms the correctness of Eqs.~(\ref{Usadel-s2})-(\ref{Usadel-t2}).
In the rest of the article we study in detail the physical consequences
of the interference of the two singlet-triplet conversion channels
and their connection with the theory of $\varphi_{0}$-Josephson junctions.

\section{The model and basic equations\label{sec:Model}}

In this section we introduce our model and discuss the symmetries
associated with superconducting systems in the presence of spin-orbit
coupling (SOC). We also present the derivation of the quasiclassical
equations in the presence of linear in momentum SOC.

\subsection{The Hamiltonian in the presence of generic SOC and symmetry arguments
for the appearance of an anomalous phase}

Our starting point is a general Hamiltonian describing a metal or
a semiconductor with a linear in momentum SOC, an exchange field and
superconducting correlations 
\begin{equation}
H=\int d{\bf r}\left[{\bf \Psi}^{\dagger}H_{\text{0}}{\bf \Psi}+V\psi_{\uparrow}^{\dagger}\psi_{\downarrow}^{\dagger}\psi_{\downarrow}\psi_{\uparrow}\right]\;,\label{hamiltonian}
\end{equation}
where $\psi_{\uparrow,\downarrow}\left(\mathbf{r}\right)$ are the
annihilation operators for spin up and down at position $\mathbf{r}$,
and ${\bf \Psi}^{\dagger}=(\psi_{\uparrow}^{\dagger},\psi_{\downarrow}^{\dagger})$
is the spinor of creation operators. $H_{0}$ is the free electron
part%
\footnote{Lower indices ($i$, $j$, $k$...) describe space variables, while
upper indices ($a$, $b$, ...) will describe spin variables.%
} 
\begin{equation}
H_{0}=\dfrac{\left(\hat{p}_{i}-\mathcal{A}_{i}\right)^{2}}{2m}-\mu+\mathcal{A}_{0}+V_{\text{imp}}\;,\label{eq:H}
\end{equation}
where $\mu$ the chemical potential and $V_{\text{imp}}$ the potential
induced by non-magnetic impurities. The magnetic interactions appear
in two places: as a SU(2) scalar potential $\mathcal{A}_{0}\equiv\mathcal{A}_{0}^{a}\sigma^{a}/2$,
describing for example the intrinsic exchange field in a ferromagnet
or a Zeeman field in a normal metal, and as a SU(2) vector potential
$\mathcal{A}_{i}\equiv\mathcal{A}_{i}^{a}\sigma^{a}/2$, describing
the SOC. The latter is associated to the momentum operator%
\footnote{We use the units where the Planck's constant $\hbar=1$ and the Boltzmann's
constant $k_{B}=1$.%
} $\hat{p}_{i}=-\mathbf{i}\partial_{i}$ in the form of a minimal coupling
$\hat{p}_{i}-\mathcal{A}_{i}$. In practice, all the linear-in-momentum
SOC can be represented as a gauge potential (see \textit{e.g.} \cite{Frohlich1993}
or \cite{Berche2013} and references therein). In the widely studied
case of a free electron gas with Rashba SOC, $\mathcal{A}_{x}^{y}=-\mathcal{A}_{y}^{x}=-\alpha$,
while in the presence of Dresselhaus SOC \emph{$\mathcal{A}_{x}^{x}=-\mathcal{A}_{y}^{y}=\beta$}.
Finally, $V=V\left({\bf r}\right)<0$ in the second term of the r.h.s
of Eq. (\ref{hamiltonian}) describes the coupling strength which
gives rise to superconductivity in some regions of space.

In analogy to electrodynamics one can define the four-potential $\mathcal{A}_{\mu}$,
with space components ($\mu=1,2,3$ or $\mu=x,y,z$) given by the
SOC and the time component ($\mu=0$) by the Zeeman field. Following
the analogy one can define the strength tensor 
\begin{equation}
\mathcal{F}_{\mu\nu}=\frac{1}{2}\mathcal{F}_{\mu\nu}^{a}\sigma^{a}=\partial_{\mu}\mathcal{A}_{\nu}-\partial_{\nu}\mathcal{A}_{\mu}-\mathbf{i}[\mathcal{A}_{\mu},\mathcal{A}_{\nu}],\label{field-tensor}
\end{equation}
and the electric and magnetic SU(2) fields 
\begin{equation}
\mathcal{E}_{k}^{a}=\mathcal{F}_{0k}^{a}\;\text{and}\;\mathcal{B}_{i}^{a}=\varepsilon_{ijk}\mathcal{F}_{jk}^{a}\;,\label{su2fields}
\end{equation}
where $\varepsilon_{ijk}$ is the Levi-Civitta symbol.

In normal metals and semiconductors, the SHE and EE are consequences
of the existence of a finite SU(2) magnetic field. For a pure-gauge
vector potential the SOC can be gauged out \cite{Bergeret2014}, the
SU(2) magnetic field is zero, and hence the SHE and EE do not appear
\footnote{Note that, in a one dimensional system described by the Hamiltonian
Eq. (\ref{eq:H}) the gauge-potential is always a pure-gauge and therefore
1D problem should not exhibit any magneto-electric effect. Nevertheless,
the above argument does not apply for topological 1D channel, where
a $\varphi_{0}$-effect has been discussed \cite{Tanaka2009,Dolcini2015},
since the dispersion relation is not quadratic in that case \cite{Mironov2014}.
Also, the problem of spin-active interface is non-trivial when spin-orbit
effect is treated as a gauge-potential, and might eventually lead
to $\varphi_{0}$-effect as well \cite{Grein2009,Liu2010a,Margaris2010a,Kulagina2014}.%
}. Following our analogy, in the superconducting case an anomalous
phase can only appear if the SU(2) magnetic field is finite. This
explains why S-F-S junction without SOC do not present any magneto-electric
effect, or equivalently, no anomalous phase. As it is well known,
the ground state of S-F-S junctions corresponds to a phase difference
either equal to $0$ or to $\pi$ \cite{golubov_kupriyanov.2004,buzdin.2005_RMP,Bergeret2005}.

A simple way to describe qualitatively magneto-electric effects in
a superconductor is to provide simple symmetry arguments. Let us consider
a ballistic superconductor at $T$ close to its critical temperature
$T_{c}$ and focus on the Ginzburg-Landau free energy. In such an
expansion, a SOC is responsible for the presence of a first-order
derivative of the order-parameter, the so-called Lifshitz invariant
which describes most of the original phenomenology of non-centrosymmetric
superconductors \cite{Bauer2012a}. Assuming that the amplitude of
the order parameter is constant but its phase position-dependent,
the Lifshitz invariant reads $F_{L}\propto T_{i}\partial_{i}\varphi$
where $T_{i}$ is a vector which has to be odd with respect to the
time-reversal operation, and SU(2) invariant. As discussed in Ref.
\cite{Bergeret2014a}, to the lowest order in SOC the Lifshitz invariant
for a superconductor can be expressed in terms of the SU(2) fields:
\begin{equation}
F_{L}\propto\Tr\left\{ \mathcal{F}_{0j}\mathcal{F}_{ji}\right\} \partial_{i}\varphi=\left(\mathcal{E}^{a}\times\mathcal{B}^{a}\right)\cdot\nabla\varphi\;.\label{FL1}
\end{equation}
If we focus on the static case, the electric field is given by $\mathcal{F}_{0j}=-\partial_{j}\mathcal{A}_{0}$.
Moreover we define the equilibrium spin current \cite{Tokatly2008b}
in terms of the SU(2) magnetic field as $\mathcal{J}_{j}=\tilde{\nabla}_{i}\mathcal{F}_{ij}=\partial\mathcal{F}_{ij}/\partial x_{i}-\mathbf{i}\left[\mathcal{A}_{i},\mathcal{F}_{ij}\right]$.
If $\mathcal{A}_{0}$ is spatially homogenous, for example induced
by an external magnetic field, Eq. \eqref{FL1} reads \cite{Bergeret2014a}
\begin{equation}
F_{L}\propto\mathcal{A}_{0}^{a}\mathcal{J}_{i}^{a}\partial_{i}\varphi\;.\label{eq:Lifshitz}
\end{equation}
This Lifshitz invariant agrees with the ones derived from microscopic
considerations \cite{Mineev2008} or quasi-classic expansions \cite{Houzet2015}
for a particular sort of SOC.

Eq. (\ref{eq:Lifshitz}) confirms our guessed Eq. (\ref{Lifshitz-gen})
and demonstrates that the Edelstein response tensor $\chi_{k}^{a}$
behaves like the spin current tensor $\mathcal{J}_{i}^{a}$. The form
of $F_{L}$ in Eq. \eqref{eq:Lifshitz}, in terms of the equilibrium
spin current, suggests that our results remain valid for any momentum
dependence of the SOC. We now proceed to derive the quasiclassical
equations and provide a microscopic description of the magneto-electric
effects in superconductors.

\subsection{The quasiclassical equations in the presence of SOC}

In order to describe the transport properties of hybrid structures
containing superconducting, normal (N) and/or ferromagnetic (F) layers
with interfaces, arbitrary temperature and degree of disorder we have
to go beyond the Ginzburg-Landau limit. We present here the quasiclassical
equations \cite{eilenberger.1968,larkin_ovchinnikov.1969,usadel.1970,b.kopnin}
for the Green's functions in the presence of a non-Abelian gauge-field
\cite{Bergeret2013,Bergeret2014,Konschelle2014} (for a similar discussion
in normal metal, see \cite{Gorini2010}).

We follow here the derivation presented in Ref. \cite{Bergeret2014}.
The basic transport equation derived from Hamiltonian \eqref{hamiltonian}
for the Wigner-transformed covariant Green functions $\check{G}\left({\bf p},{\bf r}\right)$
in the time-independent limit reads : 
\begin{multline}
\dfrac{p_{i}}{m}\tilde{\nabla}_{i}\check{G}+\left[\tau_{3}\left(\omega_{n}-\mathbf{i}\mathcal{A}_{0}\right)-\mathbf{i}\check{\Delta}+\dfrac{\left\langle \check{g}\right\rangle }{2\tau},\check{G}\right]\\
-\dfrac{1}{2}\left\{ \tau_{3}\mathcal{F}_{0j}+v_{i}\mathcal{F}_{ij},\dfrac{\partial\check{G}}{\partial p_{j}}\right\} =0\;,\label{eq:transport}
\end{multline}
where $\omega_{n}=2T\pi\left(n+1/2\right)$ is the fermionic Matsubara
frequency, $\check{\Delta}=\Delta\left(\begin{array}{cc}
0 & e^{{\bf i}\varphi}\\
-e^{-{\bf i}\varphi} & 0
\end{array}\right)$ is the ($s$-wave) gap parameter of amplitude $\Delta$ and phase
$\varphi$. The scattering at impurities is described within the Born
approximation, where $\tau$ is the elastic scattering time, $\left\langle \check{g}\right\rangle $
is the GF matrix integrated over the quasiparticle energy, and $\left\langle \cdots\right\rangle $
describes the average over the Fermi momentum direction.

After integration of \eqref{eq:transport} over the quasiparticle
energy and by using the fact that $\check{G}$ is peaked at the Fermi
level one obtains the generalized Eilenberger equation \cite{Bergeret2014,Bergeret2014a}:
\begin{multline}
v_{F}\left(n_{i}\tilde{\nabla}_{i}\right)\check{g}+\left[\tau_{3}\left(\omega_{n}-\mathbf{i}\mathcal{A}_{0}\right)-\mathbf{i}\check{\Delta},\check{g}\right]\\
-\frac{1}{2m}\left\{ n_{i}\mathcal{F}_{ij},\dfrac{\partial\check{g}}{\partial n_{j}}\right\} =-\frac{1}{2\tau}\left[\langle\check{g}\rangle,\check{g}\right]\;,\label{eq:Eilenberger}
\end{multline}
where $n_{i}$, $i=x,y,z$ are the components of the Fermi velocity
vector. When deriving \eqref{eq:Eilenberger} we have neglected corrections
to the exchange term $\mathcal{A}_{0}$ of the order of $\left|\mathcal{A}_{j}\right|/p_{F}\ll1$.
In fact, one sees from \eqref{eq:transport} that $\left\{ \tau_{3}\mathcal{F}_{0j},\partial\check{G}/\partial p_{j}\right\} $
scales like $\left\{ \mathcal{A}_{j}\partial/\partial p_{j},-\mathbf{i}\left[\tau_{3}\mathcal{A}_{0},\check{G}\right]\right\} $
since $\mathcal{F}_{0j}=-\mathbf{i}\left[\mathcal{A}_{0},\mathcal{A}_{j}\right]$,
and so it renormalizes the term $-\mathbf{i}\left[\tau_{3}\mathcal{A}_{0},\check{G}\right]$
already present in \eqref{eq:transport}. The correction to $\mathcal{A}_{0}$
is of the order $\mathcal{A}_{j}/p_{F}\ll1$ and we neglect them from
now on. 

In the Nambu space $\check{g}$ reads 
\begin{equation}
\check{g}=\left(\begin{array}{cc}
g & f\\
-\bar{f} & -\bar{g}
\end{array}\right)\;,\label{eq:g-parameterisation}
\end{equation}
where the $g$, $f$ are matrices in the spin space which depend on
the spaces coordinates $x_{i}$, the momentum direction $n_{i}$ and
the Matsubara frequency. The time-reversal conjugate $\bar{g}$ and
$\bar{f}$ are defined as $\bar{g}({\bf n})=\sigma^{y}g^{*}(-{\bf n})\sigma^{y}$
and $\bar{f}=\sigma^{y}f^{*}(-{\bf n})\sigma^{y}$. The latter is
the anomalous GF which describes the superconducting correlations.

From the knowledge of $\check{g}$ one can calculate the charge current
(density) 
\begin{equation}
\boldsymbol{j}=-\frac{\mathbf{i}\pi eN_{0}T}{2}\sum_{\omega_{n}}\Tr\left\langle \boldsymbol{v_{F}}\tau_{3}\check{g}\right\rangle \quad,\label{eq:obs-current}
\end{equation}
with $e$ the electron charge and $N_{0}$ the normal density of states
for each spin. Whereas the spin polarization is given by 
\begin{equation}
\boldsymbol{S}=\frac{\mathbf{i}\pi N_{0}T}{2}\sum_{\omega_{n}}\Tr\left\langle \tau_{3}\boldsymbol{\sigma}\check{g}\right\rangle \quad.\label{eq:obs-spin-polar}
\end{equation}

\subsection{Linearized quasiclassical equations in diffusive and pure-ballistic
limits}

In the present work we mainly consider two limiting cases: the pure
ballistic one in which $\tau\rightarrow\infty$ and the diffusive
limit where $\tau$ is a small parameter. The transport equation in
the ballistic limit is directly obtained from \eqref{eq:Eilenberger}
by neglecting the right-hand side. The diffusive limit is a bit more
puzzling. Because of the anti-commutator in the l.h.s of Eq. (\ref{eq:Eilenberger}),
the normalization condition $\check{g}^{2}=1$ does not hold directly
and therefore the usual derivation of the Usadel equations cannot
be carried out \cite{Langenberg1986}. There is, however, a way out
of this puzzle if one assumes that the amplitude of the anomalous
GF's, $f$ in \eqref{eq:g-parameterisation} is small. Then the matrix
GF \eqref{eq:g-parameterisation} can be written as $\check{g}\approx\sgn\left(\omega_{n}\right)\tau_{3}+\left(\begin{array}{cc}
0 & f\\
-\bar{f} & 0
\end{array}\right)$ and the linearized Eilenberger equation becomes an equation for $f$
\begin{multline}
\left(v_{F}n_{i}\tilde{\nabla}_{i}+2\omega_{n}\right)f-\left\{ \mathbf{i}\mathcal{A}_{0},f\right\} +2{\bf i}\Delta\sgn\left(\omega_{n}\right)+\\
-\frac{1}{2m}\left\{ n_{i}\mathcal{F}_{ij},\dfrac{\partial f}{\partial n_{j}}\right\} =-\frac{\sgn\left(\omega_{n}\right)}{\tau}\left(f-\langle f\rangle\right)\label{eq:transport-linearised}
\end{multline}
This linearization procedure is justified in two cases: either for
temperatures close to the critical temperature $T_{c}$ when the amplitude
of the order parameter $\Delta$ is small, or in S-X structures when
the proximity effect is weak due to a finite interface resistance
for arbitrary temperature.

In the diffusive limit one can expand $f\approx f_{0}+n_{k}f_{k}+\cdots$,
in angular harmonics where $\left\langle f\right\rangle =f_{0}\gg f_{k}$.
We first average \eqref{eq:transport-linearised} over the momentum
direction: 
\begin{equation}
\frac{\boldsymbol{v_{F}}}{\dim}\tilde{\nabla}_{k}f_{k}+\left\{ \omega_{n}-\mathbf{i}\mathcal{A}_{0},f_{0}\right\} =-2{\bf i}\Delta\sgn\left(\omega_{n}\right)\;,\label{eq:diff-f0}
\end{equation}
where $\dim=1,2,3$ is the dimension of the system. Next we multiply
Eq.\eqref{eq:transport-linearised} by $n_{k}$ and average over the
momentum direction to obtain 
\begin{multline}
v_{F}\tilde{\nabla}_{k}f_{0}+\left\{ \omega_{n}-\mathbf{i}\mathcal{A}_{0},f_{k}\right\} -\dfrac{1}{2m}\left\{ \mathcal{F}_{kj},f_{j}\right\} =\\
-\dfrac{\sgn\left(\omega_{n}\right)}{\tau}f_{k}\;.\label{eq:diff-fk}
\end{multline}
Eqs. \eqref{eq:diff-f0} and \eqref{eq:diff-fk} constitute a closed
set of coupled differential equations for $f_{0}$ and $f_{k}$. In
particular from Eq. (\ref{eq:diff-fk}) we can write $f_{k}$ in terms
of $f_{0}$ up to terms of second order in $\tau$: 
\begin{multline}
f_{k}\approx-\tau\sgn\left(\omega_{n}\right)v_{F}\tilde{\nabla}_{k}f_{0}-\tau^{2}\dfrac{v_{F}}{2m}\left\{ \mathcal{F}_{kj},\tilde{\nabla}_{j}f_{0}\right\} \\
+\tau^{2}v_{F}\left\{ \omega_{n}-\mathbf{i}\mathcal{A}_{0},\tilde{\nabla}_{k}f_{0}\right\} +\cdots\;.\label{eq:diff-fk-truncated}
\end{multline}
Note that the Usadel equation was obtained in several works in the
absence of gauge-fields, where one skipped the terms of the order
$\tau^{2}$. We keep here these terms since they are crucial for the
description of magneto-electric effects \cite{Malshukov2008,Malshukov2010,Bergeret2014a}.

The equations can be further simplified by noticing that the anti-commutator
in the second line of Eq. \eqref{eq:diff-fk-truncated} can be written
as 
\begin{multline}
\left\{ \omega_{n}-\mathbf{i}\mathcal{A}_{0},\tilde{\nabla}_{k}f_{0}\right\} =\tilde{\nabla}_{k}\left\{ \omega_{n}-\mathbf{i}\mathcal{A}_{0},f_{0}\right\} \\
+\mathbf{i}\left\{ \tilde{\nabla}_{k}\mathcal{A}_{0},f_{0}\right\} \;.
\end{multline}
In virtue of \eqref{eq:diff-f0}, the first term in the right-hand-side
of the last equation is in fact of order $\tau$ and so this term
in \eqref{eq:diff-fk-truncated} is of order $\tau^{3}$ and can be
neglected. The second term reads $\tilde{\nabla}_{k}\mathcal{A}_{0}=-\mathbf{i}\left[\mathcal{A}_{k},\mathcal{A}_{0}\right]=\mathcal{F}_{k0}$
for a space-independent gauge-potential. This electric field renormalizes
the paramagnetic effects $\mathcal{A}_{0}$, and is neglected in the
following. Finally, we replace \eqref{eq:diff-fk-truncated} into
\eqref{eq:diff-f0} to obtain the Usadel equation for $f_{0}$: 
\begin{multline}
-\sgn\left(\omega_{n}\right)D\tilde{\nabla}^{2}f_{0}+\left\{ \omega_{n}-\mathbf{i}\mathcal{A}_{0},f_{0}\right\} \\
-\dfrac{\tau D}{2m}\left\{ \tilde{\nabla}_{i}\mathcal{F}_{ij},\tilde{\nabla}_{j}f_{0}\right\} =-2\mathbf{i}\Delta\sgn\left(\omega_{n}\right)\label{eq:transport-diffusive}
\end{multline}
with $D=v_{F}^{2}\tau/\dim$ the diffusion constant. This equation
is supplemented by the generalized Kupriyanov-Lukichev boundary condition
\cite{Kupriyanov1988} 
\begin{equation}
\mathcal{N}_{i}\left[\tilde{\nabla}_{i}f_{0}+\dfrac{\tau\sgn\left(\omega_{n}\right)}{2m}\left\{ \mathcal{F}_{ij},\tilde{\nabla}_{j}f_{0}\right\} \right]_{x_{0}}=-\gamma f_{\text{BCS}}\label{eq:boundary-diffusive}
\end{equation}
at an interface located at position $x_{0}$ between a bulk superconductor
described by the anomalous GF $f_{\text{BCS}}$ and the X bridge.
The interface is characterized by the transparency $\gamma$ and normal
vector of component $\mathcal{N}_{i}$. For a fully transparent interface,
we impose the continuity of the GFs.

We now need to write the current and spin density in terms of the
isotropic anomalous GFs. It is easy to verify, by checking its conservation,
that in the linearized case the electric current, Eq. (\ref{eq:obs-current}),
is given by: 
\begin{equation}
\boldsymbol{j}=\frac{\mathbf{i}\pi eN_{0}T}{2}\sum_{\omega_{n}}\Tr\left\langle \boldsymbol{v_{F}}f\bar{f}\right\rangle \sgn\omega_{n}\;,\label{eq:obs-current-linearised}
\end{equation}
and correspondingly in the diffusive limit 
\begin{multline}
j_{i}={\mathbf{i}\pi eN_{0}DT}\sum_{\omega_{n}}\Tr\left\{ f_{0}\tilde{\nabla}_{i}\bar{f}_{0}-\bar{f}_{0}\tilde{\nabla}_{i}f_{0}\right.+\\
+\left.\frac{\tau\sgn\left(\omega_{n}\right)}{2m}\left(f_{0}\left\{ \mathcal{F}_{ij},\tilde{\nabla}_{j}\bar{f}_{0}\right\} +\bar{f}_{0}\left\{ \mathcal{F}_{ij},\tilde{\nabla}_{j}f_{0}\right\} \right)\right\} \;.\label{eq:obs-current-diffusive}
\end{multline}

The spin polarization \eqref{eq:obs-spin-polar} is more subtle to
deal with in the linearized approximation, since the normalization
condition do not apply in our case. In accordance with the case without
SOC, one may assume that it can be expressed in terms of the isotropic
anomalous $f$ as: 
\begin{equation}
S^{a}=\mathbf{i}\pi N_{0}T\sum_{\omega_{n}}\Tr\left\langle \sigma^{a}f\bar{f}\right\rangle \sgn\left(\omega_{n}\right)\label{eq:obs-spin-polar-linearised}
\end{equation}
with $\left\langle \sigma^{a}f\bar{f}\right\rangle =\sigma^{a}f_{0}f_{0}$
in the diffusive limit. In the next section we will show a posteriori
that these expressions leads to the known results in bulk systems
in the presence of Rashba SOC.

For the following discussions it is convenient to write the anomalous
GF $f$ as the sum of singlet (scalar) and triplet (vector in spin
space) $f=f_{s}+f_{t}^{a}\sigma^{a}$, and to expand all the spin
variables in term of Pauli matrices: $\mathcal{F}_{ij}=\mathcal{F}_{ij}^{a}\sigma^{a}/2$,
$\mathcal{A}_{\mu}=\mathcal{A}_{\mu}^{a}\sigma^{a}/2$. From Eqs.
(\ref{eq:transport-linearised}) we obtain the equations for the singlet
and triplet components in the ballistic case: 
\begin{multline}
\left(v_{F}n_{i}\partial_{i}+2\omega_{n}\right)f_{s}=-2\mathbf{i}\sgn\left(\omega_{n}\right)\Delta\\
+\left(\mathbf{i}\mathcal{A}_{0}^{a}+\dfrac{n_{i}\mathcal{F}_{ij}^{a}}{2m}\dfrac{\partial}{\partial n_{j}}\right)f_{t}^{a}\label{eq:fs-ball}
\end{multline}
\begin{equation}
v_{F}n_{i}\left(\tilde{\nabla}_{i}f_{t}\right)^{a}+2\omega_{n}f_{t}^{a}=\left(\mathbf{i}\mathcal{A}_{0}^{a}+\dfrac{n_{i}\mathcal{F}_{ij}^{a}}{2m}\dfrac{\partial}{\partial n_{j}}\right)f_{s}\;.\label{eq:ft-ball}
\end{equation}

Equivalently, from Eq. (\ref{eq:transport-diffusive}) one obtains
the equations for the isotropic part of the singlet $f_{s0}$ and
triplet $f_{t0}$ components in the diffusive case (for simplicity
we skip the subindex $0$) : 
\begin{multline}
\left(\partial_{i}^{2}-\kappa_{\omega}^{2}\right)f_{s}-2\mathbf{i}\dfrac{\Delta}{D}+\\
+\sgn\left(\omega_{n}\right)\left[\mathbf{i}\dfrac{\mathcal{A}_{0}^{a}}{D}+\dfrac{\tau}{2m}\left(\mathcal{J}_{i}{\partial}_{i}\right)^{a}\right]f_{t}^{a}=0\label{eq:fs-diff}
\end{multline}
\begin{multline}
\left(\tilde{\nabla}_{i}\tilde{\nabla}_{i}f_{t}\right)^{a}-\kappa_{\omega}^{2}f_{t}^{a}\\
+\sgn\left(\omega_{n}\right)\left(\mathbf{i}\dfrac{\mathcal{A}_{0}^{a}}{D}+\dfrac{\tau}{2m}\mathcal{J}_{i}^{a}\partial_{i}\right)f_{s}=0\label{eq:ft-diff}
\end{multline}

We write the covariant derivative as $\tilde{\nabla}_{i}=\partial_{i}-{\bf i}[\Acal_{i},\dot{]}\equiv\partial_{i}+\hat{P}_{i}$,
where $\hat{P}_{i}$ is a tensor dual to $\mathcal{A}_{i}$ with components
$P_{i}^{ab}=\varepsilon^{abc}\mathcal{A}_{i}^{c}$. Thus, $\tilde{\nabla}_{i}\tilde{\nabla}_{i}=\partial_{i}^{2}+2\hat{P}_{i}\partial_{i}-\hat{\Gamma}$,
where $\hat{P}_{i}\hat{P}_{i}=-\hat{\Gamma}$. By noticing that $(\tau/2m){\cal J}_{k}^{a}=C_{k}^{a}$,
it is easy to verify that the diffusive equations Eqs. (\ref{eq:fs-diff}-\ref{eq:ft-diff})
are identical to those derived in section II from heuristic arguments
{[}Eqs.(\ref{Usadel-s2}-\ref{Usadel-t2}){]}. One should emphasize
though that while Eqs. (\ref{eq:fs-diff}-\ref{eq:ft-diff}) are derived
for the particular case of linear in momentum SOC, Eqs.(\ref{Usadel-s2}-\ref{Usadel-t2})
suggest that the form of the diffusion equations remain the same for
arbitrary momentum dependence.

In particular the form of Eq. (\ref{eq:ft-diff}) proves the full
analogy between singlet-triplet and charge-spin coupling in diffusive
systems. {[}\textit{cf. } Eqs. (\ref{C-diffusion2}-\ref{S-diffusion2}){]}.
In Ref. \cite{Bergeret2014}, the analogy between the diffusion of
spin in normal systems and the triplet components was discussed. Here
we can extend this result and find that the tensor $C_{k}^{a}$, responsible
for the SHE in normal systems, is an additional source for the singlet-triplet
conversion and, as we will see in the next sections, is at the root
of magneto-electric effects and the anomalous phase. Equations (\ref{eq:fs-ball}
-\ref{eq:ft-ball}) and (\ref{eq:fs-diff}-\ref{eq:ft-diff}) are
the central equations of this work, which we now solve for different
situations. In section \ref{sub:All-T-ball} we go beyond this linear
approximation.

\section{The Edelstein effect in bulk superconductors for $T\rightarrow T_{c}$\label{sec:Edelstein-effect}}

In order to illustrate the usefulness of the SU(2) covariant quasiclassical
equations presented above, we study here the magneto-electric effect
and its inverse in bulk superconductors with an intrinsic SOC linear
in momentum and derive the response coefficients in \eqref{eq_SEE1}
and \eqref{eq_SEE2}.

We assume that the superconducting order parameter $\Delta$ is constant
in magnitude but has a spatially dependent phase $\Delta\left({\bf r}\right)=|\Delta|e^{\mathbf{i}\varphi\left({\bf r}\right)}$,
where $\nabla\varphi$ is assumed to be a constant vector.

Let us first consider a diffusive superconductor. From \eqref{eq:fs-diff}
in the lowest order of $\nabla\varphi$ one obtains 
\begin{equation}
f_{s}\approx-\mathbf{i}\dfrac{|\Delta|}{\left|\omega_{n}\right|}e^{{\bf i}\varphi}\;.\label{eq:fs-bulk-diff}
\end{equation}
and hence one can easily obtain the lowest order correction to the
triplet component from \eqref{eq:ft-diff} : 
\begin{equation}
f_{t}^{a}=\dfrac{|\Delta|}{\left|\omega_{n}\right|}\dfrac{\tau}{2m}\sgn\left(\omega_{n}\right)\left[\left(\hat{\Gamma}+\kappa_{\omega}^{2}\right)^{-1}\right]^{ab}\mathcal{J}_{j}^{b}\partial_{j}\varphi\;.\label{eq:ft-bulk-diff}
\end{equation}

From Eq. (\ref{eq:obs-spin-polar-linearised}) it becomes clear that
the spin density is determined by the product of the singlet (\ref{eq:fs-bulk-diff})
and triplet (\ref{eq:ft-bulk-diff}) components which results in $S^{a}=\chi_{i}^{a}\partial_{i}\varphi$
with 
\begin{equation}
\chi_{i}^{a}=4\pi N_{0}\dfrac{\tau}{2m}T\sum_{\omega_{n}>0}\dfrac{\Delta^{2}}{\omega_{n}^{2}}\left[\left(\hat{\Gamma}+\kappa_{\omega}^{2}\right)^{-1}\right]^{ab}\mathcal{J}_{i}^{b}\label{eq:chi-diff}
\end{equation}
This is the Edelstein result generalized for arbitrary linear in momentum
SOC.

With the help of Eqs. (\ref{eq:fs-diff}-\ref{eq:ft-diff}) we can
also describe the inverse EE, the so-called spin-galvanic effect.
We now assume a finite and spatially homogenous $\mathcal{A}_{0}^{a}$
and a zero phase gradient. In such a case one can obtain $f_{t}$
directly from Eq. (\ref{eq:ft-diff}), which is now proportional to
$\mathcal{A}_{0}^{a}$. By substitution of this result into the expression
for the current, Eq. (\ref{eq:obs-current-diffusive}), and by noticing
that only the second line contributes to the current we obtain $j_{i}=e\chi_{i}^{a}\mathcal{A}_{0}^{a}$,
with $\chi_{i}^{a}$ given by Eq. (\ref{eq:chi-diff}) in agreement
with Onsager reciprocity.

In short, we are able to derive in a few lines the tensor \eqref{eq:chi-diff},
which describes the EE and inverse EE in superconductors. Moreover,
the expression \eqref{eq:chi-diff} is valid for arbitrary linear
in momentum spin-orbit effect and generalizes the result obtained
in Ref. \cite{Edelstein2005} for the particular case of a Rashba
SOC. If one assumes the same here, \textit{i.e.} $\Acal_{x}^{y}=-\alpha=-\Acal_{y}^{x}$
and all the other components equal to zero, one obtains from Eq. \eqref{eq:chi-diff}
\begin{equation}
\chi_{i}^{a}=\left(\delta_{ix}^{ay}-\delta_{iy}^{ax}\right)4\pi N_{0}\dfrac{D\tau}{2m}T\sum_{\omega_{n}>0}\dfrac{\Delta^{2}}{\omega_{n}^{2}}\frac{\alpha^{3}}{2\left|\omega_{n}\right|+D\alpha^{2}}\label{EE-bulk-diff}
\end{equation}
that coincides with the expression obtained in Ref. \cite{Edelstein2005}. 

If we neglect in Eq. \eqref{eq:ft-bulk-diff} the Dyakonov-Perel relaxation,
then the triplet component is simply proportional to $f_{t}^{a}\sim\mathcal{A}_{0}^{a}\left({\bf r}\right)$.
By substituting this into the expression for the current Eq. \eqref{eq:obs-current-diffusive}
one can easily show that 
\begin{equation}
j_{i}=4e\pi N_{0}\frac{\tau}{2m}T\sum_{\omega_{n}>0}\frac{\Delta^{2}}{\omega_{n}^{2}\kappa_{\omega}^{2}}{\cal F}_{0j}^{a}{\cal F}_{ji}^{a}\;.\label{j_diff_gen}
\end{equation}
This expression suggests that a spatially inhomogenous magnetization
together with SOC may also induce a finite supercurrent. In this case
the spin-galvanic effect scales with the square of the SOC parameter,
in contrast to the $\alpha^{3}$ dependency found previously for spatially
uniform magnetization.


The same effects can be explored in the pure ballistic limit, for
which Eqs. (\ref{eq:fs-ball}-\ref{eq:ft-ball}) apply. The singlet
component in the lowest order in the SOC is given by 
\begin{equation}
f_{s}\approx-\mathbf{i}\dfrac{\Delta}{\left|\omega_{n}\right|}\left(1-\mathbf{i}\dfrac{v_{F}n_{i}}{2\omega_{n}}\partial_{i}\varphi\right)\;,\label{eq:fs-bulk-ball}
\end{equation}
whereas the triplet component can be obtained easily from Eq. (\ref{eq:ft-ball})
\begin{equation}
f_{t}^{a}=-\frac{\Delta v_{F}}{2\left|\omega_{n}\right|\omega_{n}}\left[\left(v_{F}n_{k}\hat{P}_{k}+2\omega_{n}\right)^{-1}\right]^{ab}\frac{n_{i}\Fcal_{ij}^{b}}{2m}\partial_{j}\varphi.
\end{equation}

By using Eq. \eqref{eq:obs-spin-polar-linearised} we obtain the Edelstein
result $S^{a}=\chi_{j}^{a}\partial_{j}\varphi$ but now for an arbitrary
linear in momentum SOC 
\begin{multline}
\chi_{i}^{a}=-2\pi\dfrac{N_{0}v_{F}}{2m}T\times\\
\sum_{\omega_{n}>0}\dfrac{\Delta^{2}}{\left|\omega_{n}\right|^{3}}\left\langle \left[\left(v_{F}n_{k}\hat{P}_{k}+2\omega_{n}\right)^{-1}\right]^{ab}n_{j}\mathcal{F}_{ji}^{b}\right\rangle \quad.\label{eq:S-bulk-ball}
\end{multline}

Identically, we find $j_{i}=e\chi_{i}^{a}\mathcal{A}_{0}^{a}$.

In the particular case of a 2D systems with Rashba SOC we recover
the Edelstein result for a ballistic superconductor \cite{Edelstein1995}:
\begin{equation}
\chi_{i}^{a}=\frac{\pi N_{0}\Delta^{2}}{4v_{F}m}T\sum_{\omega_{n}>0}\frac{\left(v_{F}\alpha\right)^{3}}{\left|\omega_{n}\right|^{3}\left[\left(2\omega_{n}\right)^{2}+\left(v_{F}\alpha\right)^{2}\right]}\left(\delta_{ix}^{ay}-\delta_{iy}^{ax}\right)\;.\label{EE-bulk-ball}
\end{equation}
The agreement between our and Edelstein results proves the validity
of the expression (\ref{eq:obs-spin-polar-linearised}) in the linearized
approximation.

To conclude this section we note that for Rashba SOC in both cases,
diffusive \eqref{EE-bulk-diff} and ballistic \eqref{EE-bulk-ball},
$\chi_{i}^{a}$ is proportional to $\alpha$ in the strong SOC limit
(see also \cite{Houzet2015}), and to $\alpha^{3}$ for weak spin-orbit
interaction (see also \cite{Bergeret2014a}). So, the quasi-classic
formalism is able to recover in an elegant way some well established
results obtained after cumbersome diagrammatic \cite{Edelstein1995,Edelstein2005},
and it also allows some easy generalizations of them.

\section{Magneto-electric effects in diffusive Josephson junctions\label{sec:Diffusive-limit}}

We now turn to the central topic of the present work which is the
description of magneto-electric effects in S-X-S Josephson junctions
and demonstrate their connection to the anomalous phase problem. We
first consider the diffusive limit and postpone the discussion of
ballistic junctions for the next section.

In particular we consider a S-X-S Josephson junction with an interlayer
X of length L. We assume that the magnetic interactions are only finite
in X and vanish in the S electrodes. Moreover, we assume that the
structure has infinite dimensions in the $y-z$ plane and therefore
the GFs only depend on the $x$ coordinate. The superconducting bulk
solutions in the leads are written as $f_{L}=f_{\text{BCS}}e^{-\mathbf{i}\varphi/2}$
and $f_{R}=f_{\text{BCS}}e^{\mathbf{i}\varphi/2}$, in the left $\left(x\leq-L/2\right)$
and right $\left(x\geq L/2\right)$ electrodes respectively, with
\begin{equation}
f_{\text{BCS}}=\dfrac{\Delta}{\sqrt{\omega_{n}^{2}+\Delta^{2}}}\quad,\label{eq:fBCS}
\end{equation}
whereas the normal metal fills the region $-L/2\leq x\leq L/2$.

We will consider both the highly resistive and the perfectly transparent
interfaces between the S and X parts. When the barrier transparency
is low, the linearized approximation is justified for all temperatures,
whereas for transparent barriers, one is limited to temperatures close
to the critical temperature of the junction.

For the particular case of Rashba SOC in the X region and an in-plane
exchange field the Josephson current has been calculated in Ref. \cite{Bergeret2014a}.
It has been shown explicitly that the current-phase relation is given
by $I=I_{c}\sin(\varphi-\varphi_{0})$. The anomalous phase $\varphi_{0}$
was calculated as a function of the strength of the spin fields, the
temperature and the junction parameters. Here instead we focus in
a generic linear-in-momentum SOC and we derive the expressions for
the anomalous Josephson current in the lowest order of the spin fields.
This will allow us to understand the link between the inverse EE and
the $\varphi_{0}$-junctions.

\subsection{Diffusive junction with low transparency interfaces\label{ss_lowt}}

We first consider a S-X-S diffusive Josephson junction with highly
resistive S-X interfaces. In this limit the linearization of the quasiclassical
equations is justified for all temperatures. Our goal here is to determine
the Josephson current through the junction, which in the linearized
regime is given by Eq. \eqref{eq:obs-current-diffusive}. The components
of the condensate function $f_{s}$ entering this expression, has
to be obtained by solving the system \eqref{eq:fs-diff}-\eqref{eq:ft-diff}
in the normal metal which couples the singlet with the triplet component.
For the specific S-X-S geometry considered here this equations read:
\begin{multline}
\left(\partial_{x}^{2}-\kappa_{\omega}^{2}\right)f_{s}+\sgn\left(\omega_{n}\right)\left[\mathbf{i}\dfrac{\mathcal{A}_{0}^{a}}{D}+\dfrac{\tau}{2m}\left(\mathcal{J}_{i}\partial_{i}\right)^{a}\right]f_{t}^{a}=0\text{\textordmasculine; ,}\label{eq:fs-diff2}
\end{multline}
\begin{multline}
\left(\tilde{\nabla}_{i}\tilde{\nabla}_{i}f_{t}\right)^{a}-\kappa_{\omega}^{2}f_{t}^{a}+\\
+\sgn\left(\omega_{n}\right)\left(\mathbf{i}\dfrac{\mathcal{A}_{0}^{a}}{D}+\dfrac{\tau}{2m}\mathcal{J}_{i}^{a}\partial_{i}\right)f_{s}=0\;,\label{eq:ft-diff2}
\end{multline}
and the boundary conditions for the resistive interface (\textit{cf.}
Eq.\eqref{eq:boundary-diffusive}): 
\begin{eqnarray}
\left(\partial_{x}f_{s}+\sgn\left(\omega_{n}\right)\frac{\tau}{2m}{\cal J}_{x}^{a}f_{t}^{a}\right)_{x=\pm L/2} & = & \pm\gamma f_{R,L}\nonumber \\
\partial f_{t}^{a}|_{x=\pm L/2} & = & 0\quad.\label{BC_Rb}
\end{eqnarray}

The expression Eq.\eqref{eq:obs-current-diffusive}, can be simplified
by calculating the current at the right interface ($x=L/2$) and by
using the boundary condition \eqref{BC_Rb}: 
\begin{equation}
j_{x}=\mathbf{i}e\pi DN_{0}T_{c}\gamma\sum_{\omega_{n}>0}\Tr\left\{ f_{s}\bar{f}_{R}-\bar{f}_{s}f_{R}\right\} _{x=L/2}\;.\label{eq:obs-current-diff-barrier}
\end{equation}

It is clear from this equation that the correction to the current
due to the spin-fields (the anomalous current) is proportional to
${\rm Im}\left[f_{R}^{*}\delta f_{s}(L/2)\right]$, where $\delta f_{s}$
is the first correction to the singlet component due to the gauge
potentials. In the absence of a phase difference between the S electrodes
$f_{R}$ is real and the anomalous current is proportional to the
imaginary part of the singlet component. According to Eq. (\ref{eq:fs-diff2}),
in the absence of spin-fields (exchange and SOC), there is no triplet
component and the singlet component is real. Therefore no supercurrent
flows at zero phase difference.

In the presence of spin-fields there are two sources for singlet-triplet
conversion, as seen from the second term in the l.h.s of Eq. (\ref{eq:ft-diff2}).
The first one is the extensively studied mechanism for singlet-triplet
conversion in S/F junctions via the intrinsic exchange field $\Acal_{0}$
\cite{buzdin.2005_RMP,bergeret_volkov_efetov_R.2005}. Inclusion of
SOC leads to an additional singlet-triplet conversion mechanism described
by the last term in the l.h.s of Eq (\ref{eq:ft-diff2}). As discussed
in section II, the singlet-triplet conversion in this case corresponds
to the charge-spin conversion in normal systems with SOC. Conversely,
once the triplet component is created, both mechanisms will convert
it back to singlet, as can be seen in Eq.\eqref{eq:fs-diff2}.

\begin{figure}[b]
\includegraphics[width=0.8\linewidth]{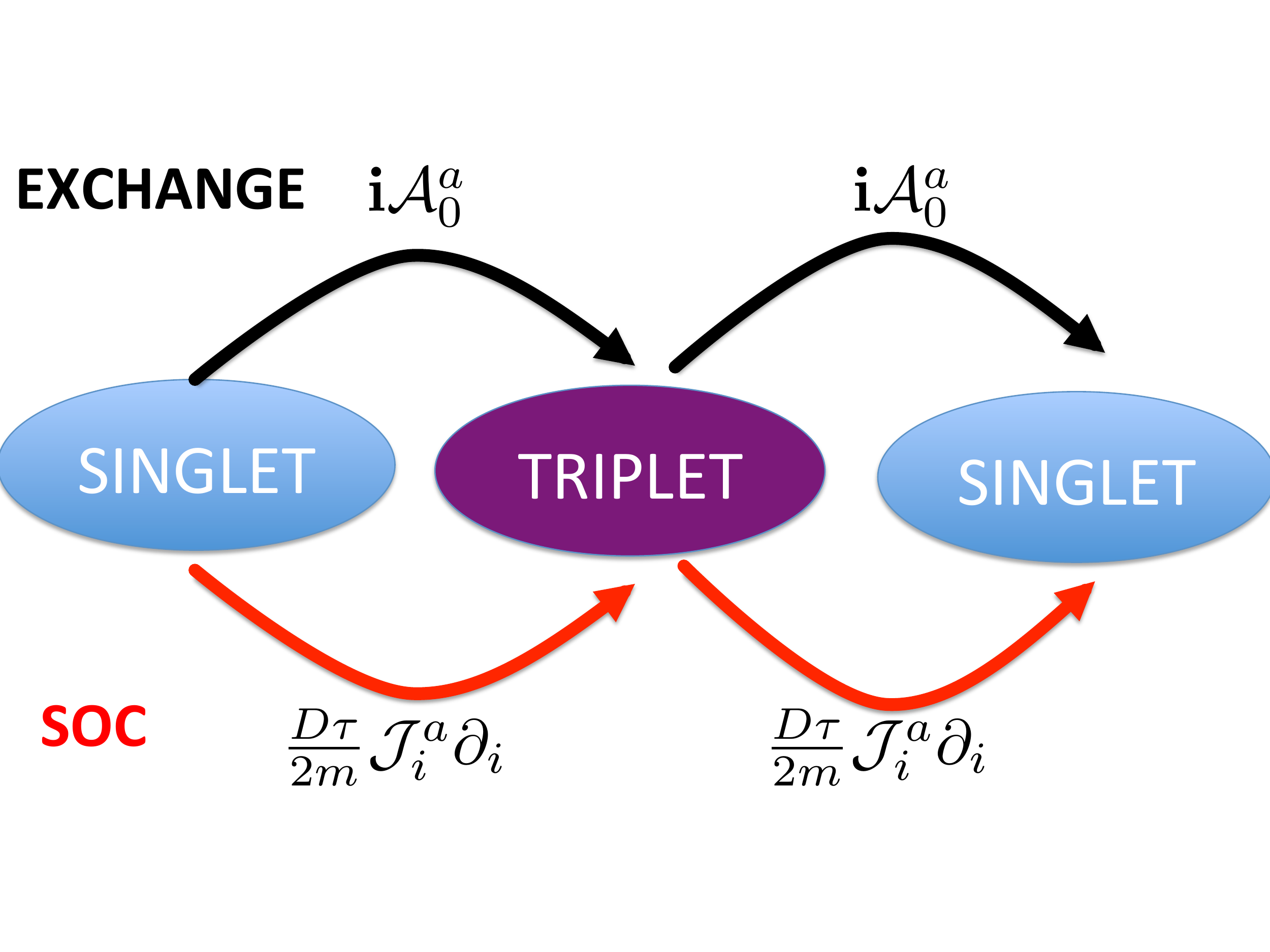}

\protect\protect\protect\caption{\label{fig_diag} Schematic representation of the singlet-triplet-singlet
conversion process at the lowest order with respect to the spin-fields.
Black arrows represent the action of the exchange field, whereas red
arrows encode the effect of the singlet-triplet coupling term due
to the SOC. Only mixed red-black paths lead to the appearance of an
anomalous phase $\varphi_{0}$ in the singlet component and hence
to a supercurrent in a S-X-S junction even without a phase bias between
the S electrodes}
\end{figure}

The singlet-triplet-singlet conversion at the lowest orders in perturbation
with respect to the spin-fields is schematized in Fig. \ref{fig_diag}.
The black arrows represent the singlet-triplet conversion due to the
exchange field which implies a $\pi/2$ phase shift due to the ${\bf i}$
factor in front of $\Acal_{0}$ in Eqs. (\ref{eq:fs-diff2}-\ref{eq:ft-diff2}).
The red arrows represent the singlet-triplet conversion due to the
SOC, specifically due to the coupling term in Eqs. (\ref{eq:fs-diff2}-\ref{eq:ft-diff2})
proportional to ${\cal J}_{i}^{a}\partial_{i}$. No additional phase
is associated with this latter process. If one follows the black path,
\textit{i.e.} the singlet-triplet-singlet conversion only due to the
exchange field, the resulting contribution to the singlet component
acquires a minus sign (a $\pi$ shift) and it is proportional to $\mathcal{A}_{0}^{2}$.
This means that there is no anomalous phase $0<\varphi_{0}<\pi$ induced
and hence no Josephson current flows when $\varphi=0$. Similarly,
if one follows the red path the resulting singlet component also remains
real with no change of sign. From Fig.\ref{fig_diag} it becomes clear
that a nontrivial $\varphi_{0}$ only appears from the \textquotedbl{}cross-term\textquotedbl{}
path that consist in one black and one red arrow. In other words,
the mutual action of exchange field and SOC leads to a finite $\varphi_{0}$
and hence to a supercurrent even at zero phase difference. In this
case the contribution to this current in the lowest order of the spin
fields, is proportional to $\mathcal{A}_{0}^{a}\mathcal{J}_{i}^{a}\partial_{i}f_{s}$
between the exchange field and the spin-current tensor, as anticipated
in the introduction. 

In order to quantify this effect and calculate $\varphi_{0}$ in the
S-X-S junctions it is convenient to introduce the singlet and triplet
propagators associated with Eqs. (\ref{eq:fs-diff2}-\ref{BC_Rb}):
\begin{align}
\left(\partial_{x}^{2}-\kappa_{\omega}^{2}\right)K_{s}\left(x,x'\right) & =-\delta\left(x-x'\right)\nonumber \\
\left.\partial_{x}K_{s}\left(x,x'\right)\right|_{x=\pm L/2} & =0\quad,\label{eq:Green-s-problem}
\end{align}
and 
\begin{align}
\left[\left(\partial_{x}+\hat{P}_{x}\right)^{2}+\hat{P}_{y}^{2}+\hat{P}_{z}^{2}-\kappa_{\omega}^{2}\right]\hat{K}_{t}\left(x,x'\right) & =-\delta\left(x-x'\right)\nonumber \\
\left(\partial_{x}+\hat{P}_{x}\right)\hat{K}_{t}\left(x,x^{\prime}\right) & =0\quad.\label{eq:Green-t-problem}
\end{align}
Thus, Eqs. (\ref{eq:fs-diff2}-\ref{BC_Rb}) can be re-written as
a set of integral equations 
\begin{multline}
f_{s}\left(x\right)=f_{s}^{\left(0\right)}\left(x\right)-\frac{\tau{\cal J}_{x}^{a}}{2m}\left[K_{s}(x,L/2)f_{t}^{a}-K_{s}(x,-L/2)\right]\\
+\sgn\left(\omega_{n}\right)\int_{-L/2}^{L/2}dx_{1}K_{s}\left(x,x_{1}\right)\times\\
\left[\mathbf{i}\dfrac{\mathcal{A}_{0}^{b}}{D}K_{s}\left(x_{1},x\right)+\dfrac{\tau}{2m}\mathcal{J}_{x}^{b}\partial_{x_{1}}\right]f_{t}^{b}(x_{1})\;,\label{eq:fs-Green}
\end{multline}
and 
\begin{multline}
f_{t}^{a}\left(x\right)=\sgn\left(\omega_{n}\right)\int_{-L/2}^{L/2}dx_{1}\times\\
\left[K_{t}^{ab}\left(x,x_{1}\right)\left(\mathbf{i}\dfrac{\mathcal{A}_{0}^{b}}{D}+\dfrac{\tau}{2m}\mathcal{J}_{1}^{a}\partial_{x_{1}}\right)f_{s}\left(x_{1}\right)\right]\;.\label{eq:ft-Green}
\end{multline}
Here $f_{s}^{\left(0\right)}=\gamma\left(K_{s}\left(x,\dfrac{L}{2}\right)f_{R}+K_{s}\left(x,-\dfrac{L}{2}\right)f_{L}\right)$
and the second term in Eq. (\ref{eq:fs-Green}) takes into account
the boundary condition (\ref{BC_Rb})

The $K_{s}$ propagator can be obtained from Eq. \eqref{eq:Green-s-problem}
\begin{multline}
K_{s}\left(x_{1},x_{2}\right)=\\
\dfrac{\cosh\kappa_{\omega}\left(L-\left|x_{1}-x_{2}\right|\right)+\cosh\kappa_{\omega}\left(x_{1}+x_{2}\right)}{2\kappa_{\omega}\sinh\kappa_{\omega}L}\;,\label{eq:Gs}
\end{multline}
whereas the equations for the triplet kernel, Eqs.(\ref{eq:Green-t-problem}),
can be written in the form of an integral equation which is convenient
for the subsequent perturbative analysis: 
\begin{multline}
\hat{K}_{t}(x_{1},x_{2})=e^{-\hat{P}_{x}x_{1}}K_{s}(x_{1},x_{2})e^{\hat{P}_{x}x_{2}}+\\
-e^{-\hat{P}_{x}x_{1}}\int_{-L/2}^{L/2}dyK_{s}(x_{1},y)e^{\hat{P}_{x}y}\hat{\Gamma}_{\perp}\hat{K}_{t}(y,x_{2})\;\label{Kt_dyson}
\end{multline}
where $\hat{\Gamma}_{\perp}=-\hat{P}_{y}^{2}-\hat{P}_{z}^{2}$.

In the lowest order of the gauge potentials one can obtain the correction
$\delta f_{s}$ to the singlet component by substituting the result
(\ref{eq:Gs}) into Eqs. \eqref{eq:fs-Green}-\eqref{eq:ft-Green}.
We consider here only the \textquotedbl{}cross-term\textquotedbl{}
correction $\delta f_{s}$ proportional to both the exchange field
$\Acal_{0}$ and the spin-current ${\cal J}_{i}$ and which is responsible
for the anomalous phase-shift: 
\begin{multline}
\delta f_{s}\left(L/2\right)=\mathbf{i}\mathcal{A}_{0}^{a}\mathcal{J}_{x}^{b}\dfrac{\tau\gamma}{2mD}\times\\
\times f_{L}\int_{-L/2}^{L/2}dy_{1}\int_{-L/2}^{L/2}dy_{2}K_{t}^{ab}(y_{2},y_{1})\frac{\cosh\left[\kappa_{\omega}\left(y_{1}-y_{2}\right)\right]}{\kappa_{\omega}\sinh\kappa_{\omega}L}\label{fs_R}
\end{multline}

In principle, one has all the elements to solve Eqs. (\ref{eq:fs-Green}-\ref{eq:ft-Green}),
for example recursively by performing a perturbative expansion in
the gauge potentials. Here, in order to get analytical compact expressions
we restrict our analysis to the short junction limit, \textit{i.e.}
$L\ll{\rm min}(\kappa_{\omega}^{-1},\left|{\cal A}_{k}\right|^{-1})$.
In this case $K_{s}\approx\kappa_{\omega}^{-2}L^{-1}$ (\textit{cf.}
Eq. \eqref{eq:Gs}) and from Eq. (\ref{Kt_dyson}) it is easy to verify
that $K_{t}$ reads 
\begin{equation}
\hat{K}_{t}\approx\dfrac{\left(\kappa_{\omega}^{2}+\hat{\Gamma}_{\perp}\right)^{-1}}{L}\;.\label{eq:Gt-short}
\end{equation}

We are interested in calculating the anomalous phase $\varphi_{0}$
which can be obtained by noticing that the current \eqref{eq:obs-current-diff-barrier}
can be written as 
\begin{equation}
j_{x}=j_{c}\sin\left(\varphi-\varphi_{0}\right)\approx j_{c}\sin\varphi-\varphi_{0}j_{c}\cos\varphi
\end{equation}
for a small $\varphi_{0}$. The anomalous phase $\varphi_{0}$ can
be obtained by setting $\varphi=0$ and dividing by the critical current
$j_{c}$ in the absence of SOC. In the short junction limit this is
given by: 
\begin{equation}
j_{c}=4e\pi DN_{0}T_{c}\gamma^{2}\sum_{\omega_{n}>0}\frac{f_{\text{BCS}}^{2}}{\kappa_{\omega}^{2}L}
\end{equation}
We follow this procedure and from Eq.\eqref{eq:obs-current-diff-barrier}
and Eqs. (\ref{fs_R}-\ref{eq:Gt-short}) we obtain 
\begin{equation}
\varphi_{0}\approx\dfrac{\tau}{2mD}L\dfrac{{\displaystyle \sum_{\omega_{n}>0}}\dfrac{f_{\text{BCS}}^{2}}{\kappa_{\omega}^{2}}\mathcal{A}_{0}^{a}\left[\left(\kappa_{\omega}^{2}+\hat{\Gamma}_{\perp}\right)^{-1}\right]^{ab}\mathcal{J}_{x}^{b}}{{\displaystyle \sum_{\omega_{n}>0}}\dfrac{f_{\text{BCS}}^{2}}{\kappa_{\omega}^{2}}}\;.\label{eq:phi-0-short-tunneling}
\end{equation}
This expression clearly shows the relation between the appearance
of the anomalous phase, $\varphi_{0}$, and the inverse Edelstein
effect in bulk systems. Both, the Josephson current (proportional
in the linearized case to $\varphi_{0}$) and the bulk supercurrent
are proportional to ${\cal A}_{0}{\cal J}_{x}$, \textit{i.e.} both
are generated from the mutual action of the exchange field and the
SOC.

It is worth noticing that in the present case of low transparent interfaces,
the anomalous phase grows linearly with $L$, the length of the junction
\eqref{eq:phi-0-short-tunneling}. In the next subsection we show
that in the case of a transparent barrier the anomalous phase behaves
like $L^{3}$.

In the particular case of a 2D situation, with a SOC coupling of Rashba
(described by the parameter $\alpha$) and Dresselhaus ($\beta$)
type we obtain from Eq. (\ref{eq:phi-0-short-tunneling}): 
\begin{equation}
\varphi_{0}\approx\dfrac{\tau L}{2m}\dfrac{{\displaystyle \sum_{\omega_{n}>0}}\dfrac{f_{\text{BCS}}^{2}}{\kappa_{\omega}^{2}}\dfrac{\left(\beta\mathcal{A}_{0}^{x}-\alpha\mathcal{A}_{0}^{y}\right)\left(\alpha^{2}-\beta^{2}\right)}{2\omega_{n}+D\left(\alpha^{2}+\beta^{2}\right)}}{{\displaystyle \sum_{\omega_{n}>0}}\dfrac{f_{\text{BCS}}^{2}}{\kappa_{\omega}^{2}}}\;.\label{eq:phi-0-short-tunneling2D}
\end{equation}
Besides controlling the anomalous phase and hence the Josephson current
by tuning the external magnetic field, this expression also suggests
that the current can be controlled by tuning the Rashba SOC by means
of an external gate. In the particular case that $\alpha=\beta$ the
anomalous phase is zero and no supercurrent will flow.

\subsection{Diffusive junction with transparent interfaces}

We now briefly consider the limit of a full transparent barrier. In
that case one assumes continuity of the quasiclassical GFs at the
S-X interfaces. The problem is then formally the same as in the previous
section, except that the second equations in \eqref{eq:Green-s-problem}
and \eqref{eq:Green-t-problem}, for the propagators $K_{s}$ and
$\hat{K}_{t}$ are replaced by: 
\begin{equation}
\left.\hat{K}_{s,t}\left(x_{1},x_{2}\right)\right|_{x_{1}=\pm L/2}=0
\end{equation}
respectively. In this case one should remove the second term in Eq.
(\ref{eq:fs-Green}) and $f_{s}^{\left(0\right)}\left(x\right)=f_{L}\sinh\left(L/2-x\right)/\sinh\left(\kappa_{\omega}L\right)+f_{R}\sinh\left(L/2+x\right)/\sinh\left(\kappa_{\omega}L\right)$.

Now the singlet propagator is given by: 
\begin{multline}
K_{s}\left(x_{1},x_{2}\right)=\\
\dfrac{\cosh\kappa_{\omega}\left(x_{1}+x_{2}\right)-\cosh\kappa_{\omega}\left(L-\left|x_{1}-x_{2}\right|\right)}{2\kappa_{\omega}\sinh\kappa_{\omega}L}\;.
\end{multline}
In the short junction limit $K_{s}$ is proportional to $L$ and it
is temperature independent. From Eq. (\ref{Kt_dyson}) $\hat{K}_{t}\sim K_{s}$.
Thus, in this case the anomalous phase-shift is also temperature independent
and proportional to 
\begin{equation}
\varphi_{0}\propto\dfrac{\tau L^{3}}{mD}\mathcal{A}_{0}^{a}\mathcal{J}_{x}^{a}\;.\label{eq:phi-0-short-transparent}
\end{equation}
In contrast to the case of finite barrier resistance, Eq. \eqref{eq:phi-0-short-tunneling},
the anomalous phase scale with $L^{3}$. This means that in short
junctions a finite barrier resistance between the S and the normal
metal favors the growth of $\varphi_{0}$. These results generalize
those presented recently in Ref.\cite{Bergeret2014a} for the particular
case of Rashba SOC.

We can then conclude that the anomalous phase, at lowest order in
the gauge potentials, is proportional to $\mathcal{A}_{0}^{a}\mathcal{J}_{x}^{a}$,
independently of the type of interface.

\section{Magneto-electric effects in ballistic Josephson junctions\label{sec:Ball-limit}}

In this section we consider a pure ballistic S-X-S junction,\textit{i.e.}
we solve \eqref{eq:Eilenberger} in the limit $\tau\rightarrow\infty$.
As before, the junction is along the $x$-axis and the two superconducting
electrodes at position $x\leq-L/2$ and $x\geq L/2$. The spin fields,
both exchange and SOC, are only finite in the X region. We assume
that the the transverse dimensions of the junction are very large,
and therefore the GFs depends on $x$and only weakly on $y,z$. We
also assume that the interfaces between X and S are perfectly transparent.

In the next subsection we first analyze the Josephson current for
temperatures close to the superconducting critical temperature $T_{c}$,
and make a connection with the diffusive structures studied in the
previous section. In the second subsection we derive analytical expressions
for the anomalous current at arbitrary temperature for the case of
small spin fields.

\subsection{Ballistic junction at $T\rightarrow T_{c}$\label{sub:Close-Tc-ball}}

In the case of large enough temperatures we analyze the linearized
Eilenberger equation The solutions for the singlet and triplet components
in equations \eqref{eq:fs-ball} and \eqref{eq:ft-ball} can be written
as propagation in two directions $f_{s,t}\left(-L/2\leq x\leq L/2\right)=f_{s,t}^{>}\left(x\right)\Theta\left(\omega_{n}/n_{x}\right)+f_{s,t}^{<}\left(x\right)\Theta\left(-\omega_{n}/n_{x}\right)$
with 
\begin{multline}
f_{s}^{<}=\dfrac{\Delta\left(L/2\right)}{\left|\omega\right|}e^{-2\omega_{n}\left(x-L/2\right)/v_{F}n_{x}}+\int_{L/2}^{x}\dfrac{dy}{v_{F}n_{x}}\times\\
e^{-2\omega_{n}\left(x-y\right)/v_{F}n_{x}}\left(\mathbf{i}\mathcal{A}_{0}^{a}+\dfrac{n_{i}\mathcal{F}_{ij}^{a}}{2m}\dfrac{\partial}{\partial n_{j}}\right)f_{t}^{a<}\left(y\right)\label{eq:fs-Dyson-ball}
\end{multline}
and 
\begin{multline}
f_{t}^{a<}=\int_{L/2}^{x}\dfrac{dy}{v_{F}n_{x}}e^{-2\omega_{n}\left(x-y\right)/v_{F}n_{x}}\times\\
\left(e^{-\hat{P}_{i}n_{i}\left(x-y\right)/n_{x}}\right)^{ab}\left(\mathbf{i}\mathcal{A}_{0}^{b}+\dfrac{n_{i}\mathcal{F}_{ij}^{b}}{2m}\dfrac{\partial}{\partial n_{j}}\right)f_{s}^{<}\left(y\right)\;.\label{eq:ft-Dyson-ball}
\end{multline}
In the opposite propagation direction $f_{s,t}^{>}$ are found from
$f_{s,t}^{<}$ by substituting $L/2\rightarrow-L/2$.

In analogy with the diffusive case (\textit{cf.} Fig \ref{fig_diag}),
expressions \eqref{eq:fs-Dyson-ball} and \eqref{eq:ft-Dyson-ball}
show explicitly the effect of the SOC on the condensate function.
In the absence of SOC the exchange field $\mathcal{A}_{0}$ is the
only source for singlet-triplet conversion. The manifestation of the
triplet component in S-F-S junctions has been extensively studied
in the past (see \cite{buzdin.2005_RMP,Bergeret2005} for reviews).
As discussed in section \ref{sec:Diffusion-of-superconducting}, the
imaginary unit ${\bf i}$ in front of the $\mathcal{A}_{0}$ terms
leads to a $\pi/2$ phase shift. In the case of a finite SOC the gauge-field,
$\mathcal{F}_{ij}$, is an additional source of triplet correlations.
Notice that in the ballistic case, $\mathcal{F}_{ij}$ not only couples
the singlet and triplet components, but also the $s$-$p$-wave components
of the condensate \cite{Gorkov2001}. Moreover, the term $e^{-\hat{P}_{i}n_{i}x/n_{x}}$
in \eqref{eq:ft-Dyson-ball} leads to a momentum dependent rotation
of the triplet component in the spin-space. 

The origin of the anomalous phase $\varphi_{0}$ can be easily understood
in the lowest order in the spin fields. Assuming a vanishing phase
difference between the superconductors and combining Eqs.\eqref{eq:fs-Dyson-ball}-\eqref{eq:ft-Dyson-ball}
with the expression for the current \eqref{eq:obs-current-linearised},
one obtains for the first nontrivial contribution to the current:
$\mathcal{F}_{ij}^{a}\partial_{n_{j}}\left(e^{-\hat{P}_{i}n_{i}x/n_{x}}\right)^{ab}\mathcal{A}_{0}^{b}\propto\mathcal{F}_{ij}^{a}P_{j}^{ab}\mathcal{A}_{0}^{b}$.
This correction is proportional to $\mathcal{J}_{i}^{b}\mathcal{A}_{0}^{b}$
and coincides with those obtained in bulk superconductors with SOC(section
\ref{sec:Edelstein-effect}) and a diffusive S-X-S junctions (section
\ref{sec:Diffusive-limit}).

Quantitatively, a compact analytical solution for the current at zero-phase
difference can be obtained from Eqs. (\ref{eq:fs-Dyson-ball}-\ref{eq:ft-Dyson-ball})
in the short junction limit, \textit{i.e.} for $L\ll v_{F}/2\omega_{n}$:
\begin{multline}
j_{x}\left(\varphi=0\right)=-\dfrac{e\pi N_{0}v_{F}}{E_{F}}\dfrac{L^{3}}{3}\Delta^{2}T_{c}\mathcal{F}_{xi}^{a}\mathcal{F}_{i0}^{a}\times\\
{\displaystyle \sum_{\omega_{n}>0}}\left\langle \dfrac{e^{-2\omega_{n}L/v_{F}\left|n_{x}\right|}}{\omega_{n}^{2}\left|n_{x}\right|}\left(1+2\dfrac{n_{i}^{2}}{n_{x}^{2}}\right)\right\rangle \;.\label{eq:j0-ball}
\end{multline}
where $\mathcal{F}_{xi}^{a}\mathcal{F}_{i0}^{a}=\mathcal{J}_{x}^{a}\mathcal{A}_{0}^{a}$.
Thus the anomalous current is generated by the spin-polarization $\mathcal{A}_{0}^{a}$
via the spin-current $\mathcal{J}_{i}^{a}$. This is the spin-galvanic
effect, discussed in the previous sections, for a ballistic S-X- S
junction.

\subsection{Arbitrary temperatures\label{sub:All-T-ball}}

The previous result for the current has been obtained at temperatures
close to the critical one. We now consider an arbitrary temperature
and calculate the current up to the lowest order in the gauge field
$\mathcal{F}_{ij}$. In order to calculate the current from Eq.\eqref{eq:obs-current}
to the lowest order in $\mathcal{F}_{ij}$ we need to compute the
first two components matrix $\check{g}=\check{g}^{\left(0\right)}+\check{g}^{\left(1\right)}+\cdots$
.

At zeroth order in $\mathcal{F}_{ij}$ the ballistic equation reduces
to 
\begin{equation}
v_{F}n_{x}\dfrac{\partial\check{g}^{\left(0\right)}}{\partial x}=\mathbf{i}\left[\tau_{3}\left(\mathbf{i}\omega_{n}+\mathcal{A}_{0}\right)+v_{F}n_{j}\mathcal{A}_{j}+\check{\Delta},\check{g}^{\left(0\right)}\right]\;,
\end{equation}
which admits for solution 
\begin{equation}
\check{g}^{\left(0\right)}=\check{u}\left(x\right)\check{g}_{0}^{\left(0\right)}\check{u}^{-1}\left(x\right)+\check{g}_{\infty}\;,\label{eq:g-0-ansatz}
\end{equation}
with $\check{g}_{0}^{\left(0\right)}$ a constant matrix found from
the boundary conditions. The propagator $\check{u}\left(x\right)$
in Eq. \eqref{eq:g-0-ansatz} is given by 
\begin{equation}
\check{u}\left(x\right)=\exp\left[\mathbf{i}\dfrac{\tau_{3}\left(\mathbf{i}\omega_{n}+\mathcal{A}_{0}\right)+v_{F}n_{j}\mathcal{A}_{j}+\check{\Delta}}{v_{F}n_{x}}x\right]\;.\label{eq:u-check}
\end{equation}
when we assume that neither $\mathcal{A}_{0}$ nor $\check{\Delta}$
nor $\mathcal{A}_{j}$ depend on the position. $\check{u}$describes
how the function $\check{g_{0}}$ ``propagates'' from its value
at $x=0$, $\check{g}_{0}^{\left(0\right)}$ , to any point $x$.

The constant $\check{g}_{\infty}$ in \eqref{eq:g-0-ansatz} satisfies
\begin{equation}
\left[\tau_{3}\left(\mathbf{i}\omega_{n}+\mathcal{A}_{0}\right)+v_{F}n_{j}\mathcal{A}_{j}+\check{\Delta},\check{g}_{\infty}\right]=0\;,\label{eq:g-inf-equation}
\end{equation}
and describes the bulk contribution deep inside the superconductor.
Notice that according to Eqs.\eqref{eq:g-inf-equation}-\eqref{eq:u-check},
$\left[\check{g}_{\infty},\check{u}\right]=0$, hence $\check{g}_{\infty}$
can not be obtained by the application of \eqref{eq:u-check}. As
we will see below {[}see \eqref{eq:sol-supra}{]}, the solutions of
$\check{u}$ in a superconductor are evanescent waves, so the contribution
$\check{u}\check{g}_{0}^{\left(0\right)}\check{u}^{-1}$ vanishes
deep inside the superconductor, whereas the contribution $\check{g}_{\infty}$
remains finite.

The first-order correction with respect to the gauge-field satisfies
\begin{multline}
v_{F}n_{x}\dfrac{\partial\check{g}^{\left(1\right)}}{\partial x}-\left\{ \dfrac{n_{i}\mathcal{F}_{ij}}{2m},\dfrac{\partial\check{g}^{\left(0\right)}}{\partial n_{j}}\right\} \\
=\mathbf{i}\left[\tau_{3}\left(\mathbf{i}\omega_{n}+\mathcal{A}_{0}\right)+v_{F}n_{j}\mathcal{A}_{j}+\check{\Delta},\check{g}^{\left(0\right)}\right]
\end{multline}
and so 
\begin{equation}
\check{g}^{\left(1\right)}=\check{u}\left(x\right)\check{g}_{0}^{\left(1\right)}\left(x\right)\check{u}^{-1}\left(x\right)\;,\label{eq:g-1-ansatz}
\end{equation}
with a position-dependent $\check{g}_{0}^{\left(1\right)}$ matrix,
which reads 
\begin{multline}
\check{g}_{0}^{\left(1\right)}=\check{g}_{1}\\
+\int_{0}^{x}\dfrac{dz}{v_{F}n_{x}}\left[\check{u}^{\dagger}\left(z\right)\left\{ \dfrac{n_{i}\mathcal{F}_{ij}}{2m},\dfrac{\partial\check{g}^{\left(0\right)}}{\partial n_{j}}\right\} \check{u}\left(z\right)\right]\;,\label{eq:g-0-1}
\end{multline}
where $\check{g}_{1}$ is a constant matrix.

The current can be written in powers of $\mathcal{F}_{ij}$, $j_{x}=j_{x}^{\left(0\right)}+j_{x}^{\left(1\right)}+\cdots$
 with {[}see \eqref{eq:obs-current}{]}
\begin{equation}
j_{x}^{\left(0\right)}=\dfrac{\mathbf{i}e\pi N_{0}v_{F}}{2}\sum_{\omega_{n}}\Tr\left\langle n_{x}\check{g}_{0}^{\left(0\right)}\tau_{3}\right\rangle 
\end{equation}
and the first order correction 
\begin{equation}
j_{x}^{\left(1\right)}=\dfrac{\mathbf{i}e\pi N_{0}v_{F}}{2}\sum_{\omega_{n}}\Tr\left\langle n_{x}\check{g}_{1}\tau_{3}\right\rangle 
\end{equation}
Notice that the second line in \eqref{eq:g-0-1} vanishes after the
angular average. We then need to obtain $\check{g}_{0}^{\left(0\right)}$
and $\check{g}_{1}$ to determine the current through the S-X-S Josephson
junction.

We separate the solution of the problem in the three regions: the
two superconductors ($x\geq L/2$ and $x\leq-L/2$) and the normal
region ($-L/2\leq x\leq L/2$). One can check that in the superconductors:
\begin{align}
\check{g}\left(x\leq-\dfrac{L}{2}\right) & =e^{-\mathbf{i}\tau_{3}\frac{\varphi}{4}}\left[S_{L}g_{L}\tau_{+}S_{L}^{\dagger}+\check{g}_{\infty}\right]e^{\mathbf{i}\tau_{3}\frac{\varphi}{4}}\;,\nonumber \\
\check{g}\left(x\geq\dfrac{L}{2}\right) & =e^{\mathbf{i}\tau_{3}\frac{\varphi}{4}}\left[S_{R}g_{R}\tau_{-}S_{R}^{\dagger}+\check{g}_{\infty}\right]e^{-\mathbf{i}\tau_{3}\frac{\varphi}{4}}\;,\label{eq:sol-supra}
\end{align}
with $\tau_{\pm}=\left(\tau_{1}\pm\mathbf{i}\tau_{2}\right)/2$ and
\begin{equation}
\check{g}_{\infty}=\dfrac{\tau_{3}\omega_{n}+\tau_{2}\Delta}{\sqrt{\omega_{n}^{2}+\Delta^{2}}}=\dfrac{\tau_{3}\sinh\eta+\tau_{2}}{\cosh\eta}\;,\label{eq:g-infinity}
\end{equation}
\begin{equation}
S_{L,R}=\dfrac{e^{\eta/2}+\mathbf{i}\tau_{1}e^{-\eta/2}}{\sqrt{2\cosh\eta}}e^{\tau_{3}\cosh\eta\left(x\pm L/2\right)/\xi_{0}}\;,
\end{equation}
where $\sinh\eta=\hbar\omega_{n}/\Delta$, and $\xi_{0}=\hbar v_{F}/\Delta$
is the superconducting coherence length. The matrices $g_{L,R}\approx g_{L,R}^{\left(0\right)}+g_{L,R}^{\left(1\right)}+\cdots$
have been expanded in power of $\mathcal{F}_{ij}$ ; $g_{L,R}^{\left(0,1,\cdots\right)}$
are constant matrices found from boundary conditions order by order.
$\check{g}_{\infty}$ is present at the zeroth order only.

In the normal region, the solution reads 
\begin{multline}
\check{g}\left(-\dfrac{L}{2}\leq x\leq\dfrac{L}{2}\right)=\check{u}_{0}\left(x\right)\check{g}_{0}\check{u}_{0}^{\dagger}\left(x\right)+\\
+\check{u}_{0}\left(x\right)\check{g}_{0}^{\left(1\right)}\left(x\right)\check{u}_{0}^{\dagger}\left(x\right)+\cdots\;,\label{eq:gN}
\end{multline}
where 
\begin{equation}
\check{g}_{0}^{\left(1\right)}\left(x\right)=\check{g}_{1}+\int_{0}^{x}\check{\mathfrak{G}}\left(z\right)dz\;,\label{eq:g-0-1-0}
\end{equation}
\[
\check{\mathfrak{G}}\left(z\right)=\dfrac{1}{v_{F}n_{x}}\check{u}_{0}^{\dagger}\left(z\right)\left\{ \dfrac{n_{i}\mathcal{F}_{ij}}{2m},\dfrac{\partial\check{u}_{0}\left(z\right)\check{g}_{0}\check{u}_{0}^{\dagger}\left(z\right)}{\partial n_{j}}\right\} \check{u}_{0}\left(z\right)\;,
\]
where $\check{u}_{0}=\check{u}\left(\Delta=0\right)$ {[}see \eqref{eq:u-check}{]}
is a unitary matrix that can be written as 
\[
\check{u}_{0}\left(x\right)=e^{-\omega_{n}\tau_{3}x/v_{F}n_{x}}\left(\begin{array}{cc}
u & 0\\
0 & \bar{u}
\end{array}\right)\;.
\]

The spin matrices $u$ and $\bar{u}$ are defined as 
\[
u\left(x,\mathbf{n}\right)=\exp\left[\mathbf{i}\dfrac{\mathcal{A}_{0}+v_{F}n_{j}\mathcal{A}_{j}}{v_{F}n_{x}}x\right]
\]
\begin{align}
\bar{u}\left(x,\mathbf{n}\right) & =\sigma^{y}u^{\ast}\left(x,-\mathbf{n}\right)\sigma^{y}\nonumber \\
 & =\exp\left[\mathbf{i}\dfrac{-\mathcal{A}_{0}+v_{F}n_{j}\mathcal{A}_{j}}{v_{F}n_{x}}x\right]\;.\label{eq:u}
\end{align}

The matrices $\check{g}_{0}$ and $\check{g}_{1}$ in Eq. \eqref{eq:gN}
are obtained from the boundary conditions, assuming continuity of
the GFs at the left and right boundaries. At zeroth order we obtain
\begin{equation}
\check{g}_{0}=\left(\begin{array}{cc}
g_{0} & f_{0}\\
-\bar{f}_{0} & -\bar{g}_{0}
\end{array}\right)\label{eq:g-parameterisation-1}
\end{equation}
\begin{equation}
g_{0}=\dfrac{U\left(-\dfrac{L}{2}\right)\bar{U}\left(\dfrac{L}{2}\right)-U\left(\dfrac{L}{2}\right)\bar{U}\left(-\dfrac{L}{2}\right)+2\sinh2\chi}{2\cosh2\chi+\Tr\left\{ U\left(L\right)\right\} }\label{eq:g0}
\end{equation}
\begin{equation}
f_{0}=-2\mathbf{i}\dfrac{e^{\chi}U\left(\dfrac{L}{2}\right)+e^{-\chi}U\left(-\dfrac{L}{2}\right)}{2\cosh2\chi+\Tr\left\{ U\left(L\right)\right\} }\label{eq:f0}
\end{equation}
\begin{align}
\chi & =\dfrac{\omega_{n}L}{v_{F}n_{x}}+\arcsinh\dfrac{\omega_{n}}{\Delta}+\mathbf{i}\dfrac{\varphi}{2}\label{eq:chi}
\end{align}
\begin{equation}
U\left(x\right)=u\left(x\right)\bar{u}\left(-x\right)\label{eq:U}
\end{equation}
whereas $\bar{U}\left(x\right)=\bar{u}\left(x\right)u\left(-x\right)$
is its time-reversal conjugate. We here give only the contribution
corresponding to the positive projection of the Fermi velocity, the
negative projection can be found straightforwardly. 

The matrices entering the first-order correction, Eq. \eqref{eq:g-0-1-0},
have the following form in Nambu space 
\begin{equation}
\check{g}_{1}=\left(\begin{array}{cc}
g_{1} & f_{1}\\
-\bar{f}_{1} & -\bar{g}_{1}
\end{array}\right)\;\text{and}\;\check{\mathfrak{G}}=\left(\begin{array}{cc}
\mathfrak{G} & \mathfrak{F}\\
-\bar{\mathfrak{F}} & -\bar{\mathfrak{G}}
\end{array}\right)\;.
\end{equation}
with 
\begin{multline}
g_{1}=-\dfrac{1}{2}\left[\int_{0}^{-L/2}+\int_{0}^{L/2}\right]\mathfrak{G}dz\\
+\dfrac{1}{2}\int_{-L/2}^{L/2}\mathfrak{G}\cdot g_{0}dz-\dfrac{1}{2}\int_{-L/2}^{L/2}\mathfrak{F}\cdot\bar{f}_{0}dz\;.\label{eq:g1}
\end{multline}
After multiplication by $n_{x}$ and taking the angular average the
first line of this equation vanishes. The second line of \eqref{eq:g1}
can be simplified using the normalization condition $g_{0}^{2}-f_{0}\bar{f}_{0}=1$
available for the zeroth order correction. We obtain 
\begin{multline}
j_{x}^{\left(1\right)}=\mathbf{i}\dfrac{e\pi N_{0}Tv_{F}}{2}{\displaystyle \sum_{\omega_{n}>0}}\int_{-L/2}^{L/2}dz\sum_{\alpha=\pm}\times\\
\Tr\left\langle \dfrac{n_{i}\mathcal{F}_{ij}}{2m}\left(\mathfrak{f}_{\alpha}\dfrac{\partial\bar{\mathfrak{f}}_{\alpha}}{\partial n_{j}}-\bar{\mathfrak{f}}_{\alpha}\dfrac{\partial\mathfrak{f}_{\alpha}}{\partial n_{j}}\right)\sgn\left(n_{x}\right)\right\rangle \;,\label{eq:Trvg1}
\end{multline}
with 
\[
\mathfrak{f}_{\pm}\left(z\right)=-2\mathbf{i}\dfrac{e^{\pm\chi_{\pm}}U\left(z+\dfrac{L}{2}\right)+e^{\mp\chi_{\pm}}U\left(z-\dfrac{L}{2}\right)}{2\cosh2\chi_{\pm}+\Tr\left\{ U\left(L\right)\right\} }
\]
\begin{equation}
\chi_{\pm}=\dfrac{\omega_{n}L}{v_{F}\left|n_{x}\right|}+\arcsinh\dfrac{\omega_{n}}{\Delta}\pm\mathbf{i}\dfrac{\varphi}{2}
\end{equation}
 and $\bar{\mathfrak{f}}_{\pm}\left(z,\mathbf{n}\right)=\sigma^{y}\mathfrak{f}_{\pm}^{\ast}\left(z,-\mathbf{n}\right)\sigma^{y}$
its time reversal conjugate.

If $\mathcal{A}_{0}$ commutes with $\mathcal{A}_{i}$, then $u\left(x\right)\bar{u}\left(-x\right)=\exp\left[\mathbf{i}\mathcal{A}_{0}x/v_{F}n_{x}\right]$
becomes independent of the SOC {[}see the definitions \eqref{eq:u}{]},
and the contribution \eqref{eq:Trvg1} vanishes. Therefore we expect
\eqref{eq:Trvg1} to be proportional to $\mathcal{F}_{ij}\left[\mathcal{A}_{0},\mathcal{A}_{j}\right]\propto\mathcal{F}_{ij}\mathcal{F}_{0j}$
at the smallest order in the gauge-fields.

\begin{figure}[b]
\includegraphics[width=0.8\linewidth]{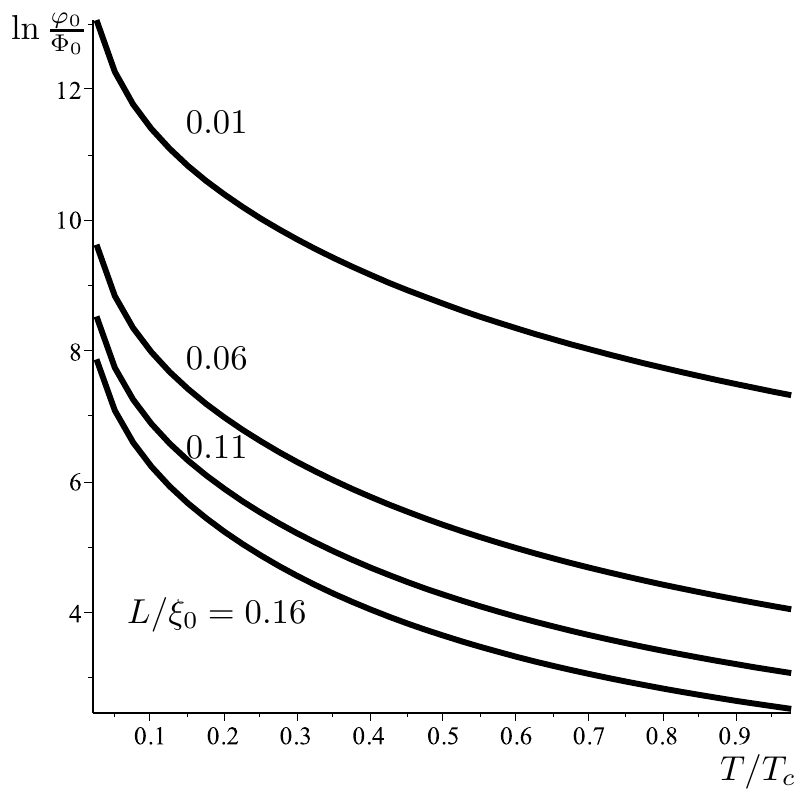}\protect\caption{\label{fig:phi-0-averaged}The temperature dependence of the averaged
anomalous phase shift \eqref{eq:phi-0-plot} in a log-scale $\ln\left(\varphi_{0}\left(T\right)/\Phi_{0}\right)$,
where $\Phi_{0}=-\hbar L^{3}\Tr\left\{ \mathcal{F}_{xy}\mathcal{F}_{y0}\right\} /6E_{F}$.
We have approximated $\Delta\left(T\right)\approx1,764T_{c}\tanh\left(1,74\sqrt{T_{c}/T-1}\right)$
which is the usual interpolation for the temperature dependence of
the superconducting gap. The curves are given for different ratio
of $L/\xi_{0}=\left\{ 0.01,\;0.06,\;0.11,\;0.16\right\} $, with $\xi_{0}=\hbar v_{F}/\Delta_{0}$
and $\Delta_{0}=1.764T_{c}$ the gap at zero temperature. Note that
$\varphi_{0}$ does not vanish when $T\rightarrow T_{c}$.}
\end{figure}

By expanding the expression \eqref{eq:Trvg1} in the gauge-potentials,
up to the term proportional to the electric-like field one obtains
\begin{widetext}
\begin{multline}
\mathfrak{f}_{\pm}\left(z\right)\approx\dfrac{-\mathbf{i}e^{\pm\chi_{\pm}}}{2\cosh^{2}\chi_{\pm}}\left(1+\dfrac{2\mathbf{i}A_{0}}{v_{x}}\left(z+\dfrac{L}{2}\right)+\dfrac{1}{2}\left(\dfrac{2\mathbf{i}A_{0}}{v_{x}}\left(z+\dfrac{L}{2}\right)\right)^{2}-\dfrac{\left(z+L/2\right)^{2}}{v_{x}^{2}}v_{k}\left[A_{k},A_{0}\right]\right)+\\
\dfrac{-\mathbf{i}e^{\mp\chi_{\pm}}}{2\cosh^{2}\chi_{\pm}}\left(1+\dfrac{2\mathbf{i}A_{0}}{v_{x}}\left(z-\dfrac{L}{2}\right)+\dfrac{1}{2}\left(\dfrac{2\mathbf{i}A_{0}}{v_{x}}\left(z-\dfrac{L}{2}\right)\right)^{2}-\dfrac{\left(z-L/2\right)^{2}}{v_{x}^{2}}v_{k}\left[A_{k},A_{0}\right]\right)+\cdots
\end{multline}
\end{widetext} where $\Tr\left\{ U\left(L\right)\right\} \approx2$
in the small gauge-field limit. Besides the terms proportional to
$\mathcal{A}_{0}$-only, responsible for the oscillations of S/F proximity
effect, the SOC $\mathcal{A}_{j}$ only appears in the electric-field
construction (the last term on each line), due to symmetry with respect
to the time-reversal. After angular averaging only the last contributions
of the two lines are non-zero. This leads to 
\begin{equation}
\dfrac{j_{x}^{\left(0\right)}}{j_{0}}=4\left\langle \left|n_{x}\right|M\right\rangle \;,\label{eq:jx0}
\end{equation}
 
\begin{multline}
\dfrac{j_{x}^{\left(1\right)}}{j_{0}}=-2\dfrac{L^{3}}{3E_{F}}\times\\
\Tr\left\{ \mathcal{F}_{xj}\mathcal{F}_{j0}\right\} \left\langle \dfrac{1}{\left|n_{x}\right|}\left(1+2\dfrac{n_{j}^{2}}{n_{x}^{2}}\right)\dfrac{\partial M}{\partial\varphi}\right\rangle \label{eq:jx1}
\end{multline}
with $j_{0}=\pi ev_{F}N_{0}T$ and
\begin{equation}
M={\displaystyle \sum_{\omega_{n}>0}}\Im\left\{ \tanh\left(\dfrac{\omega_{n}L}{v_{F}\left|n_{x}\right|}+\arcsinh\dfrac{\omega_{n}}{\Delta}+\mathbf{i}\dfrac{\varphi}{2}\right)\right\} 
\end{equation}
(note that the sum over $j$ applies inside the angular averaging
as well). As in all previous examples the anomalous current is proportional
to $\Tr\left\{ \mathcal{F}_{xk}\mathcal{F}_{k0}\right\} =\mathcal{J}_{x}^{a}\mathcal{A}_{0}^{a}$,
where the later form suggests our expressions are valid beyond the
linear-in-momentum-SOC approximation, given any spin current $\mathcal{J}_{i}^{a}$
and paramagnetic interaction $\mathcal{A}_{0}^{a}$. 

Close to the critical temperature $M\approx\Delta^{2}\sin\varphi\sum_{\omega_{n}\geq0}e^{-2\omega L/v_{F}\left|n_{x}\right|}/2\omega^{2}$
and we recover \eqref{eq:j0-ball}.

Commonly, the concept of a $\varphi_{0}$-junction is defined for
junctions with a sinusoidal current-phase relation. This is valid
at temperatures close to the critical temperature or in the case of
a weak proximity effect between the S electrodes and the X bridge.
However, in several cases the current-phase relation is more complex
and higher harmonics are involved \cite{golubov_kupriyanov.2004}.
This is the case of the ballistic junction studied here with a current-phase
relation given by sum of Eqs. \eqref{eq:jx0} and \eqref{eq:jx1}.
In such cases the $\varphi_{0}$ is defined as the phase difference
across the junction that minimize the energy, or equivalently, as
the phase difference imposed to the junction in order to get a zero
current state, \textit{i.e.} $j\left(\varphi_{0}\right)=0$. In our
perturbative analysis $\varphi_{0}$ is small and hence 
\begin{equation}
\varphi_{0}=-\dfrac{j\left(\varphi=0\right)}{\left.\partial_{\varphi}j\right|_{\varphi=0}}\;.
\end{equation}
It is clear that $j\left(\varphi=0\right)=j_{x}^{\left(1\right)}$,
whereas $\left.\partial_{\varphi}j\right|_{\varphi=0}=\left.\partial_{\varphi}j_{x}^{\left(0\right)}\right|_{\varphi=0}$,
and from Eqs. \eqref{eq:jx0} and \eqref{eq:jx1} we obtain 
\begin{equation}
\varphi_{0}=-\dfrac{\hbar L^{3}}{6E_{F}}\dfrac{\mathcal{F}_{xj}^{a}\mathcal{F}_{j0}^{a}\left.\dfrac{\partial}{\partial\varphi}\left\langle \dfrac{M}{\left|n_{x}\right|}\left(1+2\dfrac{n_{j}^{2}}{n_{x}^{2}}\right)\right\rangle \right|_{\varphi=0}}{\left.\dfrac{\partial}{\partial\varphi}\left\langle \left|n_{x}\right|M\right\rangle \right|_{\varphi=0}}\;.\label{eq:phi-0-plot}
\end{equation}

In Fig. \ref{fig:phi-0-averaged} we show the temperature dependence
of $\varphi_{0}$ for the ballistic junction for a 2D system when
only $\mathcal{F}_{xy}^{a}\mathcal{F}_{y0}^{a}$ is non-zero. We assume
a circular Fermi surface, $n_{x}=\cos\theta$ and $n_{y}=\sin\theta$.
We plot the anomalous phase for different junction lengths.

\section{Discussion and Conclusions\label{sec:Discussion-and-Conclusions}}

\begin{figure}[b]
\includegraphics[width=0.85\columnwidth]{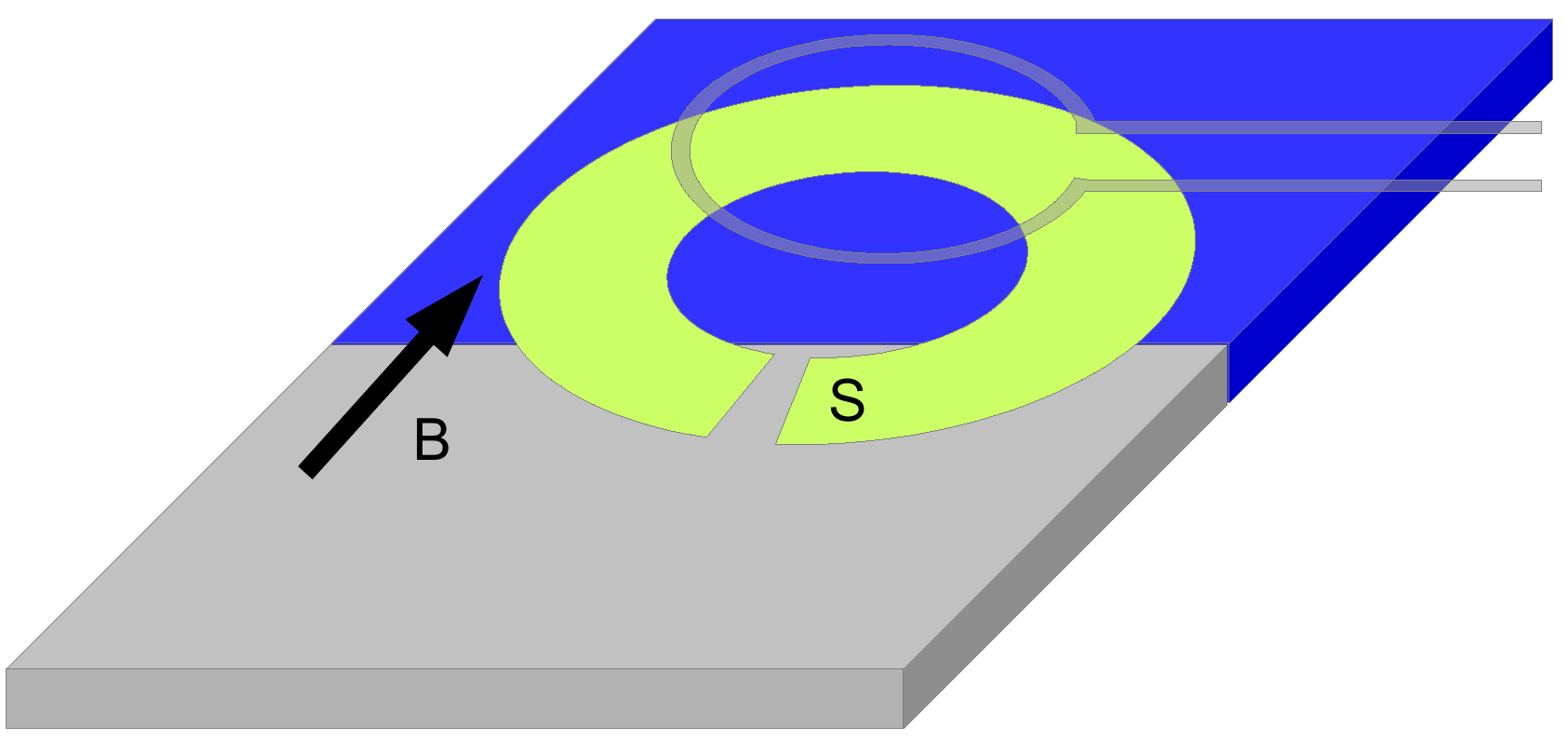}

\protect\caption{\label{fig:setup-geometries} One possible experimental setup to verify
the generation of a spontaneous current in S/X/S Josephson junction.
It consists of a truncated superconducting loop (in green) deposited
on top of a material exhibiting SOC (in grey). In order to isolate
electrically part of the S loop from the conducting substrate we assume
an insulating layer between them (in blue). By applying an in-plane
magnetic field a circulating supercurrent might be generated which
in turn induces a magnetic flux that can be measured via an extra
pick-up coil (shaded grey). For more details see discussion in the
main text. }
\end{figure}

In order to verify our findings and prove the existence of the anomalous
$\varphi_{0}$ phase one can design a superconducting ring interrupted
by a semiconducting link with a strong SOC, similar to the one used
recently in Ref. \cite{Sochnikov2014} for the characterization of
the current phase relation of a Nb/3D-HgTe/Nb junction or in \cite{bauer_bentner_aprili_etal.2004}
for the observation of a spontaneous supercurrent induced by a ferromagnetic
$\pi$-junction. A schematically view of the proposed setup is shown
in Fig.\ref{fig:setup-geometries}: It consists in a superconducting
ring (green) grown on top of a semiconductor or a metallic substrate
with strong SOC (grey). In order to isolate electrically the S ring
from the semiconductor one can for example add an insulating barrier
(blue) under the ring.

If a magnetic field is applied in the plane of the ring, it will act
as a Zeeman field and hence, according to our previous results, it
will create a spontaneous circulating supercurrent, see \cite{Bulaevskii1977,Buzdin2009}
for more details. This supercurrent will generate a magnetic flux
that in principle can be measured by a second loop \cite{Sochnikov2014}
or a micro-Hall sensor \cite{bauer_bentner_aprili_etal.2004}. 

In the case when the bridge is made of a 2D semiconductor with a generic
SOC described by a combination of Rashba and Dresselhaus terms: $\mathcal{A}_{x}=-\alpha\sigma^{y}+\beta\sigma^{x}$
and $\mathcal{A}_{y}=\alpha\sigma^{x}-\beta\sigma^{y}$ the generated
supercurrent should be proportional to 
\[
j_{s}\propto\left(\alpha^{2}-\beta^{2}\right)\left(h_{x}\beta+h_{y}\alpha\right)\;.
\]
Thus the current depends on the direction of the applied magnetic
field. In particular, for a field perpendicular to the 2D gas the
effect should vanishes. In addition by applying a gate voltage one
could modify the ratio between Dresselhaus and Rashba interactions
and hence control the supercurrent flow. We thus expect that the dependency
of the spontaneous supercurrent with respect to the orientation of
the magnetic field and/or the gate voltage realizes a clear demonstration
of the spin-galvanic effect in Josephson systems.

Instead of using a semiconducting bridge one could grow the superconducting
loop on top of a metallic substrate. Metals with strong SOC, like
Pt and Ta, are good candidates to observe the $\varphi_{0}$-junction
behavior, but also an ultra-thin layer of Pb might be used \cite{Serrier-Garcia2013}.
In such a case probably one cannot control the $\varphi_{0}$-shift
using a gate, but a spontaneous circulating current might still be
controlled by switching the in-plane external field on and off.

Eventually, the existence of a magneto-electric phase-shift $\varphi_{0}$
can be probed by measuring the Shapiro steps in S-X-S Josephson junctions
as suggested in Ref. \cite{Konschelle2009b}. 

In conclusion, we have demonstrated that the inverse Edelstein effect,
also called spin-galvanic effect, and the appearance of an anomalous
phase-shift $\varphi_{0}$ in Josephson junctions are the two sides
of the same coin. We presented a full SU(2) covariant quasi-classic
formalism that allows to study these magneto-electric phenomena in
bulk and hybrid superconducting structures with arbitrary linear-in-momentum
SOC (section \ref{sec:Model}). 

With the help of our quasi-classic transport formalism we derived
the Edelstein effect close to the critical temperature of a bulk superconductor,
recovering the Edelstein's result in a very compact way (section \ref{sec:Edelstein-effect})
and generalizing it for the case of an arbitrary linear-in-momentum
SOC. We have shown that the Edelstein effect and its inverse are reciprocal
in the sense of the Onsager relations, both in ballistic and diffusive
superconducting systems: A static supercurrent can induce a finite
magnetization due to the presence of a spin-orbit coupling, and reciprocally
a finite magnetization produces a finite supercurrent in a bulk system.
We have demonstrated that the linear-response tensor is directly proportional
to the equilibrium spin-current tensor $\mathcal{J}_{i}^{a}$.

We have also generalized this result to inhomogeneous systems. In
particular we have studied the current-phase relation of a Josephson
junction consisting of two superconductors coupled via a normal metal
with both SOC and spin-splitting field. We have demonstrated that
a supercurrent can flow even if the phase difference between the S
electrodes is zero. This current is associated to an anomalous phase-shift
$\varphi_{0}$. This result holds for both ballistic (section \ref{sec:Ball-limit})
and diffusive systems (section \ref{sec:Diffusive-limit}), for arbitrary
linear-in-momentum spin-orbit coupling, and for arbitrary barrier
resistance between the superconductor and the normal metal. For all
these situations we have demonstrated that SU(2) gauge-fields are
the only objects of relevance in the phenomenology of the $\varphi_{0}$-shift,
and in particular we have shown that $\varphi_{0}\propto\mathcal{A}_{0}^{a}\mathcal{J}_{i}^{a}=\mathcal{F}_{0j}^{a}\mathcal{F}_{ji}^{a}$,
\textit{i.e.} the anomalous phase-shift is proportional to the SU(2)
electric and magnetic fields, or equivalently to the spin-current
tensor. We thus directly linked the anomalous phase-shift in superconducting
systems to the inverse Edelstein effect (also known as the spin-galvanic
effect) extensively studied in normal systems. 
\begin{acknowledgments}
F.K. thanks F. Hassler and G. Viola for daily stimulating discussions,
as well as A.I. Buzdin for insightful discussions. F.K. is grateful
for support from the Alexander von Humboldt foundation during his
stay in the JARA-Institute for Quantum Information at RWTH-Aachen,
Germany, where part of this work was done. The work of F.S.B. was
supported by Spanish Ministerio de Econom\'{i}a y Competitividad (MINECO)
through the Project No. FIS2011-28851- C02-02 and the Basque Government
under UPV/EHU Project No. IT-756-13. I.V.T. acknowledges support from
the Spanish Grant FIS2013-46159-C3-1-P, and from the \textquotedblleft Grupos
Consolidados UPV/EHU del Gobierno Vasco\textquotedblright{} (Gant
No. IT578-13)
\end{acknowledgments}

\bibliographystyle{apsrev4-1}
\bibliography{/Users/paradis/Desktop/library,library}

\begin{thebibliography}{148}%
\makeatletter
\providecommand \@ifxundefined [1]{%
 \@ifx{#1\undefined}
}%
\providecommand \@ifnum [1]{%
 \ifnum #1\expandafter \@firstoftwo
 \else \expandafter \@secondoftwo
 \fi
}%
\providecommand \@ifx [1]{%
 \ifx #1\expandafter \@firstoftwo
 \else \expandafter \@secondoftwo
 \fi
}%
\providecommand \natexlab [1]{#1}%
\providecommand \enquote  [1]{``#1''}%
\providecommand \bibnamefont  [1]{#1}%
\providecommand \bibfnamefont [1]{#1}%
\providecommand \citenamefont [1]{#1}%
\providecommand \href@noop [0]{\@secondoftwo}%
\providecommand \href [0]{\begingroup \@sanitize@url \@href}%
\providecommand \@href[1]{\@@startlink{#1}\@@href}%
\providecommand \@@href[1]{\endgroup#1\@@endlink}%
\providecommand \@sanitize@url [0]{\catcode `\\12\catcode `\$12\catcode
  `\&12\catcode `\#12\catcode `\^12\catcode `\_12\catcode `\%12\relax}%
\providecommand \@@startlink[1]{}%
\providecommand \@@endlink[0]{}%
\providecommand \url  [0]{\begingroup\@sanitize@url \@url }%
\providecommand \@url [1]{\endgroup\@href {#1}{\urlprefix }}%
\providecommand \urlprefix  [0]{URL }%
\providecommand \Eprint [0]{\href }%
\providecommand \doibase [0]{http://dx.doi.org/}%
\providecommand \selectlanguage [0]{\@gobble}%
\providecommand \bibinfo  [0]{\@secondoftwo}%
\providecommand \bibfield  [0]{\@secondoftwo}%
\providecommand \translation [1]{[#1]}%
\providecommand \BibitemOpen [0]{}%
\providecommand \bibitemStop [0]{}%
\providecommand \bibitemNoStop [0]{.\EOS\space}%
\providecommand \EOS [0]{\spacefactor3000\relax}%
\providecommand \BibitemShut  [1]{\csname bibitem#1\endcsname}%
\let\auto@bib@innerbib\@empty
\bibitem [{\citenamefont {Buzdin}(2005{\natexlab{a}})}]{buzdin.2005_RMP}%
  \BibitemOpen
  \bibfield  {author} {\bibinfo {author} {\bibfnamefont {A.~I.}\ \bibnamefont
  {Buzdin}},\ }\href {\doibase 10.1103/RevModPhys.77.935} {\bibfield  {journal}
  {\bibinfo  {journal} {Reviews of Modern Physics}\ }\textbf {\bibinfo {volume}
  {77}},\ \bibinfo {pages} {935} (\bibinfo {year}
  {2005}{\natexlab{a}})}\BibitemShut {NoStop}%
\bibitem [{\citenamefont {Bergeret}\ \emph {et~al.}(2005)\citenamefont
  {Bergeret}, \citenamefont {Volkov},\ and\ \citenamefont
  {Efetov}}]{Bergeret2005}%
  \BibitemOpen
  \bibfield  {author} {\bibinfo {author} {\bibfnamefont {F.~S.}\ \bibnamefont
  {Bergeret}}, \bibinfo {author} {\bibfnamefont {A.~F.}\ \bibnamefont
  {Volkov}}, \ and\ \bibinfo {author} {\bibfnamefont {K.~B.}\ \bibnamefont
  {Efetov}},\ }\href {\doibase 10.1103/RevModPhys.77.1321} {\bibfield
  {journal} {\bibinfo  {journal} {Reviews of Modern Physics}\ }\textbf
  {\bibinfo {volume} {77}},\ \bibinfo {pages} {1321} (\bibinfo {year}
  {2005})}\BibitemShut {NoStop}%
\bibitem [{\citenamefont {SanGiorgio}\ \emph {et~al.}(2008)\citenamefont
  {SanGiorgio}, \citenamefont {Reymond}, \citenamefont {Beasley}, \citenamefont
  {Kwon},\ and\ \citenamefont {Char}}]{SanGiorgio2008}%
  \BibitemOpen
  \bibfield  {author} {\bibinfo {author} {\bibfnamefont {P.}~\bibnamefont
  {SanGiorgio}}, \bibinfo {author} {\bibfnamefont {S.}~\bibnamefont {Reymond}},
  \bibinfo {author} {\bibfnamefont {M.~R.}\ \bibnamefont {Beasley}}, \bibinfo
  {author} {\bibfnamefont {J.~H.}\ \bibnamefont {Kwon}}, \ and\ \bibinfo
  {author} {\bibfnamefont {K.}~\bibnamefont {Char}},\ }\href {\doibase
  10.1103/PhysRevLett.100.237002} {\bibfield  {journal} {\bibinfo  {journal}
  {Physical Review Letters}\ }\textbf {\bibinfo {volume} {100}},\ \bibinfo
  {pages} {237002} (\bibinfo {year} {2008})}\BibitemShut {NoStop}%
\bibitem [{\citenamefont {Boden}\ \emph {et~al.}(2011)\citenamefont {Boden},
  \citenamefont {Pratt},\ and\ \citenamefont {Birge}}]{Boden2011}%
  \BibitemOpen
  \bibfield  {author} {\bibinfo {author} {\bibfnamefont {K.~M.}\ \bibnamefont
  {Boden}}, \bibinfo {author} {\bibfnamefont {W.~P.}\ \bibnamefont {Pratt}}, \
  and\ \bibinfo {author} {\bibfnamefont {N.~O.}\ \bibnamefont {Birge}},\ }\href
  {\doibase 10.1103/PhysRevB.84.020510} {\bibfield  {journal} {\bibinfo
  {journal} {Physical Review B}\ }\textbf {\bibinfo {volume} {84}},\ \bibinfo
  {pages} {020510} (\bibinfo {year} {2011})},\ \Eprint
  {http://arxiv.org/abs/1108.5613} {arXiv:1108.5613} \BibitemShut {NoStop}%
\bibitem [{\citenamefont {Kontos}\ \emph {et~al.}(2001)\citenamefont {Kontos},
  \citenamefont {Aprili}, \citenamefont {Lesueur},\ and\ \citenamefont
  {Grison}}]{Kontos2001}%
  \BibitemOpen
  \bibfield  {author} {\bibinfo {author} {\bibfnamefont {T.}~\bibnamefont
  {Kontos}}, \bibinfo {author} {\bibfnamefont {M.}~\bibnamefont {Aprili}},
  \bibinfo {author} {\bibfnamefont {J.}~\bibnamefont {Lesueur}}, \ and\
  \bibinfo {author} {\bibfnamefont {X.}~\bibnamefont {Grison}},\ }\href
  {\doibase 10.1103/PhysRevLett.86.304} {\bibfield  {journal} {\bibinfo
  {journal} {Physical Review Letters}\ }\textbf {\bibinfo {volume} {86}},\
  \bibinfo {pages} {304} (\bibinfo {year} {2001})}\BibitemShut {NoStop}%
\bibitem [{\citenamefont {Cottet}(2007)}]{Cottet2007}%
  \BibitemOpen
  \bibfield  {author} {\bibinfo {author} {\bibfnamefont {A.}~\bibnamefont
  {Cottet}},\ }\href {\doibase 10.1103/PhysRevB.76.224505} {\bibfield
  {journal} {\bibinfo  {journal} {Physical Review B}\ }\textbf {\bibinfo
  {volume} {76}},\ \bibinfo {pages} {224505} (\bibinfo {year} {2007})},\
  \Eprint {http://arxiv.org/abs/0704.3975} {arXiv:0704.3975} \BibitemShut
  {NoStop}%
\bibitem [{\citenamefont {Bulaevskii}\ \emph {et~al.}(1977)\citenamefont
  {Bulaevskii}, \citenamefont {Kuzii},\ and\ \citenamefont
  {Sobyanin}}]{Bulaevskii1977}%
  \BibitemOpen
  \bibfield  {author} {\bibinfo {author} {\bibfnamefont {L.}~\bibnamefont
  {Bulaevskii}}, \bibinfo {author} {\bibfnamefont {V.~V.}\ \bibnamefont
  {Kuzii}}, \ and\ \bibinfo {author} {\bibfnamefont {A.~A.}\ \bibnamefont
  {Sobyanin}},\ }\href@noop {} {\bibfield  {journal} {\bibinfo  {journal} {JETP
  Letters}\ }\textbf {\bibinfo {volume} {25}},\ \bibinfo {pages} {290}
  (\bibinfo {year} {1977})}\BibitemShut {NoStop}%
\bibitem [{\citenamefont {Buzdin}\ \emph {et~al.}(1982)\citenamefont {Buzdin},
  \citenamefont {Bulaevskii},\ and\ \citenamefont {Panyukov}}]{Buzdin1982}%
  \BibitemOpen
  \bibfield  {author} {\bibinfo {author} {\bibfnamefont {A.~I.}\ \bibnamefont
  {Buzdin}}, \bibinfo {author} {\bibfnamefont {L.}~\bibnamefont {Bulaevskii}},
  \ and\ \bibinfo {author} {\bibfnamefont {S.~V.}\ \bibnamefont {Panyukov}},\
  }\href@noop {} {\bibfield  {journal} {\bibinfo  {journal} {Sov. Phys. JETP}\
  }\textbf {\bibinfo {volume} {35}},\ \bibinfo {pages} {178} (\bibinfo {year}
  {1982})}\BibitemShut {NoStop}%
\bibitem [{\citenamefont {Ryazanov}\ \emph {et~al.}(2001)\citenamefont
  {Ryazanov}, \citenamefont {Oboznov}, \citenamefont {Rusanov}, \citenamefont
  {Veretennikov}, \citenamefont {Golubov},\ and\ \citenamefont
  {Aarts}}]{Ryazanov2001}%
  \BibitemOpen
  \bibfield  {author} {\bibinfo {author} {\bibfnamefont {V.~V.}\ \bibnamefont
  {Ryazanov}}, \bibinfo {author} {\bibfnamefont {V.}~\bibnamefont {Oboznov}},
  \bibinfo {author} {\bibfnamefont {A.}~\bibnamefont {Rusanov}}, \bibinfo
  {author} {\bibfnamefont {A.}~\bibnamefont {Veretennikov}}, \bibinfo {author}
  {\bibfnamefont {A.}~\bibnamefont {Golubov}}, \ and\ \bibinfo {author}
  {\bibfnamefont {J.}~\bibnamefont {Aarts}},\ }\href {\doibase
  10.1103/PhysRevLett.86.2427} {\bibfield  {journal} {\bibinfo  {journal}
  {Physical Review Letters}\ }\textbf {\bibinfo {volume} {86}},\ \bibinfo
  {pages} {2427} (\bibinfo {year} {2001})}\BibitemShut {NoStop}%
\bibitem [{\citenamefont {Kontos}\ \emph {et~al.}(2002)\citenamefont {Kontos},
  \citenamefont {Aprili}, \citenamefont {Lesueur}, \citenamefont {Gen\^{e}t},
  \citenamefont {Stephanidis},\ and\ \citenamefont {Boursier}}]{Kontos2002}%
  \BibitemOpen
  \bibfield  {author} {\bibinfo {author} {\bibfnamefont {T.}~\bibnamefont
  {Kontos}}, \bibinfo {author} {\bibfnamefont {M.}~\bibnamefont {Aprili}},
  \bibinfo {author} {\bibfnamefont {J.}~\bibnamefont {Lesueur}}, \bibinfo
  {author} {\bibfnamefont {F.}~\bibnamefont {Gen\^{e}t}}, \bibinfo {author}
  {\bibfnamefont {B.}~\bibnamefont {Stephanidis}}, \ and\ \bibinfo {author}
  {\bibfnamefont {R.}~\bibnamefont {Boursier}},\ }\href {\doibase
  10.1103/PhysRevLett.89.137007} {\bibfield  {journal} {\bibinfo  {journal}
  {Physical Review Letters}\ }\textbf {\bibinfo {volume} {89}},\ \bibinfo
  {pages} {137007} (\bibinfo {year} {2002})}\BibitemShut {NoStop}%
\bibitem [{\citenamefont {Bergeret}\ \emph
  {et~al.}(2001{\natexlab{a}})\citenamefont {Bergeret}, \citenamefont
  {Volkov},\ and\ \citenamefont {Efetov}}]{Bergeret2001}%
  \BibitemOpen
  \bibfield  {author} {\bibinfo {author} {\bibfnamefont {F.~S.}\ \bibnamefont
  {Bergeret}}, \bibinfo {author} {\bibfnamefont {A.~F.}\ \bibnamefont
  {Volkov}}, \ and\ \bibinfo {author} {\bibfnamefont {K.~B.}\ \bibnamefont
  {Efetov}},\ }\href {\doibase 10.1103/PhysRevB.64.134506} {\bibfield
  {journal} {\bibinfo  {journal} {Physical Review B}\ }\textbf {\bibinfo
  {volume} {64}},\ \bibinfo {pages} {134506} (\bibinfo {year}
  {2001}{\natexlab{a}})}\BibitemShut {NoStop}%
\bibitem [{\citenamefont {Bergeret}\ \emph
  {et~al.}(2001{\natexlab{b}})\citenamefont {Bergeret}, \citenamefont
  {Volkov},\ and\ \citenamefont {Efetov}}]{Bergeret2001a}%
  \BibitemOpen
  \bibfield  {author} {\bibinfo {author} {\bibfnamefont {F.~S.}\ \bibnamefont
  {Bergeret}}, \bibinfo {author} {\bibfnamefont {A.~F.}\ \bibnamefont
  {Volkov}}, \ and\ \bibinfo {author} {\bibfnamefont {K.~B.}\ \bibnamefont
  {Efetov}},\ }\href {\doibase 10.1103/PhysRevLett.86.3140} {\bibfield
  {journal} {\bibinfo  {journal} {Physical Review Letters}\ }\textbf {\bibinfo
  {volume} {86}},\ \bibinfo {pages} {3140} (\bibinfo {year}
  {2001}{\natexlab{b}})}\BibitemShut {NoStop}%
\bibitem [{\citenamefont {Robinson}\ \emph {et~al.}(2010)\citenamefont
  {Robinson}, \citenamefont {Witt},\ and\ \citenamefont
  {Blamire}}]{Robinson2010}%
  \BibitemOpen
  \bibfield  {author} {\bibinfo {author} {\bibfnamefont {J.~W.~A.}\
  \bibnamefont {Robinson}}, \bibinfo {author} {\bibfnamefont {J.~D.~S.}\
  \bibnamefont {Witt}}, \ and\ \bibinfo {author} {\bibfnamefont {M.~G.}\
  \bibnamefont {Blamire}},\ }\href {\doibase 10.1126/science.1189246}
  {\bibfield  {journal} {\bibinfo  {journal} {Science (New York, N.Y.)}\
  }\textbf {\bibinfo {volume} {329}},\ \bibinfo {pages} {59} (\bibinfo {year}
  {2010})}\BibitemShut {NoStop}%
\bibitem [{\citenamefont {Khaire}\ \emph {et~al.}(2009)\citenamefont {Khaire},
  \citenamefont {Khasawneh}, \citenamefont {Pratt},\ and\ \citenamefont
  {Birge}}]{Khaire2010}%
  \BibitemOpen
  \bibfield  {author} {\bibinfo {author} {\bibfnamefont {T.~S.}\ \bibnamefont
  {Khaire}}, \bibinfo {author} {\bibfnamefont {M.~A.}\ \bibnamefont
  {Khasawneh}}, \bibinfo {author} {\bibfnamefont {W.~P.}\ \bibnamefont
  {Pratt}}, \ and\ \bibinfo {author} {\bibfnamefont {N.~O.}\ \bibnamefont
  {Birge}},\ }\href {\doibase 10.1103/PhysRevLett.104.137002} {\bibfield
  {journal} {\bibinfo  {journal} {Physical Review Letters}\ }\textbf {\bibinfo
  {volume} {104}},\ \bibinfo {pages} {4} (\bibinfo {year} {2009})},\ \Eprint
  {http://arxiv.org/abs/0912.0205} {arXiv:0912.0205} \BibitemShut {NoStop}%
\bibitem [{\citenamefont {Anwar}\ \emph {et~al.}(2010)\citenamefont {Anwar},
  \citenamefont {Czeschka}, \citenamefont {Hesselberth}, \citenamefont
  {Porcu},\ and\ \citenamefont {Aarts}}]{Anwar2010}%
  \BibitemOpen
  \bibfield  {author} {\bibinfo {author} {\bibfnamefont {M.}~\bibnamefont
  {Anwar}}, \bibinfo {author} {\bibfnamefont {F.}~\bibnamefont {Czeschka}},
  \bibinfo {author} {\bibfnamefont {M.}~\bibnamefont {Hesselberth}}, \bibinfo
  {author} {\bibfnamefont {M.}~\bibnamefont {Porcu}}, \ and\ \bibinfo {author}
  {\bibfnamefont {J.}~\bibnamefont {Aarts}},\ }\href {\doibase
  10.1103/PhysRevB.82.100501} {\bibfield  {journal} {\bibinfo  {journal}
  {Physical Review B}\ }\textbf {\bibinfo {volume} {82}},\ \bibinfo {pages}
  {100501} (\bibinfo {year} {2010})}\BibitemShut {NoStop}%
\bibitem [{\citenamefont {Anwar}\ \emph {et~al.}(2012)\citenamefont {Anwar},
  \citenamefont {Veldhorst}, \citenamefont {Brinkman},\ and\ \citenamefont
  {Aarts}}]{Anwar2012a}%
  \BibitemOpen
  \bibfield  {author} {\bibinfo {author} {\bibfnamefont {M.~S.}\ \bibnamefont
  {Anwar}}, \bibinfo {author} {\bibfnamefont {M.}~\bibnamefont {Veldhorst}},
  \bibinfo {author} {\bibfnamefont {A.}~\bibnamefont {Brinkman}}, \ and\
  \bibinfo {author} {\bibfnamefont {J.}~\bibnamefont {Aarts}},\ }\href
  {\doibase 10.1063/1.3681138} {\bibfield  {journal} {\bibinfo  {journal}
  {Applied Physics Letters}\ }\textbf {\bibinfo {volume} {100}},\ \bibinfo
  {pages} {052602} (\bibinfo {year} {2012})},\ \Eprint
  {http://arxiv.org/abs/1111.5809} {arXiv:1111.5809} \BibitemShut {NoStop}%
\bibitem [{\citenamefont {Usman}\ \emph {et~al.}(2011)\citenamefont {Usman},
  \citenamefont {Yates}, \citenamefont {Moore}, \citenamefont {Morrison},
  \citenamefont {Pecharsky}, \citenamefont {Gschneidner}, \citenamefont
  {Verhagen}, \citenamefont {Aarts}, \citenamefont {Zverev}, \citenamefont
  {Robinson}, \citenamefont {Witt}, \citenamefont {Blamire},\ and\
  \citenamefont {Cohen}}]{Usman2011}%
  \BibitemOpen
  \bibfield  {author} {\bibinfo {author} {\bibfnamefont {I.~T.~M.}\
  \bibnamefont {Usman}}, \bibinfo {author} {\bibfnamefont {K.~A.}\ \bibnamefont
  {Yates}}, \bibinfo {author} {\bibfnamefont {J.~D.}\ \bibnamefont {Moore}},
  \bibinfo {author} {\bibfnamefont {K.}~\bibnamefont {Morrison}}, \bibinfo
  {author} {\bibfnamefont {V.~K.}\ \bibnamefont {Pecharsky}}, \bibinfo {author}
  {\bibfnamefont {K.~A.}\ \bibnamefont {Gschneidner}}, \bibinfo {author}
  {\bibfnamefont {T.}~\bibnamefont {Verhagen}}, \bibinfo {author}
  {\bibfnamefont {J.}~\bibnamefont {Aarts}}, \bibinfo {author} {\bibfnamefont
  {V.~I.}\ \bibnamefont {Zverev}}, \bibinfo {author} {\bibfnamefont {J.~W.~A.}\
  \bibnamefont {Robinson}}, \bibinfo {author} {\bibfnamefont {J.~D.~S.}\
  \bibnamefont {Witt}}, \bibinfo {author} {\bibfnamefont {M.~G.}\ \bibnamefont
  {Blamire}}, \ and\ \bibinfo {author} {\bibfnamefont {L.~F.}\ \bibnamefont
  {Cohen}},\ }\href {\doibase 10.1103/PhysRevB.83.144518} {\bibfield  {journal}
  {\bibinfo  {journal} {Physical Review B}\ }\textbf {\bibinfo {volume} {83}},\
  \bibinfo {pages} {144518} (\bibinfo {year} {2011})}\BibitemShut {NoStop}%
\bibitem [{\citenamefont {Klose}\ \emph {et~al.}(2012)\citenamefont {Klose},
  \citenamefont {Khaire}, \citenamefont {Wang}, \citenamefont {Pratt},
  \citenamefont {Birge}, \citenamefont {McMorran}, \citenamefont {Ginley},
  \citenamefont {Borchers}, \citenamefont {Kirby}, \citenamefont {Maranville},\
  and\ \citenamefont {Unguris}}]{Klose2012}%
  \BibitemOpen
  \bibfield  {author} {\bibinfo {author} {\bibfnamefont {C.}~\bibnamefont
  {Klose}}, \bibinfo {author} {\bibfnamefont {T.~S.}\ \bibnamefont {Khaire}},
  \bibinfo {author} {\bibfnamefont {Y.}~\bibnamefont {Wang}}, \bibinfo {author}
  {\bibfnamefont {W.~P.}\ \bibnamefont {Pratt}}, \bibinfo {author}
  {\bibfnamefont {N.~O.}\ \bibnamefont {Birge}}, \bibinfo {author}
  {\bibfnamefont {B.~J.}\ \bibnamefont {McMorran}}, \bibinfo {author}
  {\bibfnamefont {T.~P.}\ \bibnamefont {Ginley}}, \bibinfo {author}
  {\bibfnamefont {J.~A.}\ \bibnamefont {Borchers}}, \bibinfo {author}
  {\bibfnamefont {B.~J.}\ \bibnamefont {Kirby}}, \bibinfo {author}
  {\bibfnamefont {B.~B.}\ \bibnamefont {Maranville}}, \ and\ \bibinfo {author}
  {\bibfnamefont {J.}~\bibnamefont {Unguris}},\ }\href {\doibase
  10.1103/PhysRevLett.108.127002} {\bibfield  {journal} {\bibinfo  {journal}
  {Physical Review Letters}\ }\textbf {\bibinfo {volume} {108}},\ \bibinfo
  {pages} {127002} (\bibinfo {year} {2012})}\BibitemShut {NoStop}%
\bibitem [{\citenamefont {Gingrich}\ \emph {et~al.}(2012)\citenamefont
  {Gingrich}, \citenamefont {Quarterman}, \citenamefont {Wang}, \citenamefont
  {Loloee}, \citenamefont {Pratt},\ and\ \citenamefont {Birge}}]{Gingrich2012}%
  \BibitemOpen
  \bibfield  {author} {\bibinfo {author} {\bibfnamefont {E.~C.}\ \bibnamefont
  {Gingrich}}, \bibinfo {author} {\bibfnamefont {P.}~\bibnamefont
  {Quarterman}}, \bibinfo {author} {\bibfnamefont {Y.}~\bibnamefont {Wang}},
  \bibinfo {author} {\bibfnamefont {R.}~\bibnamefont {Loloee}}, \bibinfo
  {author} {\bibfnamefont {W.~P.}\ \bibnamefont {Pratt}}, \ and\ \bibinfo
  {author} {\bibfnamefont {N.~O.}\ \bibnamefont {Birge}},\ }\href {\doibase
  10.1103/PhysRevB.86.224506} {\bibfield  {journal} {\bibinfo  {journal}
  {Physical Review B}\ }\textbf {\bibinfo {volume} {86}},\ \bibinfo {pages}
  {224506} (\bibinfo {year} {2012})},\ \Eprint {http://arxiv.org/abs/1208.3118}
  {arXiv:1208.3118} \BibitemShut {NoStop}%
\bibitem [{\citenamefont {Witt}\ \emph {et~al.}(2012)\citenamefont {Witt},
  \citenamefont {Robinson},\ and\ \citenamefont {Blamire}}]{Witt2012}%
  \BibitemOpen
  \bibfield  {author} {\bibinfo {author} {\bibfnamefont {J.~D.~S.}\
  \bibnamefont {Witt}}, \bibinfo {author} {\bibfnamefont {J.~W.~A.}\
  \bibnamefont {Robinson}}, \ and\ \bibinfo {author} {\bibfnamefont {M.~G.}\
  \bibnamefont {Blamire}},\ }\href {\doibase 10.1103/PhysRevB.85.184526}
  {\bibfield  {journal} {\bibinfo  {journal} {Physical Review B}\ }\textbf
  {\bibinfo {volume} {85}},\ \bibinfo {pages} {184526} (\bibinfo {year}
  {2012})}\BibitemShut {NoStop}%
\bibitem [{\citenamefont {Robinson}\ \emph {et~al.}(2012)\citenamefont
  {Robinson}, \citenamefont {Chiodi}, \citenamefont {Egilmez}, \citenamefont
  {Hal\'{a}sz},\ and\ \citenamefont {Blamire}}]{Robinson2012a}%
  \BibitemOpen
  \bibfield  {author} {\bibinfo {author} {\bibfnamefont {J.~W.~A.}\
  \bibnamefont {Robinson}}, \bibinfo {author} {\bibfnamefont {F.}~\bibnamefont
  {Chiodi}}, \bibinfo {author} {\bibfnamefont {M.}~\bibnamefont {Egilmez}},
  \bibinfo {author} {\bibfnamefont {G.~B.}\ \bibnamefont {Hal\'{a}sz}}, \ and\
  \bibinfo {author} {\bibfnamefont {M.~G.}\ \bibnamefont {Blamire}},\ }\href
  {\doibase 10.1038/srep00699} {\bibfield  {journal} {\bibinfo  {journal}
  {Scientific Reports}\ }\textbf {\bibinfo {volume} {2}},\ \bibinfo {pages}
  {699} (\bibinfo {year} {2012})}\BibitemShut {NoStop}%
\bibitem [{\citenamefont {Kalcheim}\ \emph {et~al.}(2012)\citenamefont
  {Kalcheim}, \citenamefont {Millo}, \citenamefont {Egilmez}, \citenamefont
  {Robinson},\ and\ \citenamefont {Blamire}}]{Kalcheim2012}%
  \BibitemOpen
  \bibfield  {author} {\bibinfo {author} {\bibfnamefont {Y.}~\bibnamefont
  {Kalcheim}}, \bibinfo {author} {\bibfnamefont {O.}~\bibnamefont {Millo}},
  \bibinfo {author} {\bibfnamefont {M.}~\bibnamefont {Egilmez}}, \bibinfo
  {author} {\bibfnamefont {J.~W.~A.}\ \bibnamefont {Robinson}}, \ and\ \bibinfo
  {author} {\bibfnamefont {M.~G.}\ \bibnamefont {Blamire}},\ }\href {\doibase
  10.1103/PhysRevB.85.104504} {\bibfield  {journal} {\bibinfo  {journal}
  {Physical Review B}\ }\textbf {\bibinfo {volume} {85}},\ \bibinfo {pages}
  {104504} (\bibinfo {year} {2012})}\BibitemShut {NoStop}%
\bibitem [{\citenamefont {Kalcheim}\ \emph {et~al.}(2014)\citenamefont
  {Kalcheim}, \citenamefont {Felner}, \citenamefont {Millo}, \citenamefont
  {Kirzhner}, \citenamefont {Koren}, \citenamefont {{Di Bernardo}},
  \citenamefont {Egilmez}, \citenamefont {Blamire},\ and\ \citenamefont
  {Robinson}}]{Kalcheim2014}%
  \BibitemOpen
  \bibfield  {author} {\bibinfo {author} {\bibfnamefont {Y.}~\bibnamefont
  {Kalcheim}}, \bibinfo {author} {\bibfnamefont {I.}~\bibnamefont {Felner}},
  \bibinfo {author} {\bibfnamefont {O.}~\bibnamefont {Millo}}, \bibinfo
  {author} {\bibfnamefont {T.}~\bibnamefont {Kirzhner}}, \bibinfo {author}
  {\bibfnamefont {G.}~\bibnamefont {Koren}}, \bibinfo {author} {\bibfnamefont
  {A.}~\bibnamefont {{Di Bernardo}}}, \bibinfo {author} {\bibfnamefont
  {M.}~\bibnamefont {Egilmez}}, \bibinfo {author} {\bibfnamefont {M.~G.}\
  \bibnamefont {Blamire}}, \ and\ \bibinfo {author} {\bibfnamefont {J.~W.~A.}\
  \bibnamefont {Robinson}},\ }\href {\doibase 10.1103/PhysRevB.89.180506}
  {\bibfield  {journal} {\bibinfo  {journal} {Physical Review B}\ }\textbf
  {\bibinfo {volume} {89}},\ \bibinfo {pages} {180506} (\bibinfo {year}
  {2014})}\BibitemShut {NoStop}%
\bibitem [{\citenamefont {Alidoust}\ \emph {et~al.}(2015)\citenamefont
  {Alidoust}, \citenamefont {Halterman},\ and\ \citenamefont
  {Valls}}]{Alidoust2015b}%
  \BibitemOpen
  \bibfield  {author} {\bibinfo {author} {\bibfnamefont {M.}~\bibnamefont
  {Alidoust}}, \bibinfo {author} {\bibfnamefont {K.}~\bibnamefont {Halterman}},
  \ and\ \bibinfo {author} {\bibfnamefont {O.~T.}\ \bibnamefont {Valls}},\
  }\href {\doibase 10.1103/PhysRevB.92.014508} {\bibfield  {journal} {\bibinfo
  {journal} {Physical Review B}\ }\textbf {\bibinfo {volume} {92}},\ \bibinfo
  {pages} {014508} (\bibinfo {year} {2015})},\ \Eprint
  {http://arxiv.org/abs/1506.05469} {arXiv:1506.05469} \BibitemShut {NoStop}%
\bibitem [{\citenamefont {Banerjee}\ \emph
  {et~al.}(2014{\natexlab{a}})\citenamefont {Banerjee}, \citenamefont
  {Robinson},\ and\ \citenamefont {Blamire}}]{Banerjee2014}%
  \BibitemOpen
  \bibfield  {author} {\bibinfo {author} {\bibfnamefont {N.}~\bibnamefont
  {Banerjee}}, \bibinfo {author} {\bibfnamefont {J.}~\bibnamefont {Robinson}},
  \ and\ \bibinfo {author} {\bibfnamefont {M.~G.}\ \bibnamefont {Blamire}},\
  }\href {\doibase 10.1038/ncomms5771} {\bibfield  {journal} {\bibinfo
  {journal} {Nature Communications}\ }\textbf {\bibinfo {volume} {5}},\
  \bibinfo {pages} {4771} (\bibinfo {year} {2014}{\natexlab{a}})}\BibitemShut
  {NoStop}%
\bibitem [{\citenamefont {Leksin}\ \emph {et~al.}(2012)\citenamefont {Leksin},
  \citenamefont {Garif'yanov}, \citenamefont {Garifullin}, \citenamefont
  {Fominov}, \citenamefont {Schumann}, \citenamefont {Krupskaya}, \citenamefont
  {Kataev}, \citenamefont {Schmidt},\ and\ \citenamefont
  {B\"{u}chner}}]{Leksin2012}%
  \BibitemOpen
  \bibfield  {author} {\bibinfo {author} {\bibfnamefont {P.~V.}\ \bibnamefont
  {Leksin}}, \bibinfo {author} {\bibfnamefont {N.~N.}\ \bibnamefont
  {Garif'yanov}}, \bibinfo {author} {\bibfnamefont {I.~a.}\ \bibnamefont
  {Garifullin}}, \bibinfo {author} {\bibfnamefont {Y.~V.}\ \bibnamefont
  {Fominov}}, \bibinfo {author} {\bibfnamefont {J.}~\bibnamefont {Schumann}},
  \bibinfo {author} {\bibfnamefont {Y.}~\bibnamefont {Krupskaya}}, \bibinfo
  {author} {\bibfnamefont {V.}~\bibnamefont {Kataev}}, \bibinfo {author}
  {\bibfnamefont {O.~G.}\ \bibnamefont {Schmidt}}, \ and\ \bibinfo {author}
  {\bibfnamefont {B.}~\bibnamefont {B\"{u}chner}},\ }\href {\doibase
  10.1103/PhysRevLett.109.057005} {\bibfield  {journal} {\bibinfo  {journal}
  {Physical Review Letters}\ }\textbf {\bibinfo {volume} {109}},\ \bibinfo
  {pages} {057005} (\bibinfo {year} {2012})},\ \Eprint
  {http://arxiv.org/abs/1207.0727} {arXiv:1207.0727} \BibitemShut {NoStop}%
\bibitem [{\citenamefont {Banerjee}\ \emph
  {et~al.}(2014{\natexlab{b}})\citenamefont {Banerjee}, \citenamefont {Smiet},
  \citenamefont {Smits}, \citenamefont {Ozaeta}, \citenamefont {Bergeret},
  \citenamefont {Blamire},\ and\ \citenamefont {Robinson}}]{Banerjee2014a}%
  \BibitemOpen
  \bibfield  {author} {\bibinfo {author} {\bibfnamefont {N.}~\bibnamefont
  {Banerjee}}, \bibinfo {author} {\bibfnamefont {C.~B.}\ \bibnamefont {Smiet}},
  \bibinfo {author} {\bibfnamefont {R.~G.~J.}\ \bibnamefont {Smits}}, \bibinfo
  {author} {\bibfnamefont {A.}~\bibnamefont {Ozaeta}}, \bibinfo {author}
  {\bibfnamefont {F.~S.}\ \bibnamefont {Bergeret}}, \bibinfo {author}
  {\bibfnamefont {M.~G.}\ \bibnamefont {Blamire}}, \ and\ \bibinfo {author}
  {\bibfnamefont {J.~W.~A.}\ \bibnamefont {Robinson}},\ }\href {\doibase
  10.1038/ncomms4048} {\bibfield  {journal} {\bibinfo  {journal} {Nature
  Communications}\ }\textbf {\bibinfo {volume} {5}},\ \bibinfo {pages} {3048}
  (\bibinfo {year} {2014}{\natexlab{b}})}\BibitemShut {NoStop}%
\bibitem [{\citenamefont {Wang}\ \emph {et~al.}(2014)\citenamefont {Wang},
  \citenamefont {{Di Bernardo}}, \citenamefont {Banerjee}, \citenamefont
  {Wells}, \citenamefont {Bergeret}, \citenamefont {Blamire},\ and\
  \citenamefont {Robinson}}]{Wang2014d}%
  \BibitemOpen
  \bibfield  {author} {\bibinfo {author} {\bibfnamefont {X.~L.}\ \bibnamefont
  {Wang}}, \bibinfo {author} {\bibfnamefont {A.}~\bibnamefont {{Di Bernardo}}},
  \bibinfo {author} {\bibfnamefont {N.}~\bibnamefont {Banerjee}}, \bibinfo
  {author} {\bibfnamefont {A.}~\bibnamefont {Wells}}, \bibinfo {author}
  {\bibfnamefont {F.~S.}\ \bibnamefont {Bergeret}}, \bibinfo {author}
  {\bibfnamefont {M.~G.}\ \bibnamefont {Blamire}}, \ and\ \bibinfo {author}
  {\bibfnamefont {J.~W.~A.}\ \bibnamefont {Robinson}},\ }\href {\doibase
  10.1103/PhysRevB.89.140508} {\bibfield  {journal} {\bibinfo  {journal}
  {Physical Review B}\ }\textbf {\bibinfo {volume} {89}},\ \bibinfo {pages}
  {140508} (\bibinfo {year} {2014})}\BibitemShut {NoStop}%
\bibitem [{\citenamefont {Chiodi}\ \emph {et~al.}(2013)\citenamefont {Chiodi},
  \citenamefont {Witt}, \citenamefont {Smits}, \citenamefont {Qu},
  \citenamefont {Hal\'{a}sz}, \citenamefont {Wu}, \citenamefont {Valls},
  \citenamefont {Halterman}, \citenamefont {Robinson},\ and\ \citenamefont
  {Blamire}}]{Chiodi2013}%
  \BibitemOpen
  \bibfield  {author} {\bibinfo {author} {\bibfnamefont {F.}~\bibnamefont
  {Chiodi}}, \bibinfo {author} {\bibfnamefont {J.~D.~S.}\ \bibnamefont {Witt}},
  \bibinfo {author} {\bibfnamefont {R.~G.~J.}\ \bibnamefont {Smits}}, \bibinfo
  {author} {\bibfnamefont {L.}~\bibnamefont {Qu}}, \bibinfo {author}
  {\bibfnamefont {G.~B.}\ \bibnamefont {Hal\'{a}sz}}, \bibinfo {author}
  {\bibfnamefont {C.-T.}\ \bibnamefont {Wu}}, \bibinfo {author} {\bibfnamefont
  {O.~T.}\ \bibnamefont {Valls}}, \bibinfo {author} {\bibfnamefont
  {K.}~\bibnamefont {Halterman}}, \bibinfo {author} {\bibfnamefont {J.~W.~a.}\
  \bibnamefont {Robinson}}, \ and\ \bibinfo {author} {\bibfnamefont {M.~G.}\
  \bibnamefont {Blamire}},\ }\href {\doibase 10.1209/0295-5075/101/37002}
  {\bibfield  {journal} {\bibinfo  {journal} {EPL (Europhysics Letters)}\
  }\textbf {\bibinfo {volume} {101}},\ \bibinfo {pages} {37002} (\bibinfo
  {year} {2013})},\ \Eprint {http://arxiv.org/abs/1211.1169} {arXiv:1211.1169}
  \BibitemShut {NoStop}%
\bibitem [{\citenamefont {Gu}\ \emph {et~al.}(2014)\citenamefont {Gu},
  \citenamefont {Robinson}, \citenamefont {Bianchetti}, \citenamefont
  {Stelmashenko}, \citenamefont {Astill}, \citenamefont {Grosche},
  \citenamefont {MacManus-Driscoll},\ and\ \citenamefont {Blamire}}]{Gu2014}%
  \BibitemOpen
  \bibfield  {author} {\bibinfo {author} {\bibfnamefont {Y.}~\bibnamefont
  {Gu}}, \bibinfo {author} {\bibfnamefont {J.~W.~A.}\ \bibnamefont {Robinson}},
  \bibinfo {author} {\bibfnamefont {M.}~\bibnamefont {Bianchetti}}, \bibinfo
  {author} {\bibfnamefont {N.~A.}\ \bibnamefont {Stelmashenko}}, \bibinfo
  {author} {\bibfnamefont {D.}~\bibnamefont {Astill}}, \bibinfo {author}
  {\bibfnamefont {F.~M.}\ \bibnamefont {Grosche}}, \bibinfo {author}
  {\bibfnamefont {J.~L.}\ \bibnamefont {MacManus-Driscoll}}, \ and\ \bibinfo
  {author} {\bibfnamefont {M.~G.}\ \bibnamefont {Blamire}},\ }\href {\doibase
  10.1063/1.4870141} {\bibfield  {journal} {\bibinfo  {journal} {Applied
  Physics Letters Materials}\ }\textbf {\bibinfo {volume} {2}},\ \bibinfo
  {pages} {046103} (\bibinfo {year} {2014})}\BibitemShut {NoStop}%
\bibitem [{\citenamefont {Jara}\ \emph {et~al.}(2014)\citenamefont {Jara},
  \citenamefont {Safranski}, \citenamefont {Krivorotov}, \citenamefont {Wu},
  \citenamefont {Malmi-Kakkada}, \citenamefont {Valls},\ and\ \citenamefont
  {Halterman}}]{Jara2014}%
  \BibitemOpen
  \bibfield  {author} {\bibinfo {author} {\bibfnamefont {A.~A.}\ \bibnamefont
  {Jara}}, \bibinfo {author} {\bibfnamefont {C.}~\bibnamefont {Safranski}},
  \bibinfo {author} {\bibfnamefont {I.~N.}\ \bibnamefont {Krivorotov}},
  \bibinfo {author} {\bibfnamefont {C.-T.}\ \bibnamefont {Wu}}, \bibinfo
  {author} {\bibfnamefont {A.~N.}\ \bibnamefont {Malmi-Kakkada}}, \bibinfo
  {author} {\bibfnamefont {O.~T.}\ \bibnamefont {Valls}}, \ and\ \bibinfo
  {author} {\bibfnamefont {K.}~\bibnamefont {Halterman}},\ }\href {\doibase
  10.1103/PhysRevB.89.184502} {\bibfield  {journal} {\bibinfo  {journal}
  {Physical Review B}\ }\textbf {\bibinfo {volume} {89}},\ \bibinfo {pages}
  {184502} (\bibinfo {year} {2014})},\ \Eprint {http://arxiv.org/abs/1404.2304}
  {arXiv:1404.2304} \BibitemShut {NoStop}%
\bibitem [{\citenamefont {Rusanov}\ \emph {et~al.}(2004)\citenamefont
  {Rusanov}, \citenamefont {Aarts}, \citenamefont {Hesselberth},\ and\
  \citenamefont {Buzdin}}]{Rusanov2004}%
  \BibitemOpen
  \bibfield  {author} {\bibinfo {author} {\bibfnamefont {A.}~\bibnamefont
  {Rusanov}}, \bibinfo {author} {\bibfnamefont {J.}~\bibnamefont {Aarts}},
  \bibinfo {author} {\bibfnamefont {M.}~\bibnamefont {Hesselberth}}, \ and\
  \bibinfo {author} {\bibfnamefont {A.~I.}\ \bibnamefont {Buzdin}},\ }\href
  {\doibase 10.1103/PhysRevLett.93.057002} {\bibfield  {journal} {\bibinfo
  {journal} {Physical Review Letters}\ }\textbf {\bibinfo {volume} {93}},\
  \bibinfo {pages} {57002} (\bibinfo {year} {2004})}\BibitemShut {NoStop}%
\bibitem [{\citenamefont {Rusanov}\ \emph {et~al.}(2006)\citenamefont
  {Rusanov}, \citenamefont {Habraken},\ and\ \citenamefont
  {Aarts}}]{Rusanov2006}%
  \BibitemOpen
  \bibfield  {author} {\bibinfo {author} {\bibfnamefont {A.}~\bibnamefont
  {Rusanov}}, \bibinfo {author} {\bibfnamefont {S.}~\bibnamefont {Habraken}}, \
  and\ \bibinfo {author} {\bibfnamefont {J.}~\bibnamefont {Aarts}},\ }\href
  {\doibase 10.1103/PhysRevB.73.060505} {\bibfield  {journal} {\bibinfo
  {journal} {Physical Review B}\ }\textbf {\bibinfo {volume} {73}},\ \bibinfo
  {pages} {60505} (\bibinfo {year} {2006})}\BibitemShut {NoStop}%
\bibitem [{\citenamefont {Eschrig}(2011)}]{Eschrig2011}%
  \BibitemOpen
  \bibfield  {author} {\bibinfo {author} {\bibfnamefont {M.}~\bibnamefont
  {Eschrig}},\ }\href {\doibase 10.1063/1.3541944} {\bibfield  {journal}
  {\bibinfo  {journal} {Physics Today}\ }\textbf {\bibinfo {volume} {64}},\
  \bibinfo {pages} {43} (\bibinfo {year} {2011})}\BibitemShut {NoStop}%
\bibitem [{\citenamefont {Linder}\ and\ \citenamefont
  {Robinson}(2015)}]{Linder2015}%
  \BibitemOpen
  \bibfield  {author} {\bibinfo {author} {\bibfnamefont {J.}~\bibnamefont
  {Linder}}\ and\ \bibinfo {author} {\bibfnamefont {J.~W.~A.}\ \bibnamefont
  {Robinson}},\ }\href {\doibase 10.1038/NPHYS3242} {\bibfield  {journal}
  {\bibinfo  {journal} {Nature Physics}\ }\textbf {\bibinfo {volume} {11}},\
  \bibinfo {pages} {307 } (\bibinfo {year} {2015})}\BibitemShut {NoStop}%
\bibitem [{\citenamefont {Bergeret}\ \emph {et~al.}(2012)\citenamefont
  {Bergeret}, \citenamefont {Verso},\ and\ \citenamefont
  {Volkov}}]{Bergeret2012}%
  \BibitemOpen
  \bibfield  {author} {\bibinfo {author} {\bibfnamefont {F.~S.}\ \bibnamefont
  {Bergeret}}, \bibinfo {author} {\bibfnamefont {A.}~\bibnamefont {Verso}}, \
  and\ \bibinfo {author} {\bibfnamefont {A.~F.}\ \bibnamefont {Volkov}},\
  }\href {\doibase 10.1103/PhysRevB.86.214516} {\bibfield  {journal} {\bibinfo
  {journal} {Physical Review B}\ }\textbf {\bibinfo {volume} {86}},\ \bibinfo
  {pages} {1} (\bibinfo {year} {2012})},\ \Eprint
  {http://arxiv.org/abs/1211.0870} {arXiv:1211.0870} \BibitemShut {NoStop}%
\bibitem [{\citenamefont {Pal}\ \emph {et~al.}(2014)\citenamefont {Pal},
  \citenamefont {Barber}, \citenamefont {Robinson},\ and\ \citenamefont
  {Blamire}}]{Pal2014}%
  \BibitemOpen
  \bibfield  {author} {\bibinfo {author} {\bibfnamefont {A.}~\bibnamefont
  {Pal}}, \bibinfo {author} {\bibfnamefont {Z.}~\bibnamefont {Barber}},
  \bibinfo {author} {\bibfnamefont {J.}~\bibnamefont {Robinson}}, \ and\
  \bibinfo {author} {\bibfnamefont {M.}~\bibnamefont {Blamire}},\ }\href
  {\doibase 10.1038/ncomms4340} {\bibfield  {journal} {\bibinfo  {journal}
  {Nature Communications}\ }\textbf {\bibinfo {volume} {5}},\ \bibinfo {pages}
  {3340} (\bibinfo {year} {2014})}\BibitemShut {NoStop}%
\bibitem [{\citenamefont {Massarotti}\ \emph {et~al.}(2015)\citenamefont
  {Massarotti}, \citenamefont {Pal}, \citenamefont {Rotoli}, \citenamefont
  {Longobardi}, \citenamefont {Blamire},\ and\ \citenamefont
  {Tafuri}}]{Massarotti2015}%
  \BibitemOpen
  \bibfield  {author} {\bibinfo {author} {\bibfnamefont {D.}~\bibnamefont
  {Massarotti}}, \bibinfo {author} {\bibfnamefont {A.}~\bibnamefont {Pal}},
  \bibinfo {author} {\bibfnamefont {G.}~\bibnamefont {Rotoli}}, \bibinfo
  {author} {\bibfnamefont {L.}~\bibnamefont {Longobardi}}, \bibinfo {author}
  {\bibfnamefont {M.~G.}\ \bibnamefont {Blamire}}, \ and\ \bibinfo {author}
  {\bibfnamefont {F.}~\bibnamefont {Tafuri}},\ }\href {\doibase
  10.1038/ncomms8376} {\bibfield  {journal} {\bibinfo  {journal} {Nature
  Communications}\ }\textbf {\bibinfo {volume} {6}},\ \bibinfo {pages} {7376}
  (\bibinfo {year} {2015})}\BibitemShut {NoStop}%
\bibitem [{\citenamefont {Bergeret}\ and\ \citenamefont
  {Giazotto}(2013)}]{Bergeret2013a}%
  \BibitemOpen
  \bibfield  {author} {\bibinfo {author} {\bibfnamefont {F.~S.}\ \bibnamefont
  {Bergeret}}\ and\ \bibinfo {author} {\bibfnamefont {F.}~\bibnamefont
  {Giazotto}},\ }\href {\doibase 10.1103/PhysRevB.88.014515} {\bibfield
  {journal} {\bibinfo  {journal} {Physical Review B}\ }\textbf {\bibinfo
  {volume} {88}},\ \bibinfo {pages} {014515} (\bibinfo {year} {2013})},\
  \Eprint {http://arxiv.org/abs/1305.6301} {arXiv:1305.6301} \BibitemShut
  {NoStop}%
\bibitem [{\citenamefont {Bergeret}\ and\ \citenamefont
  {Giazotto}(2014)}]{Bergeret2014b}%
  \BibitemOpen
  \bibfield  {author} {\bibinfo {author} {\bibfnamefont {F.~S.}\ \bibnamefont
  {Bergeret}}\ and\ \bibinfo {author} {\bibfnamefont {F.}~\bibnamefont
  {Giazotto}},\ }\href {\doibase 10.1103/PhysRevB.89.054505} {\bibfield
  {journal} {\bibinfo  {journal} {Physical Review B}\ }\textbf {\bibinfo
  {volume} {89}},\ \bibinfo {pages} {054505} (\bibinfo {year} {2014})},\
  \Eprint {http://arxiv.org/abs/1312.1945} {arXiv:1312.1945} \BibitemShut
  {NoStop}%
\bibitem [{\citenamefont {Ozaeta}\ \emph {et~al.}(2014)\citenamefont {Ozaeta},
  \citenamefont {Virtanen}, \citenamefont {Bergeret},\ and\ \citenamefont
  {Heikkil\"{a}}}]{Ozaeta2014a}%
  \BibitemOpen
  \bibfield  {author} {\bibinfo {author} {\bibfnamefont {A.}~\bibnamefont
  {Ozaeta}}, \bibinfo {author} {\bibfnamefont {P.}~\bibnamefont {Virtanen}},
  \bibinfo {author} {\bibfnamefont {F.~S.}\ \bibnamefont {Bergeret}}, \ and\
  \bibinfo {author} {\bibfnamefont {T.~T.}\ \bibnamefont {Heikkil\"{a}}},\
  }\href {\doibase 10.1103/PhysRevLett.112.057001} {\bibfield  {journal}
  {\bibinfo  {journal} {Physical Review Letters}\ }\textbf {\bibinfo {volume}
  {112}},\ \bibinfo {pages} {057001} (\bibinfo {year} {2014})}\BibitemShut
  {NoStop}%
\bibitem [{\citenamefont {Giazotto}\ \emph {et~al.}(2014)\citenamefont
  {Giazotto}, \citenamefont {Robinson}, \citenamefont {Moodera},\ and\
  \citenamefont {Bergeret}}]{Giazotto2014a}%
  \BibitemOpen
  \bibfield  {author} {\bibinfo {author} {\bibfnamefont {F.}~\bibnamefont
  {Giazotto}}, \bibinfo {author} {\bibfnamefont {J.~W.~A.}\ \bibnamefont
  {Robinson}}, \bibinfo {author} {\bibfnamefont {J.~S.}\ \bibnamefont
  {Moodera}}, \ and\ \bibinfo {author} {\bibfnamefont {F.~S.}\ \bibnamefont
  {Bergeret}},\ }\href {\doibase 10.1063/1.4893443} {\bibfield  {journal}
  {\bibinfo  {journal} {Applied Physics Letters}\ }\textbf {\bibinfo {volume}
  {105}},\ \bibinfo {pages} {062602} (\bibinfo {year} {2014})},\ \Eprint
  {http://arxiv.org/abs/1405.0565} {arXiv:1405.0565} \BibitemShut {NoStop}%
\bibitem [{\citenamefont {Giazotto}\ \emph {et~al.}(2015)\citenamefont
  {Giazotto}, \citenamefont {Heikkil\"{a}},\ and\ \citenamefont
  {Bergeret}}]{Giazotto2014}%
  \BibitemOpen
  \bibfield  {author} {\bibinfo {author} {\bibfnamefont {F.}~\bibnamefont
  {Giazotto}}, \bibinfo {author} {\bibfnamefont {T.}~\bibnamefont
  {Heikkil\"{a}}}, \ and\ \bibinfo {author} {\bibfnamefont {F.}~\bibnamefont
  {Bergeret}},\ }\href {\doibase 10.1103/PhysRevLett.114.067001} {\bibfield
  {journal} {\bibinfo  {journal} {Physical Review Letters}\ }\textbf {\bibinfo
  {volume} {114}},\ \bibinfo {pages} {067001} (\bibinfo {year} {2015})},\
  \Eprint {http://arxiv.org/abs/1403.1231} {arXiv:1403.1231} \BibitemShut
  {NoStop}%
\bibitem [{\citenamefont {Nasti}\ \emph {et~al.}(2015)\citenamefont {Nasti},
  \citenamefont {Parlato}, \citenamefont {Ejrnaes}, \citenamefont {Cristiano},
  \citenamefont {Taino}, \citenamefont {Myoren}, \citenamefont {Sobolewski},\
  and\ \citenamefont {Pepe}}]{Nasti2015}%
  \BibitemOpen
  \bibfield  {author} {\bibinfo {author} {\bibfnamefont {U.}~\bibnamefont
  {Nasti}}, \bibinfo {author} {\bibfnamefont {L.}~\bibnamefont {Parlato}},
  \bibinfo {author} {\bibfnamefont {M.}~\bibnamefont {Ejrnaes}}, \bibinfo
  {author} {\bibfnamefont {R.}~\bibnamefont {Cristiano}}, \bibinfo {author}
  {\bibfnamefont {T.}~\bibnamefont {Taino}}, \bibinfo {author} {\bibfnamefont
  {H.}~\bibnamefont {Myoren}}, \bibinfo {author} {\bibfnamefont
  {R.}~\bibnamefont {Sobolewski}}, \ and\ \bibinfo {author} {\bibfnamefont
  {G.}~\bibnamefont {Pepe}},\ }\href {\doibase 10.1103/PhysRevB.92.014501}
  {\bibfield  {journal} {\bibinfo  {journal} {Physical Review B}\ }\textbf
  {\bibinfo {volume} {92}},\ \bibinfo {pages} {014501} (\bibinfo {year}
  {2015})}\BibitemShut {NoStop}%
\bibitem [{\citenamefont {Bergeret}\ and\ \citenamefont
  {Tokatly}(2013)}]{Bergeret2013}%
  \BibitemOpen
  \bibfield  {author} {\bibinfo {author} {\bibfnamefont {F.~S.}\ \bibnamefont
  {Bergeret}}\ and\ \bibinfo {author} {\bibfnamefont {I.~V.}\ \bibnamefont
  {Tokatly}},\ }\href {\doibase 10.1103/PhysRevLett.110.117003} {\bibfield
  {journal} {\bibinfo  {journal} {Physical Review Letters}\ }\textbf {\bibinfo
  {volume} {110}},\ \bibinfo {pages} {117003} (\bibinfo {year} {2013})},\
  \Eprint {http://arxiv.org/abs/1211.3084} {arXiv:1211.3084} \BibitemShut
  {NoStop}%
\bibitem [{\citenamefont {Bergeret}\ and\ \citenamefont
  {Tokatly}(2014)}]{Bergeret2014}%
  \BibitemOpen
  \bibfield  {author} {\bibinfo {author} {\bibfnamefont {F.~S.}\ \bibnamefont
  {Bergeret}}\ and\ \bibinfo {author} {\bibfnamefont {I.~V.}\ \bibnamefont
  {Tokatly}},\ }\href {\doibase 10.1103/PhysRevB.89.134517} {\bibfield
  {journal} {\bibinfo  {journal} {Physical Review B}\ }\textbf {\bibinfo
  {volume} {89}},\ \bibinfo {pages} {134517} (\bibinfo {year} {2014})},\
  \Eprint {http://arxiv.org/abs/1402.1025} {arXiv:1402.1025} \BibitemShut
  {NoStop}%
\bibitem [{\citenamefont {Jacobsen}\ \emph {et~al.}(2015)\citenamefont
  {Jacobsen}, \citenamefont {Ouassou},\ and\ \citenamefont
  {Linder}}]{Jacobsen2015}%
  \BibitemOpen
  \bibfield  {author} {\bibinfo {author} {\bibfnamefont {S.~H.}\ \bibnamefont
  {Jacobsen}}, \bibinfo {author} {\bibfnamefont {J.~A.}\ \bibnamefont
  {Ouassou}}, \ and\ \bibinfo {author} {\bibfnamefont {J.}~\bibnamefont
  {Linder}},\ }\href {http://arxiv.org/abs/1503.06835} {\  (\bibinfo {year}
  {2015})},\ \Eprint {http://arxiv.org/abs/1503.06835} {arXiv:1503.06835}
  \BibitemShut {NoStop}%
\bibitem [{\citenamefont {Arjoranta}\ and\ \citenamefont
  {Heikkila}(2015)}]{Arjoranta2015}%
  \BibitemOpen
  \bibfield  {author} {\bibinfo {author} {\bibfnamefont {J.}~\bibnamefont
  {Arjoranta}}\ and\ \bibinfo {author} {\bibfnamefont {T.~T.}\ \bibnamefont
  {Heikkila}},\ }\href {http://arxiv.org/abs/1507.02320} {\  (\bibinfo {year}
  {2015})},\ \Eprint {http://arxiv.org/abs/1507.02320} {arXiv:1507.02320}
  \BibitemShut {NoStop}%
\bibitem [{\citenamefont {Alidoust}\ and\ \citenamefont
  {Halterman}(2015)}]{Alidoust2015a}%
  \BibitemOpen
  \bibfield  {author} {\bibinfo {author} {\bibfnamefont {M.}~\bibnamefont
  {Alidoust}}\ and\ \bibinfo {author} {\bibfnamefont {K.}~\bibnamefont
  {Halterman}},\ }\href {\doibase 10.1088/1367-2630/17/3/033001} {\bibfield
  {journal} {\bibinfo  {journal} {New Journal of Physics}\ }\textbf {\bibinfo
  {volume} {17}},\ \bibinfo {pages} {033001} (\bibinfo {year} {2015})},\
  \Eprint {http://arxiv.org/abs/1504.05950} {arXiv:1504.05950} \BibitemShut
  {NoStop}%
\bibitem [{\citenamefont {Maekawa}\ \emph {et~al.}(2012)\citenamefont
  {Maekawa}, \citenamefont {Valenzuela}, \citenamefont {Saitoh},\ and\
  \citenamefont {Kimura}}]{Maekawa2012}%
  \BibitemOpen
  \bibfield  {author} {\bibinfo {author} {\bibfnamefont {S.}~\bibnamefont
  {Maekawa}}, \bibinfo {author} {\bibfnamefont {S.~O.}\ \bibnamefont
  {Valenzuela}}, \bibinfo {author} {\bibfnamefont {E.}~\bibnamefont {Saitoh}},
  \ and\ \bibinfo {author} {\bibfnamefont {T.}~\bibnamefont {Kimura}},\
  }\href@noop {} {\emph {\bibinfo {title} {{Spin current}}}}\ (\bibinfo
  {publisher} {Oxford University Press},\ \bibinfo {year} {2012})\BibitemShut
  {NoStop}%
\bibitem [{\citenamefont {Kuschel}\ and\ \citenamefont
  {Reiss}(2014)}]{Kuschel2014}%
  \BibitemOpen
  \bibfield  {author} {\bibinfo {author} {\bibfnamefont {T.}~\bibnamefont
  {Kuschel}}\ and\ \bibinfo {author} {\bibfnamefont {G.}~\bibnamefont
  {Reiss}},\ }\href {\doibase 10.1038/nnano.2014.279} {\bibfield  {journal}
  {\bibinfo  {journal} {Nature Nanotechnology}\ }\textbf {\bibinfo {volume}
  {10}},\ \bibinfo {pages} {22} (\bibinfo {year} {2014})}\BibitemShut {NoStop}%
\bibitem [{\citenamefont {Manchon}(2014)}]{Manchon2014}%
  \BibitemOpen
  \bibfield  {author} {\bibinfo {author} {\bibfnamefont {A.}~\bibnamefont
  {Manchon}},\ }\href {\doibase 10.1038/nphys2957} {\bibfield  {journal}
  {\bibinfo  {journal} {Nature Physics}\ }\textbf {\bibinfo {volume} {10}},\
  \bibinfo {pages} {340} (\bibinfo {year} {2014})}\BibitemShut {NoStop}%
\bibitem [{\citenamefont {Ciccarelli}\ \emph {et~al.}(2014)\citenamefont
  {Ciccarelli}, \citenamefont {Hals}, \citenamefont {Irvine}, \citenamefont
  {Novak}, \citenamefont {Tserkovnyak}, \citenamefont {Kurebayashi},
  \citenamefont {Brataas},\ and\ \citenamefont {Ferguson}}]{Ciccarelli2014}%
  \BibitemOpen
  \bibfield  {author} {\bibinfo {author} {\bibfnamefont {C.}~\bibnamefont
  {Ciccarelli}}, \bibinfo {author} {\bibfnamefont {K.~M.~D.}\ \bibnamefont
  {Hals}}, \bibinfo {author} {\bibfnamefont {A.}~\bibnamefont {Irvine}},
  \bibinfo {author} {\bibfnamefont {V.}~\bibnamefont {Novak}}, \bibinfo
  {author} {\bibfnamefont {Y.}~\bibnamefont {Tserkovnyak}}, \bibinfo {author}
  {\bibfnamefont {H.}~\bibnamefont {Kurebayashi}}, \bibinfo {author}
  {\bibfnamefont {A.}~\bibnamefont {Brataas}}, \ and\ \bibinfo {author}
  {\bibfnamefont {A.}~\bibnamefont {Ferguson}},\ }\href {\doibase
  10.1038/nnano.2014.252} {\bibfield  {journal} {\bibinfo  {journal} {Nature
  Nanotechnology}\ }\textbf {\bibinfo {volume} {10}},\ \bibinfo {pages} {50}
  (\bibinfo {year} {2014})},\ \Eprint {http://arxiv.org/abs/1411.2779}
  {arXiv:1411.2779} \BibitemShut {NoStop}%
\bibitem [{\citenamefont {Dyakonov}\ and\ \citenamefont
  {Perel}(1971{\natexlab{a}})}]{Dyakonov1971}%
  \BibitemOpen
  \bibfield  {author} {\bibinfo {author} {\bibfnamefont {M.}~\bibnamefont
  {Dyakonov}}\ and\ \bibinfo {author} {\bibfnamefont {V.}~\bibnamefont
  {Perel}},\ }\href@noop {} {\bibfield  {journal} {\bibinfo  {journal} {JETP
  Letters}\ }\textbf {\bibinfo {volume} {13}},\ \bibinfo {pages} {467 }
  (\bibinfo {year} {1971}{\natexlab{a}})}\BibitemShut {NoStop}%
\bibitem [{\citenamefont {Dyakonov}\ and\ \citenamefont
  {Perel}(1971{\natexlab{b}})}]{Dyakonov1971a}%
  \BibitemOpen
  \bibfield  {author} {\bibinfo {author} {\bibfnamefont {M.}~\bibnamefont
  {Dyakonov}}\ and\ \bibinfo {author} {\bibfnamefont {V.}~\bibnamefont
  {Perel}},\ }\href {\doibase 10.1016/0375-9601(71)90196-4} {\bibfield
  {journal} {\bibinfo  {journal} {Physics Letters A}\ }\textbf {\bibinfo
  {volume} {35}},\ \bibinfo {pages} {459 } (\bibinfo {year}
  {1971}{\natexlab{b}})}\BibitemShut {NoStop}%
\bibitem [{\citenamefont {Chazalviel}(1975)}]{Chazalviel1975}%
  \BibitemOpen
  \bibfield  {author} {\bibinfo {author} {\bibfnamefont {J.-N.}\ \bibnamefont
  {Chazalviel}},\ }\href {\doibase 10.1103/PhysRevB.11.3918} {\bibfield
  {journal} {\bibinfo  {journal} {Physical Review B}\ }\textbf {\bibinfo
  {volume} {11}},\ \bibinfo {pages} {3918} (\bibinfo {year}
  {1975})}\BibitemShut {NoStop}%
\bibitem [{\citenamefont {Hirsch}(1999)}]{Hirsch1999}%
  \BibitemOpen
  \bibfield  {author} {\bibinfo {author} {\bibfnamefont {J.~E.}\ \bibnamefont
  {Hirsch}},\ }\href {\doibase 10.1103/PhysRevLett.83.1834} {\bibfield
  {journal} {\bibinfo  {journal} {Physical Review Letters}\ }\textbf {\bibinfo
  {volume} {83}},\ \bibinfo {pages} {1834 } (\bibinfo {year} {1999})},\ \Eprint
  {http://arxiv.org/abs/cond-mat/9906160} {arXiv:cond-mat/9906160} \BibitemShut
  {NoStop}%
\bibitem [{\citenamefont {Dyakonov}\ and\ \citenamefont
  {Khaetskii}(2006)}]{Dyakonov2008}%
  \BibitemOpen
  \bibfield  {author} {\bibinfo {author} {\bibfnamefont {M.}~\bibnamefont
  {Dyakonov}}\ and\ \bibinfo {author} {\bibfnamefont {A.}~\bibnamefont
  {Khaetskii}},\ }in\ \href {\doibase 10.1007/978-3-540-78820-1_8} {\emph
  {\bibinfo {booktitle} {Spin Physics in Semiconductors}}},\ \bibinfo {series}
  {Springer Series in Solid-State Sciences}, Vol.\ \bibinfo {volume} {157},\
  \bibinfo {editor} {edited by\ \bibinfo {editor} {\bibfnamefont
  {M.}~\bibnamefont {Dyakonov}}}\ (\bibinfo  {publisher} {Springer},\ \bibinfo
  {address} {Berlin, Heidelberg},\ \bibinfo {year} {2006})\ Chap.~\bibinfo
  {chapter} {8}, pp.\ \bibinfo {pages} {211--243}\BibitemShut {NoStop}%
\bibitem [{\citenamefont {Mishchenko}\ \emph {et~al.}(2004)\citenamefont
  {Mishchenko}, \citenamefont {Shytov},\ and\ \citenamefont
  {Halperin}}]{Mishchenko2004}%
  \BibitemOpen
  \bibfield  {author} {\bibinfo {author} {\bibfnamefont {E.~G.}\ \bibnamefont
  {Mishchenko}}, \bibinfo {author} {\bibfnamefont {A.~V.}\ \bibnamefont
  {Shytov}}, \ and\ \bibinfo {author} {\bibfnamefont {B.~I.}\ \bibnamefont
  {Halperin}},\ }\href {\doibase 10.1103/PhysRevLett.93.226602} {\bibfield
  {journal} {\bibinfo  {journal} {Physical Review Letters}\ }\textbf {\bibinfo
  {volume} {93}},\ \bibinfo {pages} {226602} (\bibinfo {year} {2004})},\
  \Eprint {http://arxiv.org/abs/cond-mat/0406730} {arXiv:cond-mat/0406730}
  \BibitemShut {NoStop}%
\bibitem [{\citenamefont {Kato}\ \emph
  {et~al.}(2004{\natexlab{a}})\citenamefont {Kato}, \citenamefont {Myers},
  \citenamefont {Gossard},\ and\ \citenamefont {Awschalom}}]{Kato2004}%
  \BibitemOpen
  \bibfield  {author} {\bibinfo {author} {\bibfnamefont {Y.~K.}\ \bibnamefont
  {Kato}}, \bibinfo {author} {\bibfnamefont {R.~C.}\ \bibnamefont {Myers}},
  \bibinfo {author} {\bibfnamefont {A.~C.}\ \bibnamefont {Gossard}}, \ and\
  \bibinfo {author} {\bibfnamefont {D.~D.}\ \bibnamefont {Awschalom}},\ }\href
  {\doibase 10.1103/PhysRevLett.93.176601} {\bibfield  {journal} {\bibinfo
  {journal} {Physical Review Letters}\ }\textbf {\bibinfo {volume} {93}},\
  \bibinfo {pages} {176601} (\bibinfo {year} {2004}{\natexlab{a}})},\ \Eprint
  {http://arxiv.org/abs/cond-mat/0403407} {arXiv:cond-mat/0403407} \BibitemShut
  {NoStop}%
\bibitem [{\citenamefont {Kato}\ \emph
  {et~al.}(2004{\natexlab{b}})\citenamefont {Kato}, \citenamefont {Myers},
  \citenamefont {Gossard},\ and\ \citenamefont {Awschalom}}]{Kato2004a}%
  \BibitemOpen
  \bibfield  {author} {\bibinfo {author} {\bibfnamefont {Y.~K.}\ \bibnamefont
  {Kato}}, \bibinfo {author} {\bibfnamefont {R.~C.}\ \bibnamefont {Myers}},
  \bibinfo {author} {\bibfnamefont {A.~C.}\ \bibnamefont {Gossard}}, \ and\
  \bibinfo {author} {\bibfnamefont {D.~D.}\ \bibnamefont {Awschalom}},\ }\href
  {\doibase 10.1126/science.1105514} {\bibfield  {journal} {\bibinfo  {journal}
  {Science (New York, N.Y.)}\ }\textbf {\bibinfo {volume} {306}},\ \bibinfo
  {pages} {1910} (\bibinfo {year} {2004}{\natexlab{b}})}\BibitemShut {NoStop}%
\bibitem [{\citenamefont {Wunderlich}\ \emph {et~al.}(2005)\citenamefont
  {Wunderlich}, \citenamefont {Kaestner}, \citenamefont {Sinova},\ and\
  \citenamefont {Jungwirth}}]{Wunderlich2005}%
  \BibitemOpen
  \bibfield  {author} {\bibinfo {author} {\bibfnamefont {J.}~\bibnamefont
  {Wunderlich}}, \bibinfo {author} {\bibfnamefont {B.}~\bibnamefont
  {Kaestner}}, \bibinfo {author} {\bibfnamefont {J.}~\bibnamefont {Sinova}}, \
  and\ \bibinfo {author} {\bibfnamefont {T.}~\bibnamefont {Jungwirth}},\ }\href
  {\doibase 10.1103/PhysRevLett.94.047204} {\bibfield  {journal} {\bibinfo
  {journal} {Physical Review Letters}\ }\textbf {\bibinfo {volume} {94}},\
  \bibinfo {pages} {047204} (\bibinfo {year} {2005})},\ \Eprint
  {http://arxiv.org/abs/cond-mat/0410295} {arXiv:cond-mat/0410295} \BibitemShut
  {NoStop}%
\bibitem [{\citenamefont {Raimondi}\ \emph {et~al.}(2006)\citenamefont
  {Raimondi}, \citenamefont {Gorini}, \citenamefont {Schwab},\ and\
  \citenamefont {Dzierzawa}}]{Raimondi2006}%
  \BibitemOpen
  \bibfield  {author} {\bibinfo {author} {\bibfnamefont {R.}~\bibnamefont
  {Raimondi}}, \bibinfo {author} {\bibfnamefont {C.}~\bibnamefont {Gorini}},
  \bibinfo {author} {\bibfnamefont {P.}~\bibnamefont {Schwab}}, \ and\ \bibinfo
  {author} {\bibfnamefont {M.}~\bibnamefont {Dzierzawa}},\ }\href {\doibase
  10.1103/PhysRevB.74.035340} {\bibfield  {journal} {\bibinfo  {journal}
  {Physical Review B}\ }\textbf {\bibinfo {volume} {74}},\ \bibinfo {pages}
  {035340} (\bibinfo {year} {2006})},\ \Eprint {http://arxiv.org/abs/cond-mat/0601525}
  {arXiv:cond-mat/0601525} \BibitemShut {NoStop}%
\bibitem [{\citenamefont {Raimondi}\ \emph {et~al.}(2012)\citenamefont
  {Raimondi}, \citenamefont {Schwab}, \citenamefont {Gorini},\ and\
  \citenamefont {Vignale}}]{Raimondi2011}%
  \BibitemOpen
  \bibfield  {author} {\bibinfo {author} {\bibfnamefont {R.}~\bibnamefont
  {Raimondi}}, \bibinfo {author} {\bibfnamefont {P.}~\bibnamefont {Schwab}},
  \bibinfo {author} {\bibfnamefont {C.}~\bibnamefont {Gorini}}, \ and\ \bibinfo
  {author} {\bibfnamefont {G.}~\bibnamefont {Vignale}},\ }\href {\doibase
  10.1002/andp.201100253} {\bibfield  {journal} {\bibinfo  {journal} {Annalen
  der Physik}\ }\textbf {\bibinfo {volume} {524}},\ \bibinfo {pages} {153}
  (\bibinfo {year} {2012})},\ \Eprint {http://arxiv.org/abs/1110.5279}
  {arXiv:1110.5279} \BibitemShut {NoStop}%
\bibitem [{\citenamefont {Valenzuela}\ and\ \citenamefont
  {Tinkham}(2006)}]{Valenzuela2006}%
  \BibitemOpen
  \bibfield  {author} {\bibinfo {author} {\bibfnamefont {S.~O.}\ \bibnamefont
  {Valenzuela}}\ and\ \bibinfo {author} {\bibfnamefont {M.}~\bibnamefont
  {Tinkham}},\ }\href {\doibase 10.1038/nature04937} {\bibfield  {journal}
  {\bibinfo  {journal} {Nature}\ }\textbf {\bibinfo {volume} {442}},\ \bibinfo
  {pages} {176} (\bibinfo {year} {2006})},\ \Eprint
  {http://arxiv.org/abs/cond-mat/0605423} {arXiv:cond-mat/0605423} \BibitemShut
  {NoStop}%
\bibitem [{\citenamefont {Morota}\ \emph {et~al.}(2011)\citenamefont {Morota},
  \citenamefont {Niimi}, \citenamefont {Ohnishi}, \citenamefont {Wei},
  \citenamefont {Tanaka}, \citenamefont {Kontani}, \citenamefont {Kimura},\
  and\ \citenamefont {Otani}}]{Morota2010}%
  \BibitemOpen
  \bibfield  {author} {\bibinfo {author} {\bibfnamefont {M.}~\bibnamefont
  {Morota}}, \bibinfo {author} {\bibfnamefont {Y.}~\bibnamefont {Niimi}},
  \bibinfo {author} {\bibfnamefont {K.}~\bibnamefont {Ohnishi}}, \bibinfo
  {author} {\bibfnamefont {D.~H.}\ \bibnamefont {Wei}}, \bibinfo {author}
  {\bibfnamefont {T.}~\bibnamefont {Tanaka}}, \bibinfo {author} {\bibfnamefont
  {H.}~\bibnamefont {Kontani}}, \bibinfo {author} {\bibfnamefont
  {T.}~\bibnamefont {Kimura}}, \ and\ \bibinfo {author} {\bibfnamefont
  {Y.}~\bibnamefont {Otani}},\ }\href {\doibase 10.1103/PhysRevB.83.174405}
  {\bibfield  {journal} {\bibinfo  {journal} {Physical Review B}\ }\textbf
  {\bibinfo {volume} {83}},\ \bibinfo {pages} {174405} (\bibinfo {year}
  {2011})},\ \Eprint {http://arxiv.org/abs/1008.0158} {arXiv:1008.0158}
  \BibitemShut {NoStop}%
\bibitem [{\citenamefont {Isasa}\ \emph {et~al.}(2014)\citenamefont {Isasa},
  \citenamefont {Villamor}, \citenamefont {Hueso}, \citenamefont {Gradhand},\
  and\ \citenamefont {Casanova}}]{Isasa2014}%
  \BibitemOpen
  \bibfield  {author} {\bibinfo {author} {\bibfnamefont {M.}~\bibnamefont
  {Isasa}}, \bibinfo {author} {\bibfnamefont {E.}~\bibnamefont {Villamor}},
  \bibinfo {author} {\bibfnamefont {L.~E.}\ \bibnamefont {Hueso}}, \bibinfo
  {author} {\bibfnamefont {M.}~\bibnamefont {Gradhand}}, \ and\ \bibinfo
  {author} {\bibfnamefont {F.}~\bibnamefont {Casanova}},\ }\href {\doibase
  10.1103/PhysRevB.91.024402} {\bibfield  {journal} {\bibinfo  {journal}
  {Physical Review B}\ }\textbf {\bibinfo {volume} {91}},\ \bibinfo {pages}
  {024402} (\bibinfo {year} {2014})},\ \Eprint {http://arxiv.org/abs/1407.4770}
  {arXiv:1407.4770} \BibitemShut {NoStop}%
\bibitem [{\citenamefont {Saitoh}\ \emph {et~al.}(2006)\citenamefont {Saitoh},
  \citenamefont {Ueda}, \citenamefont {Miyajima},\ and\ \citenamefont
  {Tatara}}]{Saitoh2006}%
  \BibitemOpen
  \bibfield  {author} {\bibinfo {author} {\bibfnamefont {E.}~\bibnamefont
  {Saitoh}}, \bibinfo {author} {\bibfnamefont {M.}~\bibnamefont {Ueda}},
  \bibinfo {author} {\bibfnamefont {H.}~\bibnamefont {Miyajima}}, \ and\
  \bibinfo {author} {\bibfnamefont {G.}~\bibnamefont {Tatara}},\ }\href
  {\doibase 10.1063/1.2199473} {\bibfield  {journal} {\bibinfo  {journal}
  {Applied Physics Letters}\ }\textbf {\bibinfo {volume} {88}},\ \bibinfo
  {pages} {182509} (\bibinfo {year} {2006})}\BibitemShut {NoStop}%
\bibitem [{\citenamefont {Ando}\ \emph {et~al.}(2008)\citenamefont {Ando},
  \citenamefont {Takahashi}, \citenamefont {Harii}, \citenamefont {Sasage},
  \citenamefont {Ieda}, \citenamefont {Maekawa},\ and\ \citenamefont
  {Saitoh}}]{Ando2008}%
  \BibitemOpen
  \bibfield  {author} {\bibinfo {author} {\bibfnamefont {K.}~\bibnamefont
  {Ando}}, \bibinfo {author} {\bibfnamefont {S.}~\bibnamefont {Takahashi}},
  \bibinfo {author} {\bibfnamefont {K.}~\bibnamefont {Harii}}, \bibinfo
  {author} {\bibfnamefont {K.}~\bibnamefont {Sasage}}, \bibinfo {author}
  {\bibfnamefont {J.}~\bibnamefont {Ieda}}, \bibinfo {author} {\bibfnamefont
  {S.}~\bibnamefont {Maekawa}}, \ and\ \bibinfo {author} {\bibfnamefont
  {E.}~\bibnamefont {Saitoh}},\ }\href {\doibase
  10.1103/PhysRevLett.101.036601} {\bibfield  {journal} {\bibinfo  {journal}
  {Physical Review Letters}\ }\textbf {\bibinfo {volume} {101}},\ \bibinfo
  {pages} {036601} (\bibinfo {year} {2008})}\BibitemShut {NoStop}%
\bibitem [{\citenamefont {Kimura}\ \emph {et~al.}(2007)\citenamefont {Kimura},
  \citenamefont {Otani}, \citenamefont {Sato}, \citenamefont {Takahashi},\ and\
  \citenamefont {Maekawa}}]{Kimura2007}%
  \BibitemOpen
  \bibfield  {author} {\bibinfo {author} {\bibfnamefont {T.}~\bibnamefont
  {Kimura}}, \bibinfo {author} {\bibfnamefont {Y.}~\bibnamefont {Otani}},
  \bibinfo {author} {\bibfnamefont {T.}~\bibnamefont {Sato}}, \bibinfo {author}
  {\bibfnamefont {S.}~\bibnamefont {Takahashi}}, \ and\ \bibinfo {author}
  {\bibfnamefont {S.}~\bibnamefont {Maekawa}},\ }\href {\doibase
  10.1103/PhysRevLett.98.156601} {\bibfield  {journal} {\bibinfo  {journal}
  {Physical Review Letters}\ }\textbf {\bibinfo {volume} {98}},\ \bibinfo
  {pages} {156601} (\bibinfo {year} {2007})},\ \Eprint
  {http://arxiv.org/abs/cond-mat/0609304} {arXiv:cond-mat/0609304} \BibitemShut
  {NoStop}%
\bibitem [{\citenamefont {Uchida}\ \emph {et~al.}(2010)\citenamefont {Uchida},
  \citenamefont {Xiao}, \citenamefont {Adachi}, \citenamefont {Ohe},
  \citenamefont {Takahashi}, \citenamefont {Ieda}, \citenamefont {Ota},
  \citenamefont {Kajiwara}, \citenamefont {Umezawa}, \citenamefont {Kawai},
  \citenamefont {Bauer}, \citenamefont {Maekawa},\ and\ \citenamefont
  {Saitoh}}]{Uchida2010}%
  \BibitemOpen
  \bibfield  {author} {\bibinfo {author} {\bibfnamefont {K.}~\bibnamefont
  {Uchida}}, \bibinfo {author} {\bibfnamefont {J.}~\bibnamefont {Xiao}},
  \bibinfo {author} {\bibfnamefont {H.}~\bibnamefont {Adachi}}, \bibinfo
  {author} {\bibfnamefont {J.}~\bibnamefont {Ohe}}, \bibinfo {author}
  {\bibfnamefont {S.}~\bibnamefont {Takahashi}}, \bibinfo {author}
  {\bibfnamefont {J.}~\bibnamefont {Ieda}}, \bibinfo {author} {\bibfnamefont
  {T.}~\bibnamefont {Ota}}, \bibinfo {author} {\bibfnamefont {Y.}~\bibnamefont
  {Kajiwara}}, \bibinfo {author} {\bibfnamefont {H.}~\bibnamefont {Umezawa}},
  \bibinfo {author} {\bibfnamefont {H.}~\bibnamefont {Kawai}}, \bibinfo
  {author} {\bibfnamefont {G.~E.~W.}\ \bibnamefont {Bauer}}, \bibinfo {author}
  {\bibfnamefont {S.}~\bibnamefont {Maekawa}}, \ and\ \bibinfo {author}
  {\bibfnamefont {E.}~\bibnamefont {Saitoh}},\ }\href {\doibase
  10.1038/NMAT2856} {\bibfield  {journal} {\bibinfo  {journal} {Nature
  materials}\ }\textbf {\bibinfo {volume} {9}},\ \bibinfo {pages} {894}
  (\bibinfo {year} {2010})},\ \Eprint {http://arxiv.org/abs/1009.5766}
  {arXiv:1009.5766} \BibitemShut {NoStop}%
\bibitem [{\citenamefont {Kajiwara}\ \emph {et~al.}(2010)\citenamefont
  {Kajiwara}, \citenamefont {Harii}, \citenamefont {Takahashi}, \citenamefont
  {Ohe}, \citenamefont {Uchida}, \citenamefont {Mizuguchi}, \citenamefont
  {Umezawa}, \citenamefont {Kawai}, \citenamefont {Ando}, \citenamefont
  {Takanashi}, \citenamefont {Maekawa},\ and\ \citenamefont
  {Saitoh}}]{Kajiwara2010}%
  \BibitemOpen
  \bibfield  {author} {\bibinfo {author} {\bibfnamefont {Y.}~\bibnamefont
  {Kajiwara}}, \bibinfo {author} {\bibfnamefont {K.}~\bibnamefont {Harii}},
  \bibinfo {author} {\bibfnamefont {S.}~\bibnamefont {Takahashi}}, \bibinfo
  {author} {\bibfnamefont {J.}~\bibnamefont {Ohe}}, \bibinfo {author}
  {\bibfnamefont {K.}~\bibnamefont {Uchida}}, \bibinfo {author} {\bibfnamefont
  {M.}~\bibnamefont {Mizuguchi}}, \bibinfo {author} {\bibfnamefont
  {H.}~\bibnamefont {Umezawa}}, \bibinfo {author} {\bibfnamefont
  {H.}~\bibnamefont {Kawai}}, \bibinfo {author} {\bibfnamefont
  {K.}~\bibnamefont {Ando}}, \bibinfo {author} {\bibfnamefont {K.}~\bibnamefont
  {Takanashi}}, \bibinfo {author} {\bibfnamefont {S.}~\bibnamefont {Maekawa}},
  \ and\ \bibinfo {author} {\bibfnamefont {E.}~\bibnamefont {Saitoh}},\ }\href
  {\doibase 10.1038/nature08876} {\bibfield  {journal} {\bibinfo  {journal}
  {Nature}\ }\textbf {\bibinfo {volume} {464}},\ \bibinfo {pages} {262}
  (\bibinfo {year} {2010})}\BibitemShut {NoStop}%
\bibitem [{\citenamefont {Aronov}\ and\ \citenamefont
  {Lyanda-Geller}(1989)}]{Aronov1989}%
  \BibitemOpen
  \bibfield  {author} {\bibinfo {author} {\bibfnamefont {A.}~\bibnamefont
  {Aronov}}\ and\ \bibinfo {author} {\bibfnamefont {Y.}~\bibnamefont
  {Lyanda-Geller}},\ }\href@noop {} {\bibfield  {journal} {\bibinfo  {journal}
  {JETP Letters}\ }\textbf {\bibinfo {volume} {50}},\ \bibinfo {pages} {431}
  (\bibinfo {year} {1989})}\BibitemShut {NoStop}%
\bibitem [{\citenamefont {Edelstein}(1990)}]{Edelstein1990}%
  \BibitemOpen
  \bibfield  {author} {\bibinfo {author} {\bibfnamefont {V.}~\bibnamefont
  {Edelstein}},\ }\href {\doibase 10.1016/0038-1098(90)90963-C} {\bibfield
  {journal} {\bibinfo  {journal} {Solid State Communications}\ }\textbf
  {\bibinfo {volume} {73}},\ \bibinfo {pages} {233} (\bibinfo {year}
  {1990})}\BibitemShut {NoStop}%
\bibitem [{\citenamefont {Silov}\ \emph {et~al.}(2004)\citenamefont {Silov},
  \citenamefont {Blajnov}, \citenamefont {Wolter}, \citenamefont {Hey},
  \citenamefont {Ploog},\ and\ \citenamefont {Averkiev}}]{Silov2004}%
  \BibitemOpen
  \bibfield  {author} {\bibinfo {author} {\bibfnamefont {A.~Y.}\ \bibnamefont
  {Silov}}, \bibinfo {author} {\bibfnamefont {P.~A.}\ \bibnamefont {Blajnov}},
  \bibinfo {author} {\bibfnamefont {J.~H.}\ \bibnamefont {Wolter}}, \bibinfo
  {author} {\bibfnamefont {R.}~\bibnamefont {Hey}}, \bibinfo {author}
  {\bibfnamefont {K.~H.}\ \bibnamefont {Ploog}}, \ and\ \bibinfo {author}
  {\bibfnamefont {N.~S.}\ \bibnamefont {Averkiev}},\ }\href {\doibase
  10.1063/1.1833565} {\bibfield  {journal} {\bibinfo  {journal} {Applied
  Physics Letters}\ }\textbf {\bibinfo {volume} {85}},\ \bibinfo {pages} {5929}
  (\bibinfo {year} {2004})}\BibitemShut {NoStop}%
\bibitem [{\citenamefont {Shen}\ \emph {et~al.}(2014)\citenamefont {Shen},
  \citenamefont {Vignale},\ and\ \citenamefont {Raimondi}}]{Shen2014a}%
  \BibitemOpen
  \bibfield  {author} {\bibinfo {author} {\bibfnamefont {K.}~\bibnamefont
  {Shen}}, \bibinfo {author} {\bibfnamefont {G.}~\bibnamefont {Vignale}}, \
  and\ \bibinfo {author} {\bibfnamefont {R.}~\bibnamefont {Raimondi}},\ }\href
  {\doibase 10.1103/PhysRevLett.112.096601} {\bibfield  {journal} {\bibinfo
  {journal} {Physical Review Letters}\ }\textbf {\bibinfo {volume} {112}},\
  \bibinfo {pages} {096601} (\bibinfo {year} {2014})},\ \Eprint
  {http://arxiv.org/abs/1311.6516} {arXiv:1311.6516} \BibitemShut {NoStop}%
\bibitem [{\citenamefont {Ganichev}\ \emph {et~al.}(2002)\citenamefont
  {Ganichev}, \citenamefont {Ivchenko}, \citenamefont {Bel'kov}, \citenamefont
  {Tarasenko}, \citenamefont {Sollinger}, \citenamefont {Weiss}, \citenamefont
  {Wegscheider},\ and\ \citenamefont {Prettl}}]{Ganichev2002}%
  \BibitemOpen
  \bibfield  {author} {\bibinfo {author} {\bibfnamefont {S.}~\bibnamefont
  {Ganichev}}, \bibinfo {author} {\bibfnamefont {E.}~\bibnamefont {Ivchenko}},
  \bibinfo {author} {\bibfnamefont {V.}~\bibnamefont {Bel'kov}}, \bibinfo
  {author} {\bibfnamefont {S.}~\bibnamefont {Tarasenko}}, \bibinfo {author}
  {\bibfnamefont {M.}~\bibnamefont {Sollinger}}, \bibinfo {author}
  {\bibfnamefont {D.}~\bibnamefont {Weiss}}, \bibinfo {author} {\bibfnamefont
  {W.}~\bibnamefont {Wegscheider}}, \ and\ \bibinfo {author} {\bibfnamefont
  {W.}~\bibnamefont {Prettl}},\ }\href {\doibase 10.1038/417153a} {\bibfield
  {journal} {\bibinfo  {journal} {Nature}\ }\textbf {\bibinfo {volume} {417}},\
  \bibinfo {pages} {153} (\bibinfo {year} {2002})}\BibitemShut {NoStop}%
\bibitem [{\citenamefont {{Rojas S\'{a}nchez}}\ \emph
  {et~al.}(2013)\citenamefont {{Rojas S\'{a}nchez}}, \citenamefont {Vila},
  \citenamefont {Desfonds}, \citenamefont {Gambarelli}, \citenamefont
  {Attan\'{e}}, \citenamefont {{De Teresa}}, \citenamefont {Mag\'{e}n},\ and\
  \citenamefont {Fert}}]{Sanchez2013}%
  \BibitemOpen
  \bibfield  {author} {\bibinfo {author} {\bibfnamefont {J.~C.}\ \bibnamefont
  {{Rojas S\'{a}nchez}}}, \bibinfo {author} {\bibfnamefont {L.}~\bibnamefont
  {Vila}}, \bibinfo {author} {\bibfnamefont {G.}~\bibnamefont {Desfonds}},
  \bibinfo {author} {\bibfnamefont {S.}~\bibnamefont {Gambarelli}}, \bibinfo
  {author} {\bibfnamefont {J.~P.}\ \bibnamefont {Attan\'{e}}}, \bibinfo
  {author} {\bibfnamefont {J.~M.}\ \bibnamefont {{De Teresa}}}, \bibinfo
  {author} {\bibfnamefont {C.}~\bibnamefont {Mag\'{e}n}}, \ and\ \bibinfo
  {author} {\bibfnamefont {A.}~\bibnamefont {Fert}},\ }\href {\doibase
  10.1038/ncomms3944} {\bibfield  {journal} {\bibinfo  {journal} {Nature
  communications}\ }\textbf {\bibinfo {volume} {4}},\ \bibinfo {pages} {2944}
  (\bibinfo {year} {2013})}\BibitemShut {NoStop}%
\bibitem [{\citenamefont {Edelstein}(1995)}]{Edelstein1995}%
  \BibitemOpen
  \bibfield  {author} {\bibinfo {author} {\bibfnamefont {V.~M.}\ \bibnamefont
  {Edelstein}},\ }\href {\doibase 10.1103/PhysRevLett.75.2004} {\bibfield
  {journal} {\bibinfo  {journal} {Physical Review Letters}\ }\textbf {\bibinfo
  {volume} {75}},\ \bibinfo {pages} {2004} (\bibinfo {year}
  {1995})}\BibitemShut {NoStop}%
\bibitem [{\citenamefont {Edelstein}(2005)}]{Edelstein2005}%
  \BibitemOpen
  \bibfield  {author} {\bibinfo {author} {\bibfnamefont {V.~M.}\ \bibnamefont
  {Edelstein}},\ }\href {\doibase 10.1103/PhysRevB.72.172501} {\bibfield
  {journal} {\bibinfo  {journal} {Physical Review B}\ }\textbf {\bibinfo
  {volume} {72}},\ \bibinfo {pages} {1} (\bibinfo {year} {2005})}\BibitemShut
  {NoStop}%
\bibitem [{\citenamefont {Yip}(2002)}]{Yip2001}%
  \BibitemOpen
  \bibfield  {author} {\bibinfo {author} {\bibfnamefont {S.~K.}\ \bibnamefont
  {Yip}},\ }\href {\doibase 10.1103/PhysRevB.65.144508} {\bibfield  {journal}
  {\bibinfo  {journal} {Physical Review B}\ }\textbf {\bibinfo {volume} {65}},\
  \bibinfo {pages} {144508} (\bibinfo {year} {2002})},\ \Eprint
  {http://arxiv.org/abs/cond-mat/0110140} {arXiv:cond-mat/0110140} \BibitemShut
  {NoStop}%
\bibitem [{\citenamefont {Bauer}\ and\ \citenamefont
  {Sigrist}(2012)}]{Bauer2012a}%
  \BibitemOpen
  \bibfield  {author} {\bibinfo {author} {\bibfnamefont {E.}~\bibnamefont
  {Bauer}}\ and\ \bibinfo {author} {\bibfnamefont {M.}~\bibnamefont
  {Sigrist}},\ }\href {\doibase 10.1007/978-3-642-24624-1} {\emph {\bibinfo
  {title} {{Non-Centrosymmetric Superconductors}}}},\ \bibinfo {series}
  {Lecture Notes in Physics}, Vol.\ \bibinfo {volume} {847}\ (\bibinfo
  {publisher} {Springer},\ \bibinfo {address} {Berlin, Heidelberg},\ \bibinfo
  {year} {2012})\BibitemShut {NoStop}%
\bibitem [{\citenamefont {Krive}\ \emph {et~al.}(2004)\citenamefont {Krive},
  \citenamefont {Kadigrobov}, \citenamefont {Shekhter},\ and\ \citenamefont
  {Jonson}}]{krive_kadigrobov_shekhter_jonson.2005}%
  \BibitemOpen
  \bibfield  {author} {\bibinfo {author} {\bibfnamefont {I.~V.}\ \bibnamefont
  {Krive}}, \bibinfo {author} {\bibfnamefont {A.~M.}\ \bibnamefont
  {Kadigrobov}}, \bibinfo {author} {\bibfnamefont {R.~I.}\ \bibnamefont
  {Shekhter}}, \ and\ \bibinfo {author} {\bibfnamefont {M.}~\bibnamefont
  {Jonson}},\ }\href {\doibase 10.1103/PhysRevB.71.214516} {\bibfield
  {journal} {\bibinfo  {journal} {Physical Review B}\ }\textbf {\bibinfo
  {volume} {71}},\ \bibinfo {pages} {214516} (\bibinfo {year} {2004})},\
  \Eprint {http://arxiv.org/abs/cond-mat/0409063} {arXiv:cond-mat/0409063}
  \BibitemShut {NoStop}%
\bibitem [{\citenamefont {Buzdin}(2008)}]{buzdin:107005.2008}%
  \BibitemOpen
  \bibfield  {author} {\bibinfo {author} {\bibfnamefont {A.~I.}\ \bibnamefont
  {Buzdin}},\ }\href {\doibase 10.1103/PhysRevLett.101.107005} {\bibfield
  {journal} {\bibinfo  {journal} {Physical Review Letters}\ }\textbf {\bibinfo
  {volume} {101}},\ \bibinfo {pages} {107005} (\bibinfo {year} {2008})},\
  \Eprint {http://arxiv.org/abs/0808.0299} {arXiv:0808.0299} \BibitemShut
  {NoStop}%
\bibitem [{\citenamefont {Liu}\ and\ \citenamefont
  {Chan}(2010{\natexlab{a}})}]{Liu2010}%
  \BibitemOpen
  \bibfield  {author} {\bibinfo {author} {\bibfnamefont {J.-F.}\ \bibnamefont
  {Liu}}\ and\ \bibinfo {author} {\bibfnamefont {K.}~\bibnamefont {Chan}},\
  }\href {\doibase 10.1103/PhysRevB.82.125305} {\bibfield  {journal} {\bibinfo
  {journal} {Physical Review B}\ }\textbf {\bibinfo {volume} {82}},\ \bibinfo
  {pages} {125305} (\bibinfo {year} {2010}{\natexlab{a}})}\BibitemShut
  {NoStop}%
\bibitem [{\citenamefont {Mal'shukov}\ \emph {et~al.}(2010)\citenamefont
  {Mal'shukov}, \citenamefont {Sadjina},\ and\ \citenamefont
  {Brataas}}]{Malshukov2010}%
  \BibitemOpen
  \bibfield  {author} {\bibinfo {author} {\bibfnamefont {A.~G.}\ \bibnamefont
  {Mal'shukov}}, \bibinfo {author} {\bibfnamefont {S.}~\bibnamefont {Sadjina}},
  \ and\ \bibinfo {author} {\bibfnamefont {A.}~\bibnamefont {Brataas}},\ }\href
  {\doibase 10.1103/PhysRevB.81.060502} {\bibfield  {journal} {\bibinfo
  {journal} {Physical Review B}\ }\textbf {\bibinfo {volume} {81}},\ \bibinfo
  {pages} {060502} (\bibinfo {year} {2010})}\BibitemShut {NoStop}%
\bibitem [{\citenamefont {Reynoso}\ \emph {et~al.}(2012)\citenamefont
  {Reynoso}, \citenamefont {Usaj}, \citenamefont {Balseiro}, \citenamefont
  {Feinberg},\ and\ \citenamefont {Avignon}}]{Reynoso2012}%
  \BibitemOpen
  \bibfield  {author} {\bibinfo {author} {\bibfnamefont {A.~A.}\ \bibnamefont
  {Reynoso}}, \bibinfo {author} {\bibfnamefont {G.}~\bibnamefont {Usaj}},
  \bibinfo {author} {\bibfnamefont {C.~A.}\ \bibnamefont {Balseiro}}, \bibinfo
  {author} {\bibfnamefont {D.}~\bibnamefont {Feinberg}}, \ and\ \bibinfo
  {author} {\bibfnamefont {M.}~\bibnamefont {Avignon}},\ }\href {\doibase
  10.1103/PhysRevB.86.214519} {\bibfield  {journal} {\bibinfo  {journal}
  {Physical Review B}\ }\textbf {\bibinfo {volume} {86}},\ \bibinfo {pages}
  {214519} (\bibinfo {year} {2012})},\ \Eprint {http://arxiv.org/abs/1212.2786}
  {arXiv:1212.2786} \BibitemShut {NoStop}%
\bibitem [{\citenamefont {Yokoyama}\ \emph {et~al.}(2013)\citenamefont
  {Yokoyama}, \citenamefont {Eto},\ and\ \citenamefont {{V.
  Nazarov}}}]{Yokoyama2012}%
  \BibitemOpen
  \bibfield  {author} {\bibinfo {author} {\bibfnamefont {T.}~\bibnamefont
  {Yokoyama}}, \bibinfo {author} {\bibfnamefont {M.}~\bibnamefont {Eto}}, \
  and\ \bibinfo {author} {\bibfnamefont {Y.}~\bibnamefont {{V. Nazarov}}},\
  }\href {\doibase 10.7566/JPSJ.82.054703} {\bibfield  {journal} {\bibinfo
  {journal} {Journal of the Physical Society of Japan}\ }\textbf {\bibinfo
  {volume} {82}},\ \bibinfo {pages} {054703} (\bibinfo {year} {2013})},\
  \Eprint {http://arxiv.org/abs/1212.5390} {arXiv:1212.5390} \BibitemShut
  {NoStop}%
\bibitem [{\citenamefont {Yokoyama}\ \emph {et~al.}(2014)\citenamefont
  {Yokoyama}, \citenamefont {Eto},\ and\ \citenamefont
  {Nazarov}}]{Yokoyama2014}%
  \BibitemOpen
  \bibfield  {author} {\bibinfo {author} {\bibfnamefont {T.}~\bibnamefont
  {Yokoyama}}, \bibinfo {author} {\bibfnamefont {M.}~\bibnamefont {Eto}}, \
  and\ \bibinfo {author} {\bibfnamefont {Y.~V.}\ \bibnamefont {Nazarov}},\
  }\href {http://arxiv.org/abs/1402.0305} {\  (\bibinfo {year} {2014})},\
  \Eprint {http://arxiv.org/abs/1402.0305} {arXiv:1402.0305} \BibitemShut
  {NoStop}%
\bibitem [{\citenamefont {Yokoyama}\ and\ \citenamefont
  {Nazarov}(2014)}]{Yokoyama2014b}%
  \BibitemOpen
  \bibfield  {author} {\bibinfo {author} {\bibfnamefont {T.}~\bibnamefont
  {Yokoyama}}\ and\ \bibinfo {author} {\bibfnamefont {Y.~V.}\ \bibnamefont
  {Nazarov}},\ }\href {\doibase 10.1209/0295-5075/108/47009} {\bibfield
  {journal} {\bibinfo  {journal} {Europhysics Letters (EPL)}\ }\textbf
  {\bibinfo {volume} {108}},\ \bibinfo {pages} {47009} (\bibinfo {year}
  {2014})},\ \Eprint {http://arxiv.org/abs/1408.4034} {arXiv:1408.4034}
  \BibitemShut {NoStop}%
\bibitem [{\citenamefont {Campagnano}\ \emph {et~al.}(2014)\citenamefont
  {Campagnano}, \citenamefont {Lucignano}, \citenamefont {Giuliano},\ and\
  \citenamefont {Tagliacozzo}}]{Campagnano2014}%
  \BibitemOpen
  \bibfield  {author} {\bibinfo {author} {\bibfnamefont {G.}~\bibnamefont
  {Campagnano}}, \bibinfo {author} {\bibfnamefont {P.}~\bibnamefont
  {Lucignano}}, \bibinfo {author} {\bibfnamefont {D.}~\bibnamefont {Giuliano}},
  \ and\ \bibinfo {author} {\bibfnamefont {A.}~\bibnamefont {Tagliacozzo}},\
  }\href {http://arxiv.org/abs/1408.3953} {\  (\bibinfo {year} {2014})},\
  \Eprint {http://arxiv.org/abs/1408.3953} {arXiv:1408.3953} \BibitemShut
  {NoStop}%
\bibitem [{\citenamefont {Bergeret}\ and\ \citenamefont
  {Tokatly}(2015)}]{Bergeret2014a}%
  \BibitemOpen
  \bibfield  {author} {\bibinfo {author} {\bibfnamefont {F.~S.}\ \bibnamefont
  {Bergeret}}\ and\ \bibinfo {author} {\bibfnamefont {I.~V.}\ \bibnamefont
  {Tokatly}},\ }\href {http://arxiv.org/abs/1409.4563} {\bibfield  {journal}
  {\bibinfo  {journal} {Europhysics Letters (EPL)}\ }\textbf {\bibinfo {volume}
  {110}},\ \bibinfo {pages} {57005} (\bibinfo {year} {2015})},\ \Eprint
  {http://arxiv.org/abs/1409.4563} {arXiv:1409.4563} \BibitemShut {NoStop}%
\bibitem [{\citenamefont {Konschelle}(2014{\natexlab{a}})}]{Konschelle2014a}%
  \BibitemOpen
  \bibfield  {author} {\bibinfo {author} {\bibfnamefont {F.}~\bibnamefont
  {Konschelle}},\ }\href {http://arxiv.org/abs/1408.4533} {\  (\bibinfo {year}
  {2014}{\natexlab{a}})},\ \Eprint {http://arxiv.org/abs/1408.4533}
  {arXiv:1408.4533} \BibitemShut {NoStop}%
\bibitem [{\citenamefont {Kulagina}\ and\ \citenamefont
  {Linder}(2014)}]{Kulagina2014}%
  \BibitemOpen
  \bibfield  {author} {\bibinfo {author} {\bibfnamefont {I.}~\bibnamefont
  {Kulagina}}\ and\ \bibinfo {author} {\bibfnamefont {J.}~\bibnamefont
  {Linder}},\ }\href {http://arxiv.org/abs/1406.7016} {\  (\bibinfo {year}
  {2014})},\ \Eprint {http://arxiv.org/abs/1406.7016} {arXiv:1406.7016}
  \BibitemShut {NoStop}%
\bibitem [{\citenamefont {Mironov}\ \emph {et~al.}(2014)\citenamefont
  {Mironov}, \citenamefont {Mel'nikov},\ and\ \citenamefont
  {Buzdin}}]{Mironov2014}%
  \BibitemOpen
  \bibfield  {author} {\bibinfo {author} {\bibfnamefont {S.~V.}\ \bibnamefont
  {Mironov}}, \bibinfo {author} {\bibfnamefont {A.~S.}\ \bibnamefont
  {Mel'nikov}}, \ and\ \bibinfo {author} {\bibfnamefont {A.~I.}\ \bibnamefont
  {Buzdin}},\ }\href {http://arxiv.org/abs/1411.1626} {\  (\bibinfo {year}
  {2014})},\ \Eprint {http://arxiv.org/abs/1411.1626} {arXiv:1411.1626}
  \BibitemShut {NoStop}%
\bibitem [{\citenamefont {Braude}\ and\ \citenamefont
  {Nazarov}(2007)}]{Braude2007}%
  \BibitemOpen
  \bibfield  {author} {\bibinfo {author} {\bibfnamefont {V.}~\bibnamefont
  {Braude}}\ and\ \bibinfo {author} {\bibfnamefont {Y.~V.}\ \bibnamefont
  {Nazarov}},\ }\href {\doibase 10.1103/PhysRevLett.98.077003} {\bibfield
  {journal} {\bibinfo  {journal} {Physical Review Letters}\ }\textbf {\bibinfo
  {volume} {98}},\ \bibinfo {pages} {077003} (\bibinfo {year} {2007})},\
  \Eprint {http://arxiv.org/abs/cond-mat/0610037} {arXiv:cond-mat/0610037}
  \BibitemShut {NoStop}%
\bibitem [{\citenamefont {Tanaka}\ \emph {et~al.}(2007)\citenamefont {Tanaka},
  \citenamefont {Golubov}, \citenamefont {Kashiwaya},\ and\ \citenamefont
  {Ueda}}]{Tanaka2007}%
  \BibitemOpen
  \bibfield  {author} {\bibinfo {author} {\bibfnamefont {Y.}~\bibnamefont
  {Tanaka}}, \bibinfo {author} {\bibfnamefont {A.}~\bibnamefont {Golubov}},
  \bibinfo {author} {\bibfnamefont {S.}~\bibnamefont {Kashiwaya}}, \ and\
  \bibinfo {author} {\bibfnamefont {M.}~\bibnamefont {Ueda}},\ }\href {\doibase
  10.1103/PhysRevLett.99.037005} {\bibfield  {journal} {\bibinfo  {journal}
  {Physical Review Letters}\ }\textbf {\bibinfo {volume} {99}},\ \bibinfo
  {pages} {037005} (\bibinfo {year} {2007})}\BibitemShut {NoStop}%
\bibitem [{\citenamefont {Eschrig}\ \emph {et~al.}(2007)\citenamefont
  {Eschrig}, \citenamefont {L\"{o}fwander}, \citenamefont {Champel},
  \citenamefont {Cuevas}, \citenamefont {Kopu},\ and\ \citenamefont
  {Sch\"{o}n}}]{Eschrig2007}%
  \BibitemOpen
  \bibfield  {author} {\bibinfo {author} {\bibfnamefont {M.}~\bibnamefont
  {Eschrig}}, \bibinfo {author} {\bibfnamefont {T.}~\bibnamefont
  {L\"{o}fwander}}, \bibinfo {author} {\bibfnamefont {T.}~\bibnamefont
  {Champel}}, \bibinfo {author} {\bibfnamefont {J.~C.}\ \bibnamefont {Cuevas}},
  \bibinfo {author} {\bibfnamefont {J.}~\bibnamefont {Kopu}}, \ and\ \bibinfo
  {author} {\bibfnamefont {G.}~\bibnamefont {Sch\"{o}n}},\ }\href {\doibase
  10.1007/s10909-007-9329-6} {\bibfield  {journal} {\bibinfo  {journal}
  {Journal of Low Temperature Physics}\ }\textbf {\bibinfo {volume} {147}},\
  \bibinfo {pages} {457} (\bibinfo {year} {2007})},\ \Eprint
  {http://arxiv.org/abs/cond-mat/0610212} {arXiv:cond-mat/0610212} \BibitemShut
  {NoStop}%
\bibitem [{\citenamefont {Grein}\ \emph {et~al.}(2009)\citenamefont {Grein},
  \citenamefont {Eschrig}, \citenamefont {Metalidis},\ and\ \citenamefont
  {Sch\"{o}n}}]{Grein2009}%
  \BibitemOpen
  \bibfield  {author} {\bibinfo {author} {\bibfnamefont {R.}~\bibnamefont
  {Grein}}, \bibinfo {author} {\bibfnamefont {M.}~\bibnamefont {Eschrig}},
  \bibinfo {author} {\bibfnamefont {G.}~\bibnamefont {Metalidis}}, \ and\
  \bibinfo {author} {\bibfnamefont {G.}~\bibnamefont {Sch\"{o}n}},\ }\href
  {\doibase 10.1103/PhysRevLett.102.227005} {\bibfield  {journal} {\bibinfo
  {journal} {Physical Review Letters}\ }\textbf {\bibinfo {volume} {102}},\
  \bibinfo {pages} {227005} (\bibinfo {year} {2009})},\ \Eprint
  {http://arxiv.org/abs/0904.0149} {arXiv:0904.0149} \BibitemShut {NoStop}%
\bibitem [{\citenamefont {Liu}\ and\ \citenamefont
  {Chan}(2010{\natexlab{b}})}]{Liu2010a}%
  \BibitemOpen
  \bibfield  {author} {\bibinfo {author} {\bibfnamefont {J.-F.}\ \bibnamefont
  {Liu}}\ and\ \bibinfo {author} {\bibfnamefont {K.}~\bibnamefont {Chan}},\
  }\href {\doibase 10.1103/PhysRevB.82.184533} {\bibfield  {journal} {\bibinfo
  {journal} {Physical Review B}\ }\textbf {\bibinfo {volume} {82}},\ \bibinfo
  {pages} {184533} (\bibinfo {year} {2010}{\natexlab{b}})}\BibitemShut
  {NoStop}%
\bibitem [{\citenamefont {Margaris}\ \emph {et~al.}(2010)\citenamefont
  {Margaris}, \citenamefont {Paltoglou},\ and\ \citenamefont
  {Flytzanis}}]{Margaris2010a}%
  \BibitemOpen
  \bibfield  {author} {\bibinfo {author} {\bibfnamefont {I.}~\bibnamefont
  {Margaris}}, \bibinfo {author} {\bibfnamefont {V.}~\bibnamefont {Paltoglou}},
  \ and\ \bibinfo {author} {\bibfnamefont {N.}~\bibnamefont {Flytzanis}},\
  }\href {\doibase 10.1088/0953-8984/22/44/445701} {\bibfield  {journal}
  {\bibinfo  {journal} {Journal of physics. Condensed matter : an Institute of
  Physics journal}\ }\textbf {\bibinfo {volume} {22}},\ \bibinfo {pages}
  {445701} (\bibinfo {year} {2010})}\BibitemShut {NoStop}%
\bibitem [{\citenamefont {Tanaka}\ \emph {et~al.}(2009)\citenamefont {Tanaka},
  \citenamefont {Yokoyama},\ and\ \citenamefont {Nagaosa}}]{Tanaka2009}%
  \BibitemOpen
  \bibfield  {author} {\bibinfo {author} {\bibfnamefont {Y.}~\bibnamefont
  {Tanaka}}, \bibinfo {author} {\bibfnamefont {T.}~\bibnamefont {Yokoyama}}, \
  and\ \bibinfo {author} {\bibfnamefont {N.}~\bibnamefont {Nagaosa}},\ }\href
  {\doibase 10.1103/PhysRevLett.103.107002} {\bibfield  {journal} {\bibinfo
  {journal} {Physical Review Letters}\ }\textbf {\bibinfo {volume} {103}},\
  \bibinfo {pages} {107002} (\bibinfo {year} {2009})},\ \Eprint
  {http://arxiv.org/abs/0907.2088} {arXiv:0907.2088} \BibitemShut {NoStop}%
\bibitem [{\citenamefont {Dolcini}\ \emph {et~al.}(2015)\citenamefont
  {Dolcini}, \citenamefont {Houzet},\ and\ \citenamefont
  {Meyer}}]{Dolcini2015}%
  \BibitemOpen
  \bibfield  {author} {\bibinfo {author} {\bibfnamefont {F.}~\bibnamefont
  {Dolcini}}, \bibinfo {author} {\bibfnamefont {M.}~\bibnamefont {Houzet}}, \
  and\ \bibinfo {author} {\bibfnamefont {J.~S.}\ \bibnamefont {Meyer}},\ }\href
  {\doibase 10.1103/PhysRevB.92.035428} {\bibfield  {journal} {\bibinfo
  {journal} {Physical Review B}\ }\textbf {\bibinfo {volume} {92}},\ \bibinfo
  {pages} {035428} (\bibinfo {year} {2015})},\ \Eprint
  {http://arxiv.org/abs/1503.01949} {arXiv:1503.01949} \BibitemShut {NoStop}%
\bibitem [{\citenamefont {Geshkenbeim}\ and\ \citenamefont
  {Larkin}(1986)}]{Geshkenbeim1986}%
  \BibitemOpen
  \bibfield  {author} {\bibinfo {author} {\bibfnamefont {V.}~\bibnamefont
  {Geshkenbeim}}\ and\ \bibinfo {author} {\bibfnamefont {A.~I.}\ \bibnamefont
  {Larkin}},\ }\href@noop {} {\bibfield  {journal} {\bibinfo  {journal} {JETP
  Letters}\ }\textbf {\bibinfo {volume} {43}},\ \bibinfo {pages} {395}
  (\bibinfo {year} {1986})}\BibitemShut {NoStop}%
\bibitem [{\citenamefont {Yip}(1995)}]{Yip1995}%
  \BibitemOpen
  \bibfield  {author} {\bibinfo {author} {\bibfnamefont {S.}~\bibnamefont
  {Yip}},\ }\href {\doibase 10.1103/PhysRevB.52.3087} {\bibfield  {journal}
  {\bibinfo  {journal} {Physical Review B}\ }\textbf {\bibinfo {volume} {52}},\
  \bibinfo {pages} {3087} (\bibinfo {year} {1995})}\BibitemShut {NoStop}%
\bibitem [{\citenamefont {Tanaka}\ and\ \citenamefont
  {Kashiwaya}(1997)}]{Tanaka1997}%
  \BibitemOpen
  \bibfield  {author} {\bibinfo {author} {\bibfnamefont {Y.}~\bibnamefont
  {Tanaka}}\ and\ \bibinfo {author} {\bibfnamefont {S.}~\bibnamefont
  {Kashiwaya}},\ }\href {\doibase 10.1103/PhysRevB.56.892} {\bibfield
  {journal} {\bibinfo  {journal} {Physical Review B}\ }\textbf {\bibinfo
  {volume} {56}},\ \bibinfo {pages} {892} (\bibinfo {year} {1997})}\BibitemShut
  {NoStop}%
\bibitem [{\citenamefont {Sigrist}(1998)}]{Sigrist1998}%
  \BibitemOpen
  \bibfield  {author} {\bibinfo {author} {\bibfnamefont {M.}~\bibnamefont
  {Sigrist}},\ }\href {\doibase 10.1143/PTP.99.899} {\bibfield  {journal}
  {\bibinfo  {journal} {Progress of Theoretical Physics}\ }\textbf {\bibinfo
  {volume} {99}},\ \bibinfo {pages} {899} (\bibinfo {year} {1998})}\BibitemShut
  {NoStop}%
\bibitem [{\citenamefont {Kashiwaya}\ and\ \citenamefont
  {Tanaka}(2000)}]{Kashiwaya2000}%
  \BibitemOpen
  \bibfield  {author} {\bibinfo {author} {\bibfnamefont {S.}~\bibnamefont
  {Kashiwaya}}\ and\ \bibinfo {author} {\bibfnamefont {Y.}~\bibnamefont
  {Tanaka}},\ }\href {\doibase 10.1088/0034-4885/63/10/202} {\bibfield
  {journal} {\bibinfo  {journal} {Reports on Progress in Physics}\ }\textbf
  {\bibinfo {volume} {63}},\ \bibinfo {pages} {1641} (\bibinfo {year}
  {2000})}\BibitemShut {NoStop}%
\bibitem [{\citenamefont {Asano}\ \emph {et~al.}(2003)\citenamefont {Asano},
  \citenamefont {Tanaka}, \citenamefont {Sigrist},\ and\ \citenamefont
  {Kashiwaya}}]{Asano2003}%
  \BibitemOpen
  \bibfield  {author} {\bibinfo {author} {\bibfnamefont {Y.}~\bibnamefont
  {Asano}}, \bibinfo {author} {\bibfnamefont {Y.}~\bibnamefont {Tanaka}},
  \bibinfo {author} {\bibfnamefont {M.}~\bibnamefont {Sigrist}}, \ and\
  \bibinfo {author} {\bibfnamefont {S.}~\bibnamefont {Kashiwaya}},\ }\href
  {\doibase 10.1103/PhysRevB.67.184505} {\bibfield  {journal} {\bibinfo
  {journal} {Physical Review B}\ }\textbf {\bibinfo {volume} {67}},\ \bibinfo
  {pages} {184505} (\bibinfo {year} {2003})}\BibitemShut {NoStop}%
\bibitem [{\citenamefont {Asano}\ \emph {et~al.}(2005)\citenamefont {Asano},
  \citenamefont {Tanaka}, \citenamefont {Sigrist},\ and\ \citenamefont
  {Kashiwaya}}]{Asano2005}%
  \BibitemOpen
  \bibfield  {author} {\bibinfo {author} {\bibfnamefont {Y.}~\bibnamefont
  {Asano}}, \bibinfo {author} {\bibfnamefont {Y.}~\bibnamefont {Tanaka}},
  \bibinfo {author} {\bibfnamefont {M.}~\bibnamefont {Sigrist}}, \ and\
  \bibinfo {author} {\bibfnamefont {S.}~\bibnamefont {Kashiwaya}},\ }\href
  {\doibase 10.1103/PhysRevB.71.214501} {\bibfield  {journal} {\bibinfo
  {journal} {Physical Review B}\ }\textbf {\bibinfo {volume} {71}},\ \bibinfo
  {pages} {214501} (\bibinfo {year} {2005})}\BibitemShut {NoStop}%
\bibitem [{\citenamefont {Brydon}\ \emph {et~al.}(2008)\citenamefont {Brydon},
  \citenamefont {Kastening}, \citenamefont {Morr},\ and\ \citenamefont
  {Manske}}]{Brydon2008}%
  \BibitemOpen
  \bibfield  {author} {\bibinfo {author} {\bibfnamefont {P.}~\bibnamefont
  {Brydon}}, \bibinfo {author} {\bibfnamefont {B.}~\bibnamefont {Kastening}},
  \bibinfo {author} {\bibfnamefont {D.}~\bibnamefont {Morr}}, \ and\ \bibinfo
  {author} {\bibfnamefont {D.}~\bibnamefont {Manske}},\ }\href {\doibase
  10.1103/PhysRevB.77.104504} {\bibfield  {journal} {\bibinfo  {journal}
  {Physical Review B}\ }\textbf {\bibinfo {volume} {77}},\ \bibinfo {pages}
  {104504} (\bibinfo {year} {2008})},\ \Eprint {http://arxiv.org/abs/0709.2918}
  {arXiv:0709.2918} \BibitemShut {NoStop}%
\bibitem [{\citenamefont {Rahnavard}\ \emph {et~al.}(2014)\citenamefont
  {Rahnavard}, \citenamefont {Manske},\ and\ \citenamefont
  {Annunziata}}]{Rahnavard2014}%
  \BibitemOpen
  \bibfield  {author} {\bibinfo {author} {\bibfnamefont {Y.}~\bibnamefont
  {Rahnavard}}, \bibinfo {author} {\bibfnamefont {D.}~\bibnamefont {Manske}}, \
  and\ \bibinfo {author} {\bibfnamefont {G.}~\bibnamefont {Annunziata}},\
  }\href {\doibase 10.1103/PhysRevB.89.214501} {\bibfield  {journal} {\bibinfo
  {journal} {Physical Review B}\ }\textbf {\bibinfo {volume} {89}},\ \bibinfo
  {pages} {214501} (\bibinfo {year} {2014})}\BibitemShut {NoStop}%
\bibitem [{\citenamefont {Reynoso}\ \emph {et~al.}(2008)\citenamefont
  {Reynoso}, \citenamefont {Usaj}, \citenamefont {Balseiro}, \citenamefont
  {Feinberg},\ and\ \citenamefont {Avignon}}]{reynoso_etal:107001.2008}%
  \BibitemOpen
  \bibfield  {author} {\bibinfo {author} {\bibfnamefont {A.~A.}\ \bibnamefont
  {Reynoso}}, \bibinfo {author} {\bibfnamefont {G.}~\bibnamefont {Usaj}},
  \bibinfo {author} {\bibfnamefont {C.}~\bibnamefont {Balseiro}}, \bibinfo
  {author} {\bibfnamefont {D.}~\bibnamefont {Feinberg}}, \ and\ \bibinfo
  {author} {\bibfnamefont {M.}~\bibnamefont {Avignon}},\ }\href {\doibase
  10.1103/PhysRevLett.101.107001} {\bibfield  {journal} {\bibinfo  {journal}
  {Physical Review Letters}\ }\textbf {\bibinfo {volume} {101}},\ \bibinfo
  {pages} {107001} (\bibinfo {year} {2008})},\ \Eprint
  {http://arxiv.org/abs/0808.1516} {arXiv:0808.1516} \BibitemShut {NoStop}%
\bibitem [{\citenamefont {Zazunov}\ \emph {et~al.}(2009)\citenamefont
  {Zazunov}, \citenamefont {Egger}, \citenamefont {Martin},\ and\ \citenamefont
  {Jonckheere}}]{Zazunov2009}%
  \BibitemOpen
  \bibfield  {author} {\bibinfo {author} {\bibfnamefont {A.}~\bibnamefont
  {Zazunov}}, \bibinfo {author} {\bibfnamefont {R.}~\bibnamefont {Egger}},
  \bibinfo {author} {\bibfnamefont {T.}~\bibnamefont {Martin}}, \ and\ \bibinfo
  {author} {\bibfnamefont {T.}~\bibnamefont {Jonckheere}},\ }\href {\doibase
  10.1103/PhysRevLett.103.147004} {\bibfield  {journal} {\bibinfo  {journal}
  {Physical Review Letters}\ }\textbf {\bibinfo {volume} {103}},\ \bibinfo
  {pages} {147004} (\bibinfo {year} {2009})},\ \Eprint
  {http://arxiv.org/abs/0909.3036} {arXiv:0909.3036} \BibitemShut {NoStop}%
\bibitem [{\citenamefont {Brunetti}\ \emph {et~al.}(2013)\citenamefont
  {Brunetti}, \citenamefont {Zazunov}, \citenamefont {Kundu},\ and\
  \citenamefont {Egger}}]{Brunetti2013}%
  \BibitemOpen
  \bibfield  {author} {\bibinfo {author} {\bibfnamefont {A.}~\bibnamefont
  {Brunetti}}, \bibinfo {author} {\bibfnamefont {A.}~\bibnamefont {Zazunov}},
  \bibinfo {author} {\bibfnamefont {A.}~\bibnamefont {Kundu}}, \ and\ \bibinfo
  {author} {\bibfnamefont {R.}~\bibnamefont {Egger}},\ }\href {\doibase
  10.1103/PhysRevB.88.144515} {\bibfield  {journal} {\bibinfo  {journal}
  {Physical Review B}\ }\textbf {\bibinfo {volume} {88}},\ \bibinfo {pages}
  {144515} (\bibinfo {year} {2013})},\ \Eprint {http://arxiv.org/abs/1305.3816}
  {arXiv:1305.3816} \BibitemShut {NoStop}%
\bibitem [{\citenamefont {Buzdin}\ and\ \citenamefont
  {Koshelev}(2003)}]{buzdin_koshelev.2003}%
  \BibitemOpen
  \bibfield  {author} {\bibinfo {author} {\bibfnamefont {A.~I.}\ \bibnamefont
  {Buzdin}}\ and\ \bibinfo {author} {\bibfnamefont {A.}~\bibnamefont
  {Koshelev}},\ }\href {\doibase 10.1103/PhysRevB.67.220504} {\bibfield
  {journal} {\bibinfo  {journal} {Physical Review B}\ }\textbf {\bibinfo
  {volume} {67}},\ \bibinfo {pages} {220504} (\bibinfo {year}
  {2003})}\BibitemShut {NoStop}%
\bibitem [{\citenamefont {Goldobin}\ \emph {et~al.}(2007)\citenamefont
  {Goldobin}, \citenamefont {Koelle}, \citenamefont {Kleiner},\ and\
  \citenamefont {Buzdin}}]{goldobin_koelle_kleiner_buzdin.2007}%
  \BibitemOpen
  \bibfield  {author} {\bibinfo {author} {\bibfnamefont {E.}~\bibnamefont
  {Goldobin}}, \bibinfo {author} {\bibfnamefont {D.}~\bibnamefont {Koelle}},
  \bibinfo {author} {\bibfnamefont {R.}~\bibnamefont {Kleiner}}, \ and\
  \bibinfo {author} {\bibfnamefont {A.~I.}\ \bibnamefont {Buzdin}},\ }\href
  {\doibase 10.1103/PhysRevB.76.224523} {\bibfield  {journal} {\bibinfo
  {journal} {Physical Review B}\ }\textbf {\bibinfo {volume} {76}},\ \bibinfo
  {pages} {224523} (\bibinfo {year} {2007})}\BibitemShut {NoStop}%
\bibitem [{\citenamefont {Goldobin}\ \emph {et~al.}(2011)\citenamefont
  {Goldobin}, \citenamefont {Koelle}, \citenamefont {Kleiner},\ and\
  \citenamefont {Mints}}]{Goldobin2011}%
  \BibitemOpen
  \bibfield  {author} {\bibinfo {author} {\bibfnamefont {E.}~\bibnamefont
  {Goldobin}}, \bibinfo {author} {\bibfnamefont {D.}~\bibnamefont {Koelle}},
  \bibinfo {author} {\bibfnamefont {R.}~\bibnamefont {Kleiner}}, \ and\
  \bibinfo {author} {\bibfnamefont {R.~G.}\ \bibnamefont {Mints}},\ }\href
  {\doibase 10.1103/PhysRevLett.107.227001} {\bibfield  {journal} {\bibinfo
  {journal} {Physical Review Letters}\ }\textbf {\bibinfo {volume} {107}},\
  \bibinfo {pages} {227001} (\bibinfo {year} {2011})},\ \Eprint
  {http://arxiv.org/abs/1110.2326} {arXiv:1110.2326} \BibitemShut {NoStop}%
\bibitem [{\citenamefont {Sickinger}\ \emph {et~al.}(2012)\citenamefont
  {Sickinger}, \citenamefont {Lipman}, \citenamefont {Weides}, \citenamefont
  {Mints}, \citenamefont {Kohlstedt}, \citenamefont {Koelle}, \citenamefont
  {Kleiner},\ and\ \citenamefont {Goldobin}}]{Sickinger2012}%
  \BibitemOpen
  \bibfield  {author} {\bibinfo {author} {\bibfnamefont {H.}~\bibnamefont
  {Sickinger}}, \bibinfo {author} {\bibfnamefont {A.}~\bibnamefont {Lipman}},
  \bibinfo {author} {\bibfnamefont {M.}~\bibnamefont {Weides}}, \bibinfo
  {author} {\bibfnamefont {R.~G.}\ \bibnamefont {Mints}}, \bibinfo {author}
  {\bibfnamefont {H.}~\bibnamefont {Kohlstedt}}, \bibinfo {author}
  {\bibfnamefont {D.}~\bibnamefont {Koelle}}, \bibinfo {author} {\bibfnamefont
  {R.}~\bibnamefont {Kleiner}}, \ and\ \bibinfo {author} {\bibfnamefont
  {E.}~\bibnamefont {Goldobin}},\ }\href {\doibase
  10.1103/PhysRevLett.109.107002} {\bibfield  {journal} {\bibinfo  {journal}
  {Physical Review Letters}\ }\textbf {\bibinfo {volume} {109}},\ \bibinfo
  {pages} {107002} (\bibinfo {year} {2012})},\ \Eprint
  {http://arxiv.org/abs/1207.3013} {arXiv:1207.3013} \BibitemShut {NoStop}%
\bibitem [{\citenamefont {Mints}(1998)}]{Mints1998}%
  \BibitemOpen
  \bibfield  {author} {\bibinfo {author} {\bibfnamefont {R.}~\bibnamefont
  {Mints}},\ }\href {\doibase 10.1103/PhysRevB.57.R3221} {\bibfield  {journal}
  {\bibinfo  {journal} {Physical Review B}\ }\textbf {\bibinfo {volume} {57}},\
  \bibinfo {pages} {R3221} (\bibinfo {year} {1998})}\BibitemShut {NoStop}%
\bibitem [{\citenamefont {Buzdin}(2005{\natexlab{b}})}]{buzdin.2005.J}%
  \BibitemOpen
  \bibfield  {author} {\bibinfo {author} {\bibfnamefont {A.~I.}\ \bibnamefont
  {Buzdin}},\ }\href {\doibase 10.1051/jp4:2005131052} {\bibfield  {journal}
  {\bibinfo  {journal} {Journal de Physique IV (Proceedings)}\ }\textbf
  {\bibinfo {volume} {131}},\ \bibinfo {pages} {213} (\bibinfo {year}
  {2005}{\natexlab{b}})}\BibitemShut {NoStop}%
\bibitem [{\citenamefont {Buzdin}(2005{\natexlab{c}})}]{buzdin.2005}%
  \BibitemOpen
  \bibfield  {author} {\bibinfo {author} {\bibfnamefont {A.~I.}\ \bibnamefont
  {Buzdin}},\ }\href {\doibase 10.1103/PhysRevB.72.100501} {\bibfield
  {journal} {\bibinfo  {journal} {Physical Review B}\ }\textbf {\bibinfo
  {volume} {72}},\ \bibinfo {pages} {100501} (\bibinfo {year}
  {2005}{\natexlab{c}})}\BibitemShut {NoStop}%
\bibitem [{\citenamefont {Konschelle}(2014{\natexlab{b}})}]{Konschelle2014}%
  \BibitemOpen
  \bibfield  {author} {\bibinfo {author} {\bibfnamefont {F.}~\bibnamefont
  {Konschelle}},\ }\href {\doibase 10.1140/epjb/e2014-50143-0} {\bibfield
  {journal} {\bibinfo  {journal} {The European Physical Journal B}\ }\textbf
  {\bibinfo {volume} {87}},\ \bibinfo {pages} {119} (\bibinfo {year}
  {2014}{\natexlab{b}})},\ \Eprint {http://arxiv.org/abs/1403.1797}
  {arXiv:1403.1797} \BibitemShut {NoStop}%
\bibitem [{\citenamefont {Rashba}(2003)}]{Rashba2003}%
  \BibitemOpen
  \bibfield  {author} {\bibinfo {author} {\bibfnamefont {E.}~\bibnamefont
  {Rashba}},\ }\href {\doibase 10.1103/PhysRevB.68.241315} {\bibfield
  {journal} {\bibinfo  {journal} {Physical Review B}\ }\textbf {\bibinfo
  {volume} {68}},\ \bibinfo {pages} {241315} (\bibinfo {year} {2003})},\
  \Eprint {http://arxiv.org/abs/cond-mat/0311110} {arXiv:cond-mat/0311110}
  \BibitemShut {NoStop}%
\bibitem [{\citenamefont {Tokatly}(2008)}]{Tokatly2008b}%
  \BibitemOpen
  \bibfield  {author} {\bibinfo {author} {\bibfnamefont {I.~V.}\ \bibnamefont
  {Tokatly}},\ }\href {\doibase 10.1103/PhysRevLett.101.106601} {\bibfield
  {journal} {\bibinfo  {journal} {Physical Review Letters}\ }\textbf {\bibinfo
  {volume} {101}},\ \bibinfo {pages} {106601} (\bibinfo {year} {2008})},\
  \Eprint {http://arxiv.org/abs/0802.1350} {arXiv:0802.1350} \BibitemShut
  {NoStop}%
\bibitem [{\citenamefont {Agterberg}\ and\ \citenamefont
  {Kaur}(2007)}]{Agterberg2006}%
  \BibitemOpen
  \bibfield  {author} {\bibinfo {author} {\bibfnamefont {D.}~\bibnamefont
  {Agterberg}}\ and\ \bibinfo {author} {\bibfnamefont {R.}~\bibnamefont
  {Kaur}},\ }\href {\doibase 10.1103/PhysRevB.75.064511} {\bibfield  {journal}
  {\bibinfo  {journal} {Physical Review B}\ }\textbf {\bibinfo {volume} {75}},\
  \bibinfo {pages} {064511} (\bibinfo {year} {2007})},\ \Eprint
  {http://arxiv.org/abs/cond-mat/0612216} {arXiv:cond-mat/0612216} \BibitemShut
  {NoStop}%
\bibitem [{\citenamefont {Mal'shukov}\ \emph {et~al.}(2005)\citenamefont
  {Mal'shukov}, \citenamefont {Wang}, \citenamefont {Chu},\ and\ \citenamefont
  {Chao}}]{Malshukov2005}%
  \BibitemOpen
  \bibfield  {author} {\bibinfo {author} {\bibfnamefont {A.~G.}\ \bibnamefont
  {Mal'shukov}}, \bibinfo {author} {\bibfnamefont {L.~Y.}\ \bibnamefont
  {Wang}}, \bibinfo {author} {\bibfnamefont {C.~S.}\ \bibnamefont {Chu}}, \
  and\ \bibinfo {author} {\bibfnamefont {K.~A.}\ \bibnamefont {Chao}},\ }\href
  {\doibase 10.1103/PhysRevLett.95.146601} {\bibfield  {journal} {\bibinfo
  {journal} {Physical Review Letters}\ }\textbf {\bibinfo {volume} {95}},\
  \bibinfo {pages} {146601} (\bibinfo {year} {2005})},\ \Eprint
  {http://arxiv.org/abs/cond-mat/0506724} {arXiv:cond-mat/0506724} \BibitemShut
  {NoStop}%
\bibitem [{\citenamefont {Stanescu}\ and\ \citenamefont
  {Galitski}(2007)}]{Stanescu2007}%
  \BibitemOpen
  \bibfield  {author} {\bibinfo {author} {\bibfnamefont {T.~D.}\ \bibnamefont
  {Stanescu}}\ and\ \bibinfo {author} {\bibfnamefont {V.}~\bibnamefont
  {Galitski}},\ }\href {\doibase 10.1103/PhysRevB.75.125307} {\bibfield
  {journal} {\bibinfo  {journal} {Physical Review B}\ }\textbf {\bibinfo
  {volume} {75}},\ \bibinfo {pages} {125307} (\bibinfo {year} {2007})},\
  \Eprint {http://arxiv.org/abs/cond-mat/0611165} {arXiv:cond-mat/0611165}
  \BibitemShut {NoStop}%
\bibitem [{\citenamefont {Duckheim}\ \emph {et~al.}(2009)\citenamefont
  {Duckheim}, \citenamefont {Maslov},\ and\ \citenamefont
  {Loss}}]{Duckheim2009}%
  \BibitemOpen
  \bibfield  {author} {\bibinfo {author} {\bibfnamefont {M.}~\bibnamefont
  {Duckheim}}, \bibinfo {author} {\bibfnamefont {D.~L.}\ \bibnamefont
  {Maslov}}, \ and\ \bibinfo {author} {\bibfnamefont {D.}~\bibnamefont
  {Loss}},\ }\href {\doibase 10.1103/PhysRevB.80.235327} {\bibfield  {journal}
  {\bibinfo  {journal} {Physical Review B}\ }\textbf {\bibinfo {volume} {80}},\
  \bibinfo {pages} {235327} (\bibinfo {year} {2009})},\ \Eprint
  {http://arxiv.org/abs/0909.1892} {arXiv:0909.1892} \BibitemShut {NoStop}%
\bibitem [{\citenamefont {Fr\"{o}hlich}\ and\ \citenamefont
  {Studer}(1993)}]{Frohlich1993}%
  \BibitemOpen
  \bibfield  {author} {\bibinfo {author} {\bibfnamefont {J.}~\bibnamefont
  {Fr\"{o}hlich}}\ and\ \bibinfo {author} {\bibfnamefont {U.}~\bibnamefont
  {Studer}},\ }\href {\doibase 10.1103/RevModPhys.65.733} {\bibfield  {journal}
  {\bibinfo  {journal} {Reviews of Modern Physics}\ }\textbf {\bibinfo {volume}
  {65}},\ \bibinfo {pages} {733} (\bibinfo {year} {1993})}\BibitemShut
  {NoStop}%
\bibitem [{\citenamefont {Berche}\ and\ \citenamefont
  {Medina}(2013)}]{Berche2013}%
  \BibitemOpen
  \bibfield  {author} {\bibinfo {author} {\bibfnamefont {B.}~\bibnamefont
  {Berche}}\ and\ \bibinfo {author} {\bibfnamefont {E.}~\bibnamefont
  {Medina}},\ }\href {\doibase 10.1088/0143-0807/34/1/161} {\bibfield
  {journal} {\bibinfo  {journal} {European Journal of Physics}\ }\textbf
  {\bibinfo {volume} {34}},\ \bibinfo {pages} {161} (\bibinfo {year} {2013})},\
  \Eprint {http://arxiv.org/abs/1210.0105} {arXiv:1210.0105} \BibitemShut
  {NoStop}%
\bibitem [{\citenamefont {Golubov}\ \emph {et~al.}(2004)\citenamefont
  {Golubov}, \citenamefont {Kupriyanov},\ and\ \citenamefont
  {Il'ichev}}]{golubov_kupriyanov.2004}%
  \BibitemOpen
  \bibfield  {author} {\bibinfo {author} {\bibfnamefont {A.}~\bibnamefont
  {Golubov}}, \bibinfo {author} {\bibfnamefont {M.}~\bibnamefont {Kupriyanov}},
  \ and\ \bibinfo {author} {\bibfnamefont {E.}~\bibnamefont {Il'ichev}},\
  }\href {\doibase 10.1103/RevModPhys.76.411} {\bibfield  {journal} {\bibinfo
  {journal} {Reviews of Modern Physics}\ }\textbf {\bibinfo {volume} {76}},\
  \bibinfo {pages} {411} (\bibinfo {year} {2004})}\BibitemShut {NoStop}%
\bibitem [{\citenamefont {Mineev}\ and\ \citenamefont
  {Samokhin}(2008)}]{Mineev2008}%
  \BibitemOpen
  \bibfield  {author} {\bibinfo {author} {\bibfnamefont {V.~P.}\ \bibnamefont
  {Mineev}}\ and\ \bibinfo {author} {\bibfnamefont {K.~V.}\ \bibnamefont
  {Samokhin}},\ }\href {\doibase 10.1103/PhysRevB.78.144503} {\bibfield
  {journal} {\bibinfo  {journal} {Physical Review B}\ }\textbf {\bibinfo
  {volume} {78}},\ \bibinfo {pages} {7} (\bibinfo {year} {2008})},\ \Eprint
  {http://arxiv.org/abs/0807.3021} {arXiv:0807.3021} \BibitemShut {NoStop}%
\bibitem [{\citenamefont {Houzet}\ and\ \citenamefont
  {Meyer}(2015)}]{Houzet2015}%
  \BibitemOpen
  \bibfield  {author} {\bibinfo {author} {\bibfnamefont {M.}~\bibnamefont
  {Houzet}}\ and\ \bibinfo {author} {\bibfnamefont {J.~S.}\ \bibnamefont
  {Meyer}},\ }\href {http://arxiv.org/abs/1502.04504} {\bibfield  {journal}
  {\bibinfo  {journal} {arXiv}\ } (\bibinfo {year} {2015})},\ \Eprint
  {http://arxiv.org/abs/1502.04504} {arXiv:1502.04504} \BibitemShut {NoStop}%
\bibitem [{\citenamefont {Eilenberger}(1968)}]{eilenberger.1968}%
  \BibitemOpen
  \bibfield  {author} {\bibinfo {author} {\bibfnamefont {G.}~\bibnamefont
  {Eilenberger}},\ }\href {\doibase 10.1007/BF01379803} {\bibfield  {journal}
  {\bibinfo  {journal} {Zeitschrift f\"{u}r Physik}\ }\textbf {\bibinfo
  {volume} {214}},\ \bibinfo {pages} {195} (\bibinfo {year}
  {1968})}\BibitemShut {NoStop}%
\bibitem [{\citenamefont {Larkin}\ and\ \citenamefont
  {Ovchinnikov}(1969)}]{larkin_ovchinnikov.1969}%
  \BibitemOpen
  \bibfield  {author} {\bibinfo {author} {\bibfnamefont {A.~I.}\ \bibnamefont
  {Larkin}}\ and\ \bibinfo {author} {\bibfnamefont {Y.~N.}\ \bibnamefont
  {Ovchinnikov}},\ }\href@noop {} {\bibfield  {journal} {\bibinfo  {journal}
  {Sov. Phys. JETP}\ }\textbf {\bibinfo {volume} {28}},\ \bibinfo {pages}
  {1200} (\bibinfo {year} {1969})}\BibitemShut {NoStop}%
\bibitem [{\citenamefont {Usadel}(1970)}]{usadel.1970}%
  \BibitemOpen
  \bibfield  {author} {\bibinfo {author} {\bibfnamefont {K.~D.}\ \bibnamefont
  {Usadel}},\ }\href {\doibase 10.1103/PhysRevLett.25.507} {\bibfield
  {journal} {\bibinfo  {journal} {Physical Review Letters}\ }\textbf {\bibinfo
  {volume} {25}},\ \bibinfo {pages} {507} (\bibinfo {year} {1970})}\BibitemShut
  {NoStop}%
\bibitem [{\citenamefont {Kopnin}(2001)}]{b.kopnin}%
  \BibitemOpen
  \bibfield  {author} {\bibinfo {author} {\bibfnamefont {N.~B.}\ \bibnamefont
  {Kopnin}},\ }\href@noop {} {\emph {\bibinfo {title} {{Theory of
  nonequilibrium superconductivity}}}}\ (\bibinfo  {publisher} {Oxford
  {U}niversity {P}ress},\ \bibinfo {year} {2001})\BibitemShut {NoStop}%
\bibitem [{\citenamefont {Gorini}\ \emph {et~al.}(2010)\citenamefont {Gorini},
  \citenamefont {Schwab}, \citenamefont {Raimondi},\ and\ \citenamefont
  {Shelankov}}]{Gorini2010}%
  \BibitemOpen
  \bibfield  {author} {\bibinfo {author} {\bibfnamefont {C.}~\bibnamefont
  {Gorini}}, \bibinfo {author} {\bibfnamefont {P.}~\bibnamefont {Schwab}},
  \bibinfo {author} {\bibfnamefont {R.}~\bibnamefont {Raimondi}}, \ and\
  \bibinfo {author} {\bibfnamefont {A.~L.}\ \bibnamefont {Shelankov}},\ }\href
  {\doibase 10.1103/PhysRevB.82.195316} {\bibfield  {journal} {\bibinfo
  {journal} {Physical Review B}\ }\textbf {\bibinfo {volume} {82}},\ \bibinfo
  {pages} {195316} (\bibinfo {year} {2010})},\ \Eprint
  {http://arxiv.org/abs/1003.5763} {arXiv:1003.5763} \BibitemShut {NoStop}%
\bibitem [{\citenamefont {Langenberg}\ and\ \citenamefont
  {Larkin}(1986)}]{Langenberg1986}%
  \BibitemOpen
  \bibfield  {author} {\bibinfo {author} {\bibfnamefont {D.~N.}\ \bibnamefont
  {Langenberg}}\ and\ \bibinfo {author} {\bibfnamefont {A.~I.}\ \bibnamefont
  {Larkin}},\ }\href@noop {} {\emph {\bibinfo {title} {{Nonequilibrium
  superconductivity}}}}\ (\bibinfo  {publisher} {North-Holland},\ \bibinfo
  {year} {1986})\BibitemShut {NoStop}%
\bibitem [{\citenamefont {Mal'shukov}\ and\ \citenamefont
  {Chu}(2008)}]{Malshukov2008}%
  \BibitemOpen
  \bibfield  {author} {\bibinfo {author} {\bibfnamefont {A.}~\bibnamefont
  {Mal'shukov}}\ and\ \bibinfo {author} {\bibfnamefont {C.}~\bibnamefont
  {Chu}},\ }\href {\doibase 10.1103/PhysRevB.78.104503} {\bibfield  {journal}
  {\bibinfo  {journal} {Physical Review B}\ }\textbf {\bibinfo {volume} {78}},\
  \bibinfo {pages} {104503} (\bibinfo {year} {2008})},\ \Eprint
  {http://arxiv.org/abs/0801.4419} {arXiv:0801.4419} \BibitemShut {NoStop}%
\bibitem [{\citenamefont {Kupriyanov}\ and\ \citenamefont
  {Lukichev}(1988)}]{Kupriyanov1988}%
  \BibitemOpen
  \bibfield  {author} {\bibinfo {author} {\bibfnamefont {M.~Y.}\ \bibnamefont
  {Kupriyanov}}\ and\ \bibinfo {author} {\bibfnamefont {V.~F.}\ \bibnamefont
  {Lukichev}},\ }\href
  {http://www.jetp.ac.ru/cgi-bin/e/index/e/67/6/p1163?a=list} {\bibfield
  {journal} {\bibinfo  {journal} {Sov. Phys. JETP}\ }\textbf {\bibinfo {volume}
  {67}},\ \bibinfo {pages} {1163} (\bibinfo {year} {1988})}\BibitemShut
  {NoStop}%
\bibitem [{\citenamefont {Gor'kov}\ and\ \citenamefont
  {Rashba}(2001)}]{Gorkov2001}%
  \BibitemOpen
  \bibfield  {author} {\bibinfo {author} {\bibfnamefont {L.~P.}\ \bibnamefont
  {Gor'kov}}\ and\ \bibinfo {author} {\bibfnamefont {E.}~\bibnamefont
  {Rashba}},\ }\href {\doibase 10.1103/PhysRevLett.87.037004} {\bibfield
  {journal} {\bibinfo  {journal} {Physical Review Letters}\ }\textbf {\bibinfo
  {volume} {87}},\ \bibinfo {pages} {37004} (\bibinfo {year}
  {2001})}\BibitemShut {NoStop}%
\bibitem [{\citenamefont {Sochnikov}\ \emph {et~al.}(2014)\citenamefont
  {Sochnikov}, \citenamefont {Maier}, \citenamefont {Watson}, \citenamefont
  {Kirtley}, \citenamefont {Gould}, \citenamefont {Tkachov}, \citenamefont
  {Hankiewicz}, \citenamefont {Br\"{u}ne}, \citenamefont {Buhmann},
  \citenamefont {Molenkamp},\ and\ \citenamefont {Moler}}]{Sochnikov2014}%
  \BibitemOpen
  \bibfield  {author} {\bibinfo {author} {\bibfnamefont {I.}~\bibnamefont
  {Sochnikov}}, \bibinfo {author} {\bibfnamefont {L.}~\bibnamefont {Maier}},
  \bibinfo {author} {\bibfnamefont {C.~A.}\ \bibnamefont {Watson}}, \bibinfo
  {author} {\bibfnamefont {J.~R.}\ \bibnamefont {Kirtley}}, \bibinfo {author}
  {\bibfnamefont {C.}~\bibnamefont {Gould}}, \bibinfo {author} {\bibfnamefont
  {G.}~\bibnamefont {Tkachov}}, \bibinfo {author} {\bibfnamefont {E.~M.}\
  \bibnamefont {Hankiewicz}}, \bibinfo {author} {\bibfnamefont
  {C.}~\bibnamefont {Br\"{u}ne}}, \bibinfo {author} {\bibfnamefont
  {H.}~\bibnamefont {Buhmann}}, \bibinfo {author} {\bibfnamefont {L.~W.}\
  \bibnamefont {Molenkamp}}, \ and\ \bibinfo {author} {\bibfnamefont {K.~A.}\
  \bibnamefont {Moler}},\ }\href {\doibase 10.1103/PhysRevLett.114.066801}
  {\bibfield  {journal} {\bibinfo  {journal} {Physical Review Letters}\
  }\textbf {\bibinfo {volume} {114}},\ \bibinfo {pages} {066801} (\bibinfo
  {year} {2014})},\ \Eprint {http://arxiv.org/abs/1410.1111} {arXiv:1410.1111}
  \BibitemShut {NoStop}%
\bibitem [{\citenamefont {Bauer}\ \emph {et~al.}(2003)\citenamefont {Bauer},
  \citenamefont {Bentner}, \citenamefont {Aprili}, \citenamefont {{Della
  Rocca}}, \citenamefont {Reinwald}, \citenamefont {Wegscheider},\ and\
  \citenamefont {Strunk}}]{bauer_bentner_aprili_etal.2004}%
  \BibitemOpen
  \bibfield  {author} {\bibinfo {author} {\bibfnamefont {A.}~\bibnamefont
  {Bauer}}, \bibinfo {author} {\bibfnamefont {J.}~\bibnamefont {Bentner}},
  \bibinfo {author} {\bibfnamefont {M.}~\bibnamefont {Aprili}}, \bibinfo
  {author} {\bibfnamefont {M.~L.}\ \bibnamefont {{Della Rocca}}}, \bibinfo
  {author} {\bibfnamefont {M.}~\bibnamefont {Reinwald}}, \bibinfo {author}
  {\bibfnamefont {W.}~\bibnamefont {Wegscheider}}, \ and\ \bibinfo {author}
  {\bibfnamefont {C.}~\bibnamefont {Strunk}},\ }\href {\doibase
  10.1103/PhysRevLett.92.217001} {\bibfield  {journal} {\bibinfo  {journal}
  {Physical Review Letters}\ }\textbf {\bibinfo {volume} {92}},\ \bibinfo
  {pages} {4} (\bibinfo {year} {2003})},\ \Eprint
  {http://arxiv.org/abs/cond-mat/0312165} {arXiv:cond-mat/0312165} \BibitemShut
  {NoStop}%
\bibitem [{\citenamefont {Buzdin}(2009)}]{Buzdin2009}%
  \BibitemOpen
  \bibfield  {author} {\bibinfo {author} {\bibfnamefont {A.}~\bibnamefont
  {Buzdin}},\ }\href {\doibase 10.1088/1742-6596/150/5/052029} {\bibfield
  {journal} {\bibinfo  {journal} {Journal of Physics: Conference Series}\
  }\textbf {\bibinfo {volume} {150}},\ \bibinfo {pages} {052029} (\bibinfo
  {year} {2009})}\BibitemShut {NoStop}%
\bibitem [{\citenamefont {Serrier-Garcia}\ \emph {et~al.}(2013)\citenamefont
  {Serrier-Garcia}, \citenamefont {Cuevas}, \citenamefont {Cren}, \citenamefont
  {Brun}, \citenamefont {Cherkez}, \citenamefont {Debontridder}, \citenamefont
  {Fokin}, \citenamefont {Bergeret},\ and\ \citenamefont
  {Roditchev}}]{Serrier-Garcia2013}%
  \BibitemOpen
  \bibfield  {author} {\bibinfo {author} {\bibfnamefont {L.}~\bibnamefont
  {Serrier-Garcia}}, \bibinfo {author} {\bibfnamefont {J.~C.}\ \bibnamefont
  {Cuevas}}, \bibinfo {author} {\bibfnamefont {T.}~\bibnamefont {Cren}},
  \bibinfo {author} {\bibfnamefont {C.}~\bibnamefont {Brun}}, \bibinfo {author}
  {\bibfnamefont {V.}~\bibnamefont {Cherkez}}, \bibinfo {author} {\bibfnamefont
  {F.}~\bibnamefont {Debontridder}}, \bibinfo {author} {\bibfnamefont
  {D.}~\bibnamefont {Fokin}}, \bibinfo {author} {\bibfnamefont {F.~S.}\
  \bibnamefont {Bergeret}}, \ and\ \bibinfo {author} {\bibfnamefont
  {D.}~\bibnamefont {Roditchev}},\ }\href {\doibase
  10.1103/PhysRevLett.110.157003} {\bibfield  {journal} {\bibinfo  {journal}
  {Physical Review Letters}\ }\textbf {\bibinfo {volume} {110}},\ \bibinfo
  {pages} {157003} (\bibinfo {year} {2013})},\ \Eprint
  {http://arxiv.org/abs/1401.8102} {arXiv:1401.8102} \BibitemShut {NoStop}%
\bibitem [{\citenamefont {Konschelle}\ and\ \citenamefont
  {Buzdin}(2009)}]{Konschelle2009b}%
  \BibitemOpen
  \bibfield  {author} {\bibinfo {author} {\bibfnamefont {F.}~\bibnamefont
  {Konschelle}}\ and\ \bibinfo {author} {\bibfnamefont {A.~I.}\ \bibnamefont
  {Buzdin}},\ }\href {\doibase 10.1103/PhysRevLett.102.017001} {\bibfield
  {journal} {\bibinfo  {journal} {Physical Review Letters}\ }\textbf {\bibinfo
  {volume} {102}},\ \bibinfo {pages} {017001} (\bibinfo {year} {2009})},\
  \Eprint {http://arxiv.org/abs/0810.4286} {arXiv:0810.4286} \BibitemShut
  {NoStop}%
\end{thebibliography}%

\end{document}